\documentclass[superscriptaddress,prc,twocolumn,showpacs,preprintnumbers,amsmath,amssymb,bibnotes,secnumarabic,altaffilletter,floatfix,aps,10pt]{revtex4-1}

\usepackage{dcolumn}
\usepackage{bm}
\usepackage{hyperref}
\usepackage[usenames]{color}
\usepackage[pdftex]{graphicx}
\usepackage{epstopdf}
\usepackage{subfigure}
\usepackage{units} 

\def\nue{$\nu_e$}
\def\nuebar{$\bar{\nu}_e$}
\def\numu{$\nu_\mu$}
\def\nutau{$\nu_\tau$}
\def\nuone{$\nu_1$}
\def\nutwo{$\nu_2$}
\def\nuthree{$\nu_3$}

\def\heavywater{D$_2$O}
\def\water {H$_2$O}
\newcommand{\iso}[2]{{}^{#1}{\rm #2}}

\def\hep{\textit{hep} neutrino}
\def\B{$\iso{8}{B}$ neutrino}
\def\fB{\Phi_{\rm B}}
\newcommand{\Pes}{P_{es}(E_\nu)}
\newcommand{\Peed}{P_{ee}^{\rm d}(E_\nu)}
\newcommand{\Peen}{P_{ee}^{\rm n}(E_\nu)}

\newcommand{\Aee}{A_{ee}(E_\nu)}
\newcommand{\Peea}{c_0}
\newcommand{\Peeb}{c_1}
\newcommand{\Peec}{c_2}
\newcommand{\Aeea}{a_0}
\newcommand{\Aeeb}{a_1}

\newcommand{\chis}{$\chi^2$}
\newcommand{\pa}{$p_a$}
\newcommand{\pb}{$p_b$}
\newcommand{\pc}{$p_c$}
\newcommand{\pd}{$p_d$}
\newcommand{\npsa}{\epsilon_{\rm PID}}

\newcommand{\flux}{\times10^6\,{\rm cm^{-2}s^{-1}}}
\newcommand{\thetaonetwo}{\theta_{12}}
\newcommand{\thetaonethree}{\theta_{13}}
\newcommand{\thetatwothree}{\theta_{23}}
\newcommand{\Dmonetwo}{\Delta m^2_{21}}
\newcommand{\Dmtwothree}{\Delta m^2_{32}}
\newcommand{\Dmonethree}{\Delta m^2_{31}}
\newcommand{\tanthetaonetwo}{\tan^2\thetaonetwo{}}
\newcommand{\sinthetaonethree}{\sin^2\thetaonethree{}}
\newcommand{\teff}{T_{\rm eff}}
\newcommand{\Encd}{E_{\rm NCD}}
\newcommand{\cts}{\cos\theta_\odot}
\newcommand{\ncapPMT}{\epsilon_{n}^{\rm PMT}}
\newcommand{\ncapNCD}{\epsilon_{n}^{\rm NCD}}
\newcommand{\ncdEscale}{$a_1^{\textrm{NCDE}}$}
\newcommand{\ncdEres}{$b_0^{\textrm{NCDE}}$}
\newcommand{\be}{\beta_{14}}

\newcommand{\errorStatSys}[3]{#1\pm #2{\rm (stat.)}\pm #3{\rm (syst.)}}

\newcommand{\errorStatASys}[4]{#1\pm #2{\rm (stat.)}^{+#3}_{-#4}{\rm (syst.)}}

\newcommand{\errorA}[3]{#1^{+#2}_{-#3}}

\newcommand{\numberSNOBflux}{$(\errorStatASys{5.25}{0.16}{0.11}{0.13})\flux$}
\newcommand{\numberSNOThetaonetwo}{$\tanthetaonetwo = \errorA{0.427}{0.033}{0.029}$}
\newcommand{\numberSNODmonetwo}{$\Dmonetwo = (\errorA{5.6}{1.9}{1.4})\times10^{-5}\,{\rm eV^2}$}
\newcommand{\numberPeea}{$\Peea{} = \errorStatSys{0.317}{0.016}{0.009}$}

\newcommand{\numberGlobalThetaonetwo}{$\tanthetaonetwo = \errorA{0.446}{0.030}{0.029}$}
\newcommand{\numberGlobalThetaonethree}{$\sinthetaonethree = (\errorA{2.5}{1.8}{1.5})\times10^{-2}$}
\newcommand{\numberGlobalThetaonethreeLimit}{$\sinthetaonethree < 0.053$}
\newcommand{\numberGlobalDmonetwo}{$\Dmonetwo = (\errorA{7.41}{0.21}{0.19})\times10^{-5}\,{\rm eV^2}$}

\newcommand{\bpsFlux}{$(5.88\pm0.65)\flux$}
\newcommand{\bpsAGSflux}{$(4.85\pm0.58)\flux$}
\newcommand{\BPSfive}{$5.69\flux$}

\newcommand{\bEndEnergy}{$15\,{\rm MeV}$}
\newcommand{\hepEndEnergy}{$18.8\,{\rm MeV}$}

\begin{document}

\title{Combined Analysis of all Three Phases of Solar Neutrino Data from the Sudbury Neutrino Observatory}


%
\newcommand{\alta}{Department of Physics, University of 
Alberta, Edmonton, Alberta, T6G 2R3, Canada}
\newcommand{\ubc}{Department of Physics and Astronomy, University of 
British Columbia, Vancouver, BC V6T 1Z1, Canada}
\newcommand{\bnl}{Chemistry Department, Brookhaven National 
Laboratory,  Upton, NY 11973-5000}
\newcommand{\carleton}{Ottawa-Carleton Institute for Physics, Department of Physics, Carleton University, Ottawa, Ontario K1S 5B6, Canada}
\newcommand{\uog}{Physics Department, University of Guelph,  
Guelph, Ontario N1G 2W1, Canada}
\newcommand{\lu}{Department of Physics and Astronomy, Laurentian 
University, Sudbury, Ontario P3E 2C6, Canada}
\newcommand{\lbnl}{Institute for Nuclear and Particle Astrophysics and 
Nuclear Science Division, Lawrence Berkeley National Laboratory, Berkeley, CA 94720}
\newcommand{\lbla}{ Lawrence Berkeley National Laboratory, Berkeley, CA}
\newcommand{\lanl}{Los Alamos National Laboratory, Los Alamos, NM 87545}
\newcommand{\llnl}{Lawrence Livermore National Laboratory, Livermore, CA}
\newcommand{\lanla}{Los Alamos National Laboratory, Los Alamos, NM 87545}
\newcommand{\oxford}{Department of Physics, University of Oxford, 
Denys Wilkinson Building, Keble Road, Oxford OX1 3RH, UK}
\newcommand{\penn}{Department of Physics and Astronomy, University of 
Pennsylvania, Philadelphia, PA 19104-6396}
\newcommand{\queens}{Department of Physics, Queen's University, 
Kingston, Ontario K7L 3N6, Canada}
\newcommand{\uw}{Center for Experimental Nuclear Physics and Astrophysics, 
and Department of Physics, University of Washington, Seattle, WA 98195}
\newcommand{\uwa}{Center for Experimental Nuclear Physics and Astrophysics, 
and Department of Physics, University of Washington, Seattle, WA}
\newcommand{\uta}{Department of Physics, University of Texas at Austin, Austin, TX 78712-0264}
\newcommand{\triumf}{TRIUMF, 4004 Wesbrook Mall, Vancouver, BC V6T 2A3, Canada}
\newcommand{\ralimp}{Rutherford Appleton Laboratory, Chilton, Didcot OX11 0QX, UK}
\newcommand{\iusb}{Department of Physics and Astronomy, Indiana University, South Bend, IN}
\newcommand{\fnal}{Fermilab, Batavia, IL}
\newcommand{\uo}{Department of Physics and Astronomy, University of Oregon, Eugene, OR}
\newcommand{\hu}{Department of Physics, Hiroshima University, Hiroshima, Japan}
\newcommand{\slac}{Stanford Linear Accelerator Center, Menlo Park, CA}
\newcommand{\mac}{Department of Physics, McMaster University, Hamilton, ON}
\newcommand{\doe}{US Department of Energy, Germantown, MD}
\newcommand{\lund}{Department of Physics, Lund University, Lund, Sweden}
\newcommand{\mpi}{Max-Planck-Institut for Nuclear Physics, Heidelberg, Germany}
\newcommand{\uom}{Ren\'{e} J.A. L\'{e}vesque Laboratory, Universit\'{e} de Montr\'{e}al, Montreal, PQ}
\newcommand{\cwru}{Department of Physics, Case Western Reserve University, Cleveland, OH}
\newcommand{\pnnl}{Pacific Northwest National Laboratory, Richland, WA}
\newcommand{\uc}{Department of Physics, University of Chicago, Chicago, IL}
\newcommand{\mitt}{Laboratory for Nuclear Science, Massachusetts Institute of Technology, Cambridge, MA 02139}
\newcommand{\ucsd}{Department of Physics, University of California at San Diego, La Jolla, CA }
\newcommand{	\lsu	}{Department of Physics and Astronomy, Louisiana State University, Baton Rouge, LA 70803}
\newcommand{\imp}{Imperial College, London, UK}
\newcommand{\uci}{Department of Physics, University of California, Irvine, CA 92717}
\newcommand{\ucia}{Department of Physics, University of California, Irvine, CA}
\newcommand{\suss}{Department of Physics and Astronomy, University of Sussex, Brighton, UK}
\newcommand{\lifep}{Laborat\'{o}rio de Instrumenta\c{c}\~{a}o e F\'{\i}sica Experimental de
Part\'{\i}culas, Av. Elias Garcia 14, 1$^{\circ}$, 1000-149 Lisboa, Portugal}
\newcommand{\hku}{Department of Physics, The University of Hong Kong, Hong Kong.}
\newcommand{\aecl}{Atomic Energy of Canada, Limited, Chalk River Laboratories, Chalk River, ON K0J 1J0, Canada}
\newcommand{\nrc}{National Research Council of Canada, Ottawa, ON K1A 0R6, Canada}
\newcommand{\princeton}{Department of Physics, Princeton University, Princeton, NJ}
\newcommand{\birkbeck}{Birkbeck College, University of London, Malet Road, London WC1E 7HX, UK}
\newcommand{\snoi}{SNOLAB, Sudbury, ON P3Y 1M3, Canada}
\newcommand{\uba}{University of Buenos Aires, Argentina}
\newcommand{\hvd}{Department of Physics, Harvard University, Cambridge, MA}
\newcommand{\pny}{Goldman Sachs, 85 Broad Street, New York, NY}
\newcommand{\pnv}{Remote Sensing Lab, PO Box 98521, Las Vegas, NV 89193}
\newcommand{\psis}{Paul Schiffer Institute, Villigen, Switzerland}
\newcommand{\liverpool}{Department of Physics, University of Liverpool, Liverpool, UK}
\newcommand{\uto}{Department of Physics, University of Toronto, Toronto, ON, Canada}
\newcommand{\uwisc}{Department of Physics, University of Wisconsin, Madison, WI}
\newcommand{\psu}{Department of Physics, Pennsylvania State University,
     University Park, PA}
\newcommand{\anl}{Deparment of Mathematics and Computer Science, Argonne
     National Laboratory, Lemont, IL}
\newcommand{\cornell}{Department of Physics, Cornell University, Ithaca, NY}
\newcommand{\tufts}{Department of Physics and Astronomy, Tufts University, Medford, MA}
\newcommand{\ucd}{Department of Physics, University of California, Davis, CA}
\newcommand{\unc}{Department of Physics, University of North Carolina, Chapel Hill, NC}
\newcommand{\dresden}{Institut f\"{u}r Kern- und Teilchenphysik, Technische Universit\"{a}t Dresden, Dresden, Germany}
\newcommand{\isu}{Department of Physics, Idaho State University, Pocatello, ID}
\newcommand{\qmul}{Dept. of Physics, Queen Mary University, London, UK}
\newcommand{\ucsb}{Dept. of Physics, University of California, Santa Barbara, CA}
\newcommand{\cern}{CERN, Geneva, Switzerland}
\newcommand{\utah}{Dept. of Physics, University of Utah, Salt Lake City, UT}
\newcommand{\casa}{Center for Astrophysics and Space Astronomy, University of Colorado, Boulder, CO}
\newcommand{\susel}{Sanford Laboratory at Homestake, Lead, SD}  
\newcommand{\ntu}{Center of Cosmology and Particle Astrophysics, National Taiwan University, Taiwan}
\newcommand{\berlin}{Institute for Space Sciences, Freie Universit\"{a}t Berlin,
Leibniz-Institute of Freshwater Ecology and Inland Fisheries, Germany}
\newcommand{\bhsu}{Black Hills State University, Spearfish, SD} 
\newcommand{\queensa}{Dept.\,of Physics, Queen's University, 
Kingston, Ontario, Canada} 
\newcommand{\aasu}{Dept.\,of Chemistry and Physics, Armstrong Atlantic State University, Savannah, GA}
\newcommand{\ucb}{Physics Department, University of California at Berkeley, and Lawrence Berkeley National Laboratory, Berkeley, CA}
\newcommand{\mcgill}{Physics Department, McGill University, Montreal, QC, Canada}
\newcommand{\columbia}{Columbia University, New York, NY}
\newcommand{\rhul}{Dept. of Physics, Royal Holloway University of London, Egham, Surrey, UK}
\newcommand{\ubama}{Department of Physics and Astronomy, University of Alabama, Tuscaloosa, AL}
\newcommand{\kit}{Instit\"{u}t f\"{u}r Experimentelle Kernphysik, Karlsruher Instit\"{u}t f\"{u}r Technologie, Karlsruhe, Germany}
\newcommand{\saclay}{CEA-Saclay, DSM/IRFU/SPP, Gif-sur-Yvette, France }


\affiliation{\alta}
\affiliation{\ubc}
\affiliation{\bnl}
\affiliation{\carleton}
\affiliation{\uog}
\affiliation{\lu}
\affiliation{\lbnl}
\affiliation{\lifep}
\affiliation{\lanl}
\affiliation{\lsu}
\affiliation{\mitt}
\affiliation{\oxford}
\affiliation{\penn}
\affiliation{\queens}
\affiliation{\ralimp}
\affiliation{\snoi}
\affiliation{\uta}
\affiliation{\triumf}
\affiliation{\uw}

\author{B.~Aharmim}\affiliation{\lu}
\author{S.\,N.~Ahmed}\affiliation{\queens}
\author{A.\,E.~Anthony}\altaffiliation{Present address: \casa}\affiliation{\uta}
\author{N.~Barros}\altaffiliation{Present address: \dresden}\affiliation{\lifep}
\author{E.\,W.~Beier}\affiliation{\penn}
\author{A.~Bellerive}\affiliation{\carleton}
\author{B.~Beltran}\affiliation{\alta}
\author{M.~Bergevin}\altaffiliation{Present address: \ucd}\affiliation{\lbnl}\affiliation{\uog}
\author{S.\,D.~Biller}\affiliation{\oxford}
\author{K.~Boudjemline}\affiliation{\carleton}\affiliation{\queens}
% ...
			
\collaboration{SNO Collaboration}
\noaffiliation



\begin{abstract}
We report results from a combined analysis of solar neutrino data from all phases of the Sudbury Neutrino Observatory. By exploiting particle identification information obtained from the proportional counters installed during the third phase, this analysis improved background rejection in that phase of the experiment. The combined analysis resulted in a total flux of active neutrino flavors from $\iso{8}{B}$ decays in the Sun of \numberSNOBflux{}. A two-flavor neutrino oscillation analysis yielded \numberSNODmonetwo{} and \numberSNOThetaonetwo{}. A three-flavor neutrino oscillation analysis combining this result with results of all other solar neutrino experiments and the KamLAND experiment yielded \numberGlobalDmonetwo{}, \numberGlobalThetaonetwo{}, and \numberGlobalThetaonethree{}. This implied an upper bound of \numberGlobalThetaonethreeLimit{} at the 95\% confidence level (C.L.).
\end{abstract}

\maketitle


\section{Introduction}
The Sudbury Neutrino Observatory (SNO) was designed to measure the flux of neutrinos produced by $\iso{8}{B}$ decays in the Sun, so-called \B{}s, and to study neutrino oscillations, as proposed by Herb Chen~\cite{cite:herb}. As a result of measurements with the SNO detector and other experiments, it is now well-established that neutrinos are massive and that the weak eigenstates (\nue{}, \numu{}, \nutau{}) are mixtures of the mass eigenstates (\nuone{}, \nutwo{}, \nuthree{}). The probability of detecting a neutrino in the same weak eigenstate in which it was created depends on the energy and propagation distance of the neutrino, the effects of matter~\cite{cite:wolf, cite:msw}, the neutrino mixing angles ($\thetaonetwo{}$, $\thetatwothree{}$, $\thetaonethree{}$), a phase ($\delta$) which can lead to charge-parity violation, and the differences between the squares of the neutrino mass eigenvalues ($\Dmonetwo{}$, $\Dmtwothree{}$, $\Dmonethree{}$)~\cite{MNS, Pontecorvo}.

The SNO detector observed \B{}s via three different reactions. By measuring the rate of neutral current (NC) reactions,
\begin{eqnarray}
\nu_x + d		&\rightarrow	&p + n + \nu_x,
\end{eqnarray}
which is equally sensitive to all three active neutrino flavors, the SNO experiment determined the total \B{} flux, $\fB{}$, independently of any specific active neutrino flavor oscillation hypothesis~\cite{cite:herb}. The predicted flux from solar model calculations~\cite{ssh:bps09} is \bpsFlux{}, BPS09(GS), or \bpsAGSflux{}, BPS09(AGSS09), using a recent measurement of the heavy-element abundance at the Sun's surface. Previous analyses of SNO data~\cite{cite:snoletaPaper, cite:snoncd} measured $\fB{}$ more precisely than the solar model predictions. A more precise measurement of $\fB{}$ would better constrain these solar models, but may not necessarily determine which metallicity is correct due to the large uncertainties at present on both predictions.

By measuring the rate of charged current (CC) reactions,
\begin{eqnarray}
\nu_e + d		&\rightarrow	&p + p + e^-,
\end{eqnarray}
which is only sensitive to \nue{}s, and comparing this to the NC reaction rate, it was possible to determine the neutrino survival probability as a function of energy. This can then constrain the neutrino oscillation parameters independently of any specific prediction of $\fB{}$.

The SNO experiment also measured the rate of elastic scattering (ES) reactions,
\begin{eqnarray}
\nu_x + e^-	&\rightarrow	&\nu_x + e^-,
\end{eqnarray}
which is sensitive to all neutrino flavors, but the cross-section for \nue{}s is approximately six times larger than that for the other flavors.

We present in this article a final combined analysis of all solar neutrino data from the SNO experiment. Section~\ref{sec:detector} gives an overview of the detector. In Section~\ref{sec:fit} we describe the method used to combine all the data in a fit which determines $\fB{}$ and a parameterized form of the \nue{} survival probability. Section~\ref{sec:phaseIII} describes a new particle identification technique that allowed us to significantly suppress the background events in the proportional counters used in the third phase of the SNO experiment. Section~\ref{sec:results} presents the results of the new analysis of data from Phase III, and the combined analysis of data from all phases. The results of this combined analysis are interpreted in the context of neutrino oscillations in Section~\ref{sec:oscillations}.


\section{The SNO Detector}
\label{sec:detector}
\begin{figure}[tbp]
\includegraphics[width=\columnwidth]{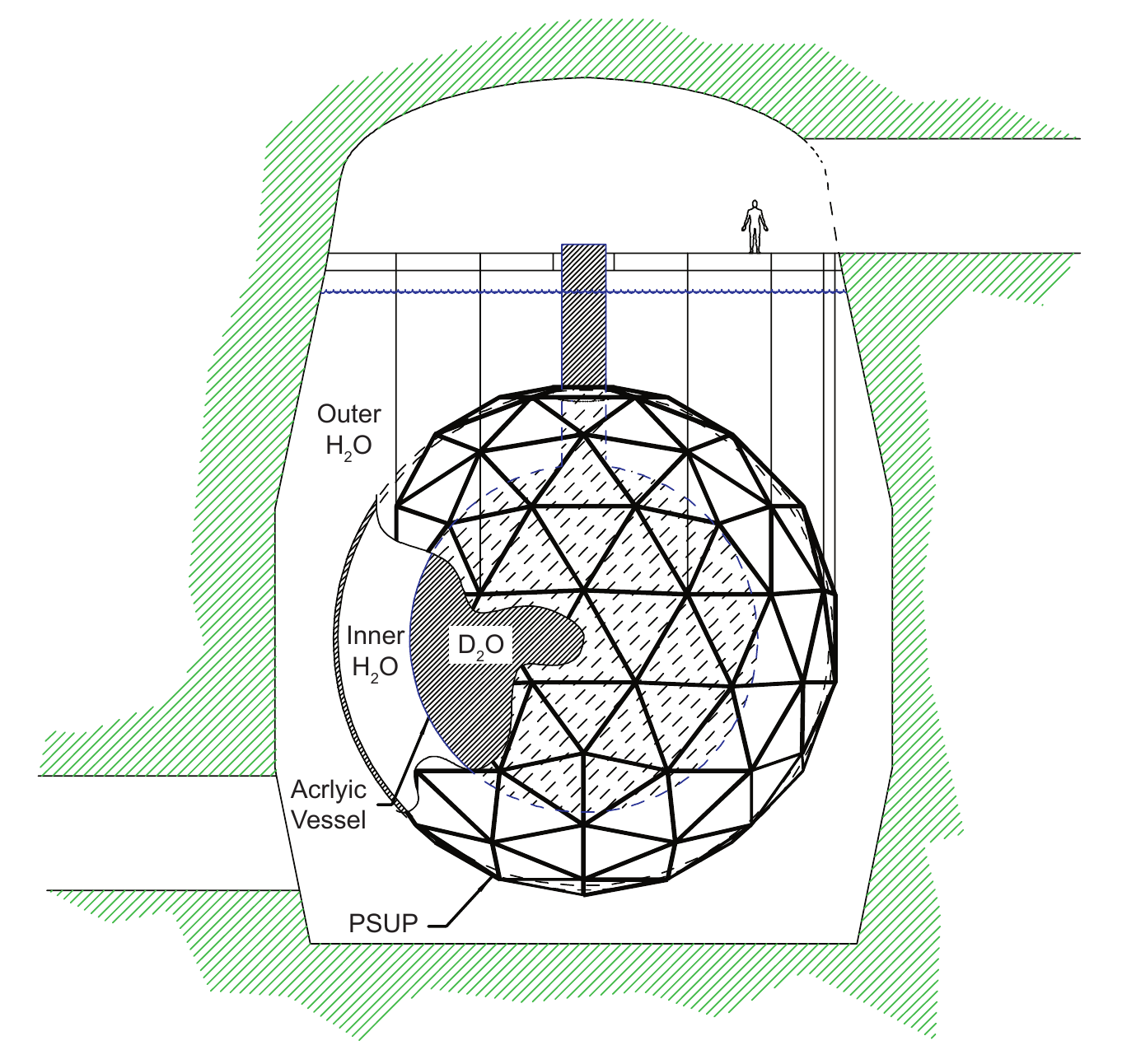}
\caption{Schematic diagram of the SNO detector. We used a coordinate system with the center of the detector as the origin, and $z$ direction as vertically upward.}
\label{fig:snodet}
\end{figure}

The SNO detector~\cite{cite:snonim}, shown schematically in Figure~\ref{fig:snodet}, consisted of an inner volume containing $10^6$\,kg of 99.92\% isotopically pure heavy water (${\rm ^2H_2O}$, hereafter referred to as \heavywater{}) within a 12 m diameter transparent acrylic vessel (AV). Over $7\times 10^6$\,kg of \water{} between the rock and the AV shielded the \heavywater{} from external radioactive backgrounds. An array of 9456 inward-facing 20 cm Hamamatsu R1408 photomultiplier tubes (PMTs), installed on an 17.8 m diameter stainless steel geodesic structure (PSUP), detected Cherenkov radiation produced in both the \heavywater{} and \water{}. A non-imaging light concentrator~\cite{Doucas1996579} mounted on each PMT increased the effective photocathode coverage to nearly 55\% of 4$\pi$. The PMT thresholds were set to 1/4 of the charge from a single photoelectron. The inner $1.7\times 10^6$\,kg of \water{} between the AV and the PSUP shielded the \heavywater{} against radioactive backgrounds from the PSUP and PMTs. Extensive purification systems removed radioactive isotopes from both the \heavywater{} and the \water{}~\cite{Aharmim2009531}.

The detector was located in Vale's Creighton mine ($46^\circ 28'30''$\,N latitude, $81^\circ 12'04''$\,W longitude) near Sudbury, Ontario, Canada, with the center of the detector at a depth of 2092\,m (5890$\pm$94 meters water equivalent). At this depth, the rate of cosmic-ray muons entering the detector was approximately three per hour. Ninety-one outward-facing PMTs attached to the PSUP detected cosmic-ray muons. An offline veto based on information from these PMTs significantly reduced cosmogenic backgrounds.

The recoil electrons from both the ES and CC reactions were detected directly through their production of Cherenkov light. The total amount of light detected by the PMT array was correlated with the energy of the interacting neutrino.

The SNO detector operated in three distinct phases distinguished by how the neutrons from the NC interactions were detected. In Phase I, the detected neutrons captured on deuterons in the \heavywater{} releasing a single 6.25 MeV $\gamma$-ray, and it was the Cherenkov light of secondary Compton electrons or $e^+e^-$ pairs that was detected. In Phase II, $2\times10^3\,{\rm kg}$ of NaCl were added to the \heavywater{}, and the neutrons captured predominantly on $\iso{35}{Cl}$ nuclei, which have a much larger neutron capture cross-section than deuterium nuclei, resulting in a higher neutron detection efficiency. Capture on chlorine also released more energy (8.6 MeV) and yielded multiple $\gamma$-rays, which aided in identifying neutron events. In Phase III, an array of proportional counters (the Neutral Current Detection, or NCD, array) was deployed in the \heavywater{}~\cite{cite:snoncdnim}.

The proportional counters were constructed of approximately 2 m long high purity nickel tubes welded together to form longer ``strings". Neutrons were detected via the reaction
\begin{eqnarray}
\iso{3}{He}+n \rightarrow \iso{3}{H} + p,
\end{eqnarray}
where the triton and proton had a total kinetic energy of 0.76\,MeV, and travelled in opposite directions. The NCD array consisted of 36 strings filled with $\iso{3}{He}$, and an additional 4 strings filled with $\iso{4}{He}$ that were insensitive to the neutron signals and were used to study backgrounds. Energetic charged particles within the proportional counters produced ionization electrons, and the induced voltage caused by these electrons was recorded as a function of time, referred to as a waveform. To increase the dynamic range, the waveform was logarithmically amplified before being digitized~\cite{cite:snoncdnim}. 


\section{Combined analysis}
\label{sec:fit}

In this article we present an analysis that combines data from all three phases of the SNO experiment. The combination accounts for any correlations in the systematic uncertainties between phases. The data were split into day and night sets in order to search for matter effects as the neutrinos propagated through the Earth.

The general form of the analysis was a fit to Monte Carlo-derived probability density functions (PDFs) for each of the possible signal and background types. As with previous analyses of SNO data, the following four variables were calculated for each event recorded with the PMT array: the effective electron kinetic energy, $\teff{}$, reconstructed under the hypothesis that the light was caused by a single electron; the cube of the radial position, $r$, divided by 600\,cm, $\rho=(\nicefrac{r{\rm [cm]}}{600})^3$; the isotropy of the detected light, $\be{}$; and the angle of the reconstructed electron propagation relative to the direction of the Sun, $\cts{}$. Different algorithms to calculate both $\teff{}$ and $\rho$ were used for the first two phases and the third phase. References~\cite{PhysRevC.72.055502,cite:snoletaPaper, ncdlongpaper} contain detailed descriptions of how these variables were calculated. The energy deposited in the gas of a proportional counter, $\Encd{}$, was calculated for each event recorded with the NCD array, and the correlated waveform was determined~\cite{ncdlongpaper}.

Although there were multiple sets of data in this fit, the result was a single $\fB{}$ and energy-dependent \nue{} survival probability as described in Section~\ref{sec:parameterization}. We summarize the event selection and backgrounds in Section~\ref{sec:backgrounds}. Sections~\ref{sec:PDFs} and \ref{sec:efficiencies}, respectively, describe the PDFs and efficiencies. The method for combining the multiple sets of data in a single analysis is presented in Section~\ref{sec:method3phase}. Finally, Section~\ref{sec:crosschecks} outlines the alternative analyses to verify the combined analysis.

\subsection{Parameterization of the {\boldmath \B{}} signal}
\label{sec:parameterization}

We fitted the neutrino signal based on an average $\fB{}$ for day and night, a \nue{} survival probability as a function of neutrino energy, $E_\nu$, during the day, $\Peed{}$, and an asymmetry between the day and night survival probabilities, $\Aee{}$, defined by
\begin{eqnarray}
\Aee{} = 2\frac{\Peen{}-\Peed{}}{\Peen{}+\Peed{}},
\end{eqnarray}
where $\Peen{}$ was the \nue{} survival probability during the night. This was the same parameterization as we used in our previous analysis of data from Phases I and II~\cite{cite:snoletaPaper}.

Monte Carlo simulations assuming the standard solar model and no neutrino oscillations were used to determine the event variables for \B{} interactions in the detector. These simulations were then scaled by the factors given in Table~\ref{tab:scalings}.

\begingroup
\squeezetable
\begin{table}[htdp]
\caption{\B{} interactions scaling factors.}
\begin{center}
\begin{tabular}{lll}
\hline\hline
Interaction&			Day/Night&	Scaling factor\\
\hline
$\rm {CC,ES_e}$&		Day&		$\fB{}\Peed{}$\\
$\rm {ES_{\mu\tau}}$&	Day&		$\fB{}[1-\Peed{}]$\\
$\rm {CC,ES_e}$&		Night&		$\fB{}\Peen{}$\\
$\rm {ES_{\mu\tau}}$&	Night&		$\fB{}[1-\Peen{}]$\\
$\rm {NC}$&			Day+Night&	$\fB{}$\\
\hline\hline
\end{tabular}
\end{center}
\label{tab:scalings}
\end{table}
\endgroup

Unlike our earlier publications~\cite{PhysRevLett.89.011302, PhysRevC.72.055502, cite:snoncd}, this parameterization included a constraint on the rate of ES interactions relative to the rate of CC interactions based on their relative cross-sections. It also had the advantage that the fitted parameters ($\fB{}$, $\Peed{}$, and $\Aee{}$) were all directly related to the scientific questions under investigation. Moreover, it disentangled the detector response from the fit result as $\Peed{}$ and $\Aee{}$ were functions of $E_\nu$ as opposed to $\teff{}$.

Appendix~\ref{apx:sterile} explains how this parameterization can be used to describe sterile neutrino models that do not predict any day/night asymmetry in the sterile neutrino flux and do not predict any distortion in the sterile $E_\nu$ spectrum~\footnote{In our previous analysis of data from Phases I and II~\cite{cite:snoletaPaper} we described this method as imposing a unitarity constraint, which was not technically correct.}. Reference~\cite{thesisPL} presents a very general sterile neutrino analysis that includes day versus night asymmetries.

Due to the broad $\teff{}$ resolution of the detector, $\Peed{}$ was not sensitive to sharp distortions and was parameterized by
\begin{eqnarray}
\label{eqn:peed}
\Peed{} & = & \Peea{} + \Peeb{} (E_\nu [{\rm MeV}] - 10)\\
\nonumber & &\; + \Peec{} (E_\nu [{\rm MeV}]- 10)^2,
\end{eqnarray}
where $\Peea{}$, $\Peeb{}$, and $\Peec{}$ were parameters defining the \nue{} survival probability. Simulations showed that the fit was not sensitive to higher order terms in the polynomial. Expanding the function around 10\,MeV, which corresponds approximately to the peak in the detectable \B{} $E_\nu$ spectrum, reduced correlations between $\Peea{}$, $\Peeb{}$, and $\Peec{}$. For the same reasons, $\Aee{}$ was parameterized by 
\begin{eqnarray}
\Aee{} & = & \Aeea{} + \Aeeb{}(E_\nu [{\rm MeV}] - 10),
\label{eqn:aee}
\end{eqnarray}
where $\Aeea{}$, and $\Aeeb{}$ were parameters defining the relative difference between the night and day \nue{} survival probability. By disallowing sharp changes in the neutrino signal that can mimic the background signal at low energies, these parameterizations reduced the covariances between the neutrino interaction and background rates.

To correctly handle ES events, we simulated $\nu_\mu$s with the same $E_\nu$ spectrum as \nue{}s, such that scaling factors for these interactions in Table~\ref{tab:scalings} were satisfied. In our previous analysis~\cite{cite:snoletaPaper} we approximated the $\nu_\mu$ and $\nu_\tau$ cross-section by 0.156 times the \nue{} cross-section, and then included an additional systematic uncertainty to account for the fact that the ratio of the \nue{} to $\nu_\mu$ ES cross-section is not constant as a function of $E_\nu$.

\subsection{Event selection and backgrounds}
\label{sec:backgrounds}

Table~\ref{tab:datasets} summarizes the data periods used in this analysis. We used the same periods of data as our most recent analyses of data from these phases~\cite{cite:snoletaPaper, cite:snoncd}.

\begingroup
\squeezetable
\begin{table}[htdp]
\caption{Dates for the data in the different phases used in this analysis.}
\begin{center}
\begin{tabular}{lllrr}
\hline\hline
Phase&	Start date&			End date&		\multicolumn{2}{c}{Total time [days]}\\
&		&					&				Day&	Night\\
\hline
I&		November 1999&		May 2001&		119.9&	157.4\\
II&		July 2001&			August 2003&		176.5&	214.9\\
III&		November 2004&		November 2006&	176.6&	208.6\\
\hline\hline
\end{tabular}
\end{center}
\label{tab:datasets}
\end{table}
\endgroup

Event cuts to select good candidates were identical to those in the previous analyses of these data~\cite{cite:snoletaPaper, cite:snoncd}. The following cuts on the event variables were applied: $\rho<\left(\nicefrac{550\,{\rm [cm]}}{600\,{\rm [cm]}} \right)^3=0.77025$, $-0.12<\be{}<0.95$, $3.5\,{\rm MeV} < \teff{}<20.0\,{\rm MeV}$ for Phases I and II, and $6.0\,{\rm MeV} < \teff{}<20.0\,{\rm MeV}$ for Phase III. After these cuts the data consisted of events from ES, CC, and NC interactions of \B{}s, and a number of different background sources.

Radioactive decays produced two main background types: ``electron-like" events, which resulted from $\beta$-particles or $\gamma$-rays with a total energy above our $\teff{}$ threshold, and neutrons produced by the photo-disintegration of deuterons by $\gamma$-rays with energies above 2.2\,MeV. During Phase III, only the neutron events were observed from radioactive background decays, due to the higher $\teff{}$ threshold for that phase.

The radioactive decays of $\iso{214}{Bi}$ and $\iso{208}{Tl}$ within the regions of the detector filled with \heavywater{} and \water{} were major sources of background events. $\iso{214}{Bi}$ is part of the $\iso{238}{U}$ decay chain, but it was most likely not in equilibrium with the early part of the decay chain. The most likely source of $\iso{214}{Bi}$ was from $\iso{222}{Rn}$ entering the \heavywater{} and \water{} from mine air. $\iso{208}{Tl}$ is part of the $\iso{232}{Th}$ decay chain. These sources of radiation produced both electron-like events and photo-disintegration neutrons. {\it Ex-situ} measurements~\cite{Andersen2003386, Andersen2003399} of background levels in the \heavywater{} and \water{} provided constraints on the rate of these decays, as given in Tables~\ref{tab:bckphasesIandII} and \ref{tab:bckphaseIII} of Appendix~\ref{apx:systematic}.

Background sources originating from the AV included decays of $\iso{208}{Tl}$ within the acrylic, which produced both electron-like events and photo-disintegration neutrons. In addition, radon progeny that accumulated on the surface of the AV during construction could create neutrons through ($\alpha$,n) reactions on isotopes of carbon and oxygen within the AV. Near the $\teff{}$ threshold in Phases I and II the majority of background events originated from radioactive decays within the PMTs.

Due to the dissolved NaCl in the \heavywater{} during Phase II, calibration sources that produced neutrons, and other sources of neutrons, led to the creation of $\iso{24}{Na}$ via neutron capture on $\iso{23}{Na}$. $\iso{24}{Na}$ decays with a half-life of approximately 15\,hours, producing a low energy $\beta$ particle and two $\gamma$-rays. One of these $\gamma$-rays has an energy of 2.75\,MeV, which could photo-disintegrate a deuteron. This resulted in additional electron-like events and photo-disintegration neutrons during Phase II. Periods after calibrations were removed from the data, but there were remaining backgrounds.

During Phase III there were additional photo-disintegration neutron backgrounds due to radioactivity in the nickel and cables of the NCD array, as well as two ``hotspots" on the strings referred to as K2 and K5.
The estimated number of these background events, given in Table~\ref{tab:bckphaseIII} of Appendix~\ref{apx:systematic}, were based on previous analyses of these data~\cite{ncdlongpaper} except for backgrounds from the K5 hotspot, which was based on a recent reanalysis~\cite{O'Keeffe:2011rg}. The previously estimated number of neutrons observed in the NCD array due to the K5 hotspot was $31.6\pm3.7$, which assumed there was Th and a small amount of U in the hotspot based on both {\it ex-situ} and {\it in-situ} measurements. Based on measurements performed on the strings after they were removed from the \heavywater{}, it was determined that the radioactivity was likely on the surface and most likely pure Th with very little U. This resulted in a new estimate of $45.5^{+7.5}_{-8.4}$ neutron background events observed in the NCD array from this hotspot.

Aside from the radioactive decay backgrounds, there were additional backgrounds to the \B{} measurement due to \nue{}s produced by the following reaction,
\begin{eqnarray}
^3{\rm He} + p	&\rightarrow	&^4{\rm He} + e^{+} + \nu_e,
\end{eqnarray}
in the Sun, so-called \hep{}s. These have a maximum energy of \hepEndEnergy{}, which is slightly above the \B{} maximum energy of \bEndEnergy{}, and the standard solar model (SSM) prediction for their flux is approximately one thousand times smaller than $\fB{}$~\cite{bs05}. Estimates of this background were based on the SSM prediction including the effects of neutrino oscillations obtained from previous analyses~\cite{cite:snoletaPaper}. There were instrumental backgrounds that reconstructed near the AV. Above $\teff{}\approx6\,{\rm MeV}$ these events formed a distinct peak at low values of $\be{}$, so they were easily removed by the cuts on $\be{}$ and $\rho$. At lower $\teff{}$, position reconstruction uncertainties increase, and the $\be{}$ distribution of these ``AV instrumental background" events broadens, resulting in incomplete removal by these cuts. This background was negligible in Phase III due to the higher $\teff{}$ threshold used for the analysis of data from that phase. Finally, there were also background events due to neutrinos produced by particle decays in the atmosphere. The estimated numbers for these background events, given in Tables~\ref{tab:bckphasesIandII} and \ref{tab:bckphaseIII} of Appendix~\ref{apx:systematic}, were based on previous analyses of these data~\cite{cite:snoletaPaper, ncdlongpaper}.

\subsection{PDFs}
\label{sec:PDFs}

For Phases I and II the event variables $\teff{}$, $\rho$, $\be{}$, and $\cts{}$ were used to construct 4-dimensional PDFs. For Phase III the reduced number of NC events observed with the PMT array made the $\be{}$ event variable unnecessary, so the PDFs were 3-dimensional in the remaining three event variables. Monte Carlo simulations were used to derive the PDFs for all signal and background classes observed with the PMT array except for backgrounds originating from radioactivity in the PMTs, which was described by an analytical function. Compared to the previous analysis of data from Phases I and II~\cite{cite:snoletaPaper}, we increased the number of Monte Carlo events for the CC and ES interactions by a factor of four, and for NC interactions and some background types by a factor of two.

The Monte Carlo simulation was verified using a variety of calibration sources. Based on these comparisons a number of systematic uncertainties were defined to represent possible variations in the event variables relative to the calibrations. In general these included differences in the offset, scale, and resolution for each of the event variables. Appendix~\ref{apx:systematic} gives the complete list of systematic uncertainties associated with the PDFs. Except where specified these uncertainties were the same as those used in the most recent analyses of these data~\cite{cite:snoletaPaper, ncdlongpaper}.

Extensive calibrations using a $\iso{16}{N}$ $\gamma$-ray source~\cite{Dragowsky2002284}, which produced electrons with kinetic energies of approximately 5.05\,MeV from Compton scattering, allowed us to calibrate $\teff{}$. In Phase I the linearity of $\teff{}$ with respect to electron kinetic energy was tested using a proton-triton fusion $\gamma$-ray source~\cite{Poon2000115}, which produced electrons with kinetic energies up to approximately 19.0\,MeV from Compton scattering and pair-production. Based on these sources, we parameterized the reconstructed electron kinetic energy including a possible non-linearity by
\begin{eqnarray}
\teff{}' &=&\teff{}\left (1+c_0^E\frac{\teff{}{\rm [MeV]}-5.05}{19.0-5.05}\right ),
\end{eqnarray}
where $c_0^E$ represents the level of non-linearity. The linearity in all phases was tested using the following two electron sources: a $\iso{8}{Li}$ calibration source~\cite{Tagg2002178} that produced electrons with a continuous distribution up to approximately 13\,MeV; and electrons with a continuous energy distribution up to approximately 50\,MeV produced by the decay of muons that stopped within the AV. These studies revealed non-linearities consistent with zero. We assumed any non-linearities below our level of sensitivity were correlated between all three phases, and we used a value of $c_0^E=0\pm0.0069$, which was equal to the value used in the previous analysis of data from Phases I and II~\cite{cite:snoletaPaper}.

During Phase III, $\iso{24}{Na}$Cl brine was injected into the \heavywater{} on two separate occasions~\cite{Boudjemline2010171}. The brine was thoroughly mixed in the \heavywater{} and provided a uniformly distributed source of $\gamma$-rays, allowing us to study possible $\teff{}$ variations in regions that were previously not sampled due to the restricted movement of the $\iso{16}{N}$ source. The observed variation in the event rate of $\iso{24}{Na}$ decays within the fiducial volume of solar neutrino analysis was consistent with what was allowed by the $\teff{}$ calibration parameters determined with the $\iso{16}{N}$ source at $\teff{}>6\,{\rm MeV}$.

While the intrinsic rate of radioactive backgrounds from solid bulk materials such as the acrylic vessel or PMT array were not expected to vary over the course of the experiment, variations in detector response make the detected rates vary over time, and because of differences in the livetime fractions between day and night, these variations were aliased into day/night differences. PDFs derived from Monte Carlo naturally include day/night detector asymmetries because the detector simulation tracks changes to the detector response. Our previous analysis derived the analytical PDFs for radioactivity originating from the PMTs using a bifurcated analysis of real data with the day and night data combined~\cite{cite:snoletaPaper}, which did not account for possible day/night asymmetries.

\begin{figure}[tbp]
\includegraphics[height=\columnwidth, angle=90]{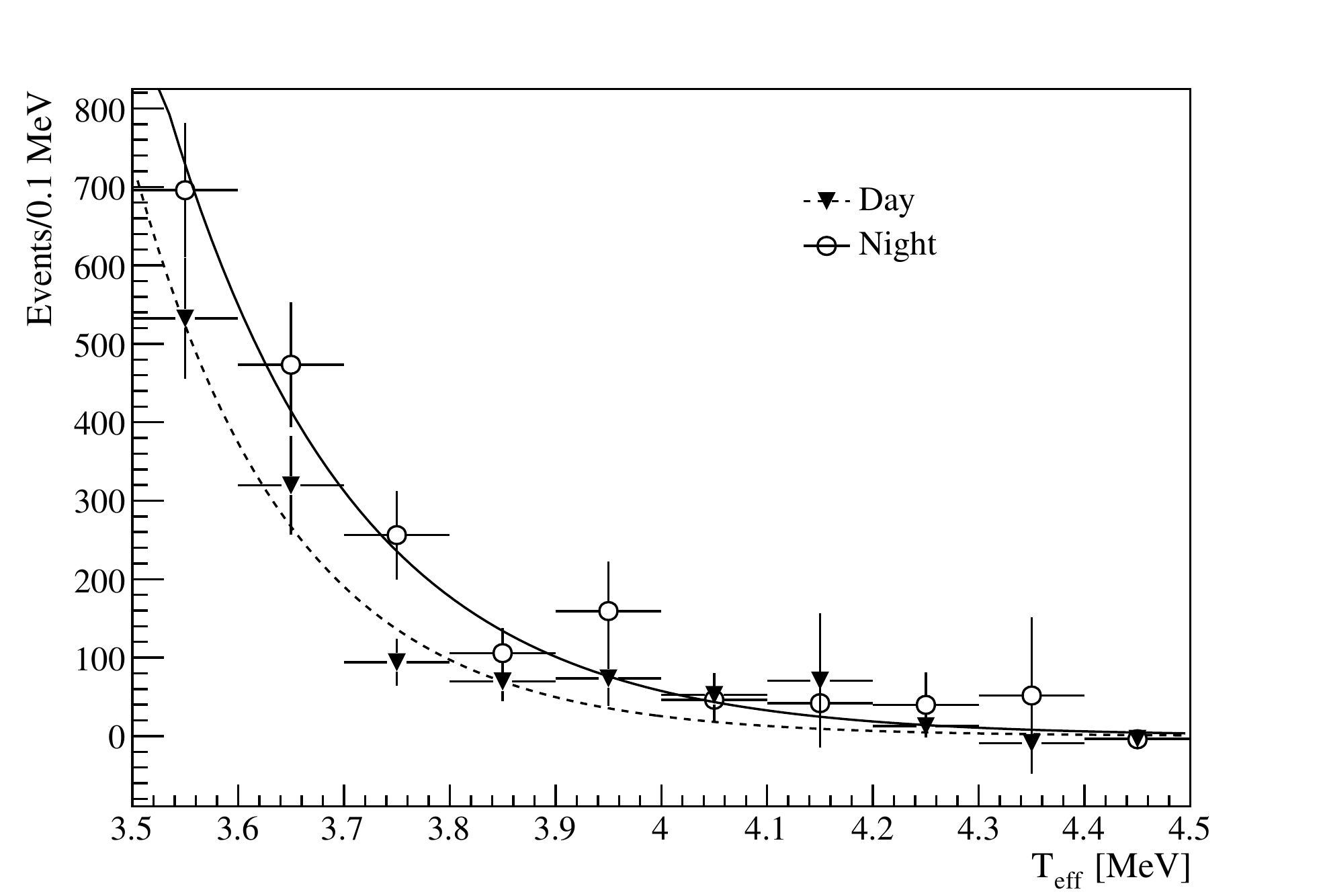}
\caption{$\teff{}$ spectra for the PMT background events obtained from a bifurcated analysis of data from Phase I including the best fits to Equation~\ref{eqn:PMTPDF}.}
\label{fig:PMTbetagamma}
\end{figure}

To accommodate such asymmetries in the present analysis we allowed different observed background rates between day and night, and we repeated the bifurcated analysis with the data separated into day and night sets. Similarly to the previous analysis~\cite{cite:snoletaPaper} the PDF was parameterized by the following function
\begin{eqnarray}
\nonumber P_{\rm PMT}(&&\teff{},\be{}, \rho) = e^{\epsilon\teff{}}\\
\nonumber &&\times \left( e^{\nu\rho}+|b+1|-1 \right)\\
&&\times \mathcal{N}\left(\be{}|\omega_0+\omega_1(\rho-0.61), \beta_s \right),
\label{eqn:PMTPDF}
\end{eqnarray}
where $\epsilon$, $\nu$, $b$, $\omega_0$, $\omega_1$, and $\beta_s$ were parameters determined from the fit to the bifurcated data. $\mathcal{N}(x|\bar{x},\sigma)$ represents a Gaussian distribution in $x$ with mean $\bar{x}$ and standard deviation $\sigma$. The uncertainties in the radial parameters were obtained from one dimensional scans of the likelihood function because the magnitude operator distorted the likelihood function at $b=-1$. Compared to the function used in the previous analysis~\cite{cite:snoletaPaper}, $\omega_1\rho$ was replaced with $\omega_1(\rho-0.61)$ to reduce the correlation between $\omega_0$ and $\omega_1$. Table~\ref{tab:PMTbetagamma} lists the PDF parameters from this analysis. We observed a weak day/night asymmetry in these results, in particular at roughly the 1$\sigma$ level in the $\teff{}$ spectrum. Figure~\ref{fig:PMTbetagamma} shows the fits to the $\teff{}$ spectrum for Phase I.

\begingroup
\squeezetable
\begin{table}[htdp]
\caption{PMT background PDF parameters as determined by a bifurcated analysis. $\rho_{\nu b}$ is the correlation between the $\nu$ and $b$ parameters.}
\begin{center}
\begin{tabular}{lrrrr}
\hline\hline
Parameter&			\multicolumn{4}{c}{Phase}\\
&			I-day&			I-night&			II-day&			II-night\\
\hline
$\epsilon$&		$-6.7\pm1.3$&	$-5.6\pm1.0$&	$-6.3\pm0.9$&	$-7.0\pm0.9$\\
$\nu$&			$6.6\pm0.9$&	$6.8\pm1.5$&	$5.3\pm1.0$&	$5.7\pm1.1$\\
$b$&			$-1.0\pm1.3$&	$3.3\pm12.0$&	$-0.3\pm2.1$&	$0.5\pm3.0$\\
$\rho_{\nu b}$&	0.60&			0.96&			0.91&			0.94\\
\hline\hline
\end{tabular}
\end{center}
\label{tab:PMTbetagamma}
\end{table}
\endgroup

The $E_\nu$ spectrum of \B{}s used to derive the PDFs for ES, CC, and NC interactions was obtained, including the uncertainties, from Reference~\cite{b8winter}. Our previous analysis of data from Phases I and II~\cite{cite:snoletaPaper} included this uncertainty for the CC and ES PDFs. In this analysis we also included the effects of this uncertainty on the normalization of the NC rate.

\subsection{Efficiencies}
\label{sec:efficiencies}

Table~\ref{tab:effSys} in Appendix~\ref{apx:systematic} lists the uncertainties associated with neutron detection. The majority of these were identical to previous analyses of these data~\cite{cite:snoletaPaper, ncdlongpaper} except where indicated below.

We recently published~\cite{ncdlongpaper} an analysis based on calibration data from Phase III that determined that the fraction of neutrons created in NC interactions that were detectable with the PMT array, $\ncapPMT{},$ was $0.0502\pm0.0014$. The previous analysis of data from Phase III~\cite{cite:snoncd} used $\ncapPMT{}=0.0485\pm0.0006$, which relied on a Monte Carlo based method to determine the uncertainty on the neutron detection efficiency. This new analysis avoids the dependence on Monte Carlo simulations.

Similarly, a calibration based method was used to calculate the fraction of neutrons created by NC interactions that were captured in the gas of the NCD array, $\ncapNCD{}$. This yielded $\ncapNCD{}=0.211\pm0.005$~\cite{ncdlongpaper}, which had slightly better precision than the value of $\ncapNCD{}=0.211\pm0.007$ used in the previous analysis of data from Phase III~\cite{cite:snoncd}. We multiplied $\ncapNCD{}$ by a correction factor of $0.862 \pm 0.004$ in order to determine the efficiency for detecting NC interactions in the NCD array. The correction factor, averaged over the duration of Phase III, included the mean live fraction of the signal processing systems, threshold efficiencies, and signal acceptance due to event selection cuts.

This analysis corrected a 1.2\% error in the normalization of the number of NC events observed in the PMT array that was in the previous analysis of data from Phase III~\cite{cite:snoncd}. Because the majority of NC events were observed in the NCD array, this normalization error had a relatively small effect on the measured NC flux. In addition we have corrected a 0.1\% error in the deuteron density used for that analysis.

\subsection{Description of the fit}
\label{sec:method3phase}

The combined fit to all phases was performed using the maximum likelihood technique, where the negative log likelihood function was minimized using MINUIT~\cite{James:1975dr}. The events observed in the PMT and NCD arrays were uncorrelated, therefore the negative log likelihood function for all data was given by
\begin{eqnarray}
-\log{L_{\rm data}} = -\log{L_{\rm PMT}} -\log{L_{\rm NCD}},
\end{eqnarray}
where $L_{\rm PMT}$ and $L_{\rm NCD}$, respectively, were the likelihood functions for the events observed in the PMT and NCD arrays.

The negative log likelihood function in the PMT array was given by
\begin{eqnarray}
\nonumber -\log{L_{\rm PMT}} &=& \sum_{j=1}^{N}\lambda_{j}(\vec{\eta})\\
&&- \sum_{i=1}^{n_{\rm PMT}}\log \left[\sum_{j=1}^{N}\lambda_{j}(\vec{\eta}) f(\vec{x}_i|j,\vec{\eta})\right],
\end{eqnarray}
where $N$ was the number of different event classes, $\vec{\eta}$ was a vector of ``nuisance" parameters associated with the systematic uncertainties, $\lambda_{j}(\vec{\eta})$ was the mean of a Poisson distribution for the $j^{\rm th}$ class, $\vec{x}_i$ was the vector of event variables for event $i$, $n_{\rm PMT}$ was the total number of events in the PMT array during the three phases, and $f(\vec{x}_i|j,\vec{\eta})$ was the PDF for events of type $j$.

The PDFs for the signal events were re-weighted based on Equations~\ref{eqn:peed} and \ref{eqn:aee}. This was a CPU intensive task that was prohibitive for the kernel based PDFs used in the previous analysis of data from Phases I and II~\cite{cite:snoletaPaper}. Therefore, in that analysis, the PDFs were also binned based on $E_\nu$, and these PDFs were then weighted by the integral of Equations~\ref{eqn:peed} and \ref{eqn:aee} within that bin. This analysis did not require this approximation when calculating the best fit, although an approximation was used when ``scanning" (described below) the systematic uncertainties. As described in Reference~\cite{thesisPL}, $\lambda(\vec{\eta})$ was re-parameterized such that the Monte Carlo-based PDFs did not need to be normalized.

In the previous analysis of the PMT array data from Phase III~\cite{cite:snoncd}, the nuisance parameters were only propagated on the PDFs for neutrino interactions, while this analysis also propagates these parameters on the PDFs for background events. Because the number of background events observed in the PMT array was small relative to the number of neutrino events, this had a relatively minor effect on that result.

The negative log likelihood function in the NCD array was given by
\begin{eqnarray}
-\log{L_{\rm NCD}} &= & \frac{1}{2}\left (\frac{\sum_{j=1}^{N}\nu_{j}(\vec{\eta}) - n_{\rm NCD}}{\sigma_{\rm NCD}}\right )^2,
\end{eqnarray}
where $\nu_{j}(\vec{\eta})$ was the mean of a Poisson distribution for the $j^{\rm th}$ class, $n_{\rm NCD}$ was the total number of neutrons observed in the NCD array based on the fit described in Section~\ref{sec:encdfit}, and $\sigma_{\rm NCD}$ was the associated uncertainty.

The mean number of events for a given class was often related to the mean number of events for another class; for instance, the number of ES, CC, and NC events in each phase were determined from the parameters in Section~\ref{sec:parameterization}.

Constraints were placed on various nuisance parameters and the rate of certain classes of background events. Including these constraints, the negative log likelihood function was given by 
\begin{eqnarray}
-\log{L} &=& -\log{L_{\rm data}} + \frac{1}{2}(\vec{\eta}-\vec{\mu})^{\rm T}{\mathbf \Sigma^{-1}}(\vec{\eta}-\vec{\mu}),
\end{eqnarray}
where $\vec{\eta}$ was the value of the nuisance parameters, $\vec{\mu}$ was the constraint on the parameters, and $\mathbf \Sigma$ was the covariance matrix for the constraints. Appendix~\ref{apx:systematic} lists all of the constraints.

Our previous analysis of data from Phases I and II~\cite{cite:snoletaPaper} imposed a physical bound on background rates, so that they were not allowed to become negative in the fit. Without these bounds the background from neutrons originating from the AV in Phase II favors a rate whose central value was negative, but consistent with zero. The uncertainty in the PDFs due to the finite Monte Carlo statistics could explain the fitted negative value. The previously reported ensemble tests~\cite{cite:snoletaPaper} used a central value for this background that was more than two statistical standard deviations above zero, such that no significant effect from the bound was observed. Using a positive bound for the backgrounds when ensemble testing with Monte Carlo data that does not contain any neutrons originating from the AV in Phase II tends to shift $\fB{}$ down on average compared to the flux used to simulate the data, as we only obtained background rates that were equal to or higher than the values used in the simulations. Removing this bound allowed closer agreement between the expectation values for the signal parameters and the values used in the simulated data. We therefore removed these bounds in this analysis to provide a more frequentist approach that facilitates the combination of the SNO results with other experiments.

In the previous analysis of data from Phases I and II~\cite{cite:snoletaPaper} the background constraints obtained for the average of the day and night rate (e.g. for the {\it ex-situ} measurements of $\iso{214}{Bi}$ and $\iso{208}{Tl}$) were applied independently to both the day and the night rates, which resulted in a narrower constraint on these backgrounds than we intended. This analysis correctly applies this as a day and night averaged constraint.

Because of the large number of constraints, and the time involved in modifying some PDFs, there were three methods for handling the nuisance parameters. Some were ``floated," i.e. allowed to vary within the MINUIT minimization of the negative log likelihood function. Others were ``scanned," where a series of fits were performed with different values of the parameter in order to find the best fit. This process was repeated for all scanned nuisance parameters multiple times to converge on the global minimum of the fit. Finally, some were ``shifted-and-refitted," where the central values of the fit parameters were not varied, but the effect of the nuisance parameter on the uncertainties was calculated. The method used for each parameter depended on the size of that constraints effect on $\fB{}$ and the \nue{} survival probability parameters. Appendix~\ref{apx:systematic} lists how each nuisance parameter was treated.

In addition to the systematic uncertainties considered in previous analyses, this analysis also included a systematic uncertainty due to the finite Monte Carlo statistics used to construct the PDFs. We performed 1000 independent fits in which the number of events in each bin of the PDF were drawn from a Poisson distribution. The uncertainty due to the finite Monte Carlo statistics was determined from the distribution of the fit parameters.

In order to calculate the total systematic uncertainty on the $\fB{}$ and the \nue{} survival probability parameters, we applied the shift-and-refit 100 times for each parameter in order to calculate the asymmetrical likelihood distribution for that parameter. We then performed one million fits with the nuisance parameters drawn randomly from these distributions. The total systematic uncertainty was obtained from the resulting distribution of the fit results. This is the first time we have applied this procedure, which correctly accounts for the combination of asymmetrical uncertainties. In order to calculate the effects of the day/night or MC systematic uncertainties, respectively, this procedure was repeated with only the nuisance parameter related to day/night differences or MC statistics varied.

In total the fit included $\fB{}$, the five \nue{} survival probability parameters described in Section~\ref{sec:parameterization}, 36 background rate parameters, 35 floated or scanned nuisance parameters, and 82 shift-and-refit nuisance parameters.

The biases and uncertainties obtained from this analysis method were tested using simulated data. The number of simulated sets of data was restricted by the amount of Monte Carlo data available. For simulated data containing neutrino interactions and some background classes, we did bias tests with 250 sets of data. For simulated data containing neutrino interactions and all background classes, we did bias tests with 14 sets of data. All of these tests showed the method was unbiased and produced uncertainties consistent with frequentist statistics.

\subsection{Crosschecks}
\label{sec:crosschecks}

As a crosscheck on the analysis method described above, we developed two independent analyses. The first crosscheck compared the results from the above method run only on data from Phases I and II. This was crosschecked against the method described in the previous analysis of this data~\cite{cite:snoletaPaper} with the most significant changes from this analysis included in that previous analysis. The results from the two methods were in agreement.

We developed an alternate Bayesian based analysis where the posterior probability distribution was sampled using a Markov Chain Monte Carlo (MCMC). This analysis was applied to data from Phase III, using the results from the maximum likelihood analysis performed on data from Phases I and II as a prior. The priors for background and neutrino interaction rates had zero probability for negative rates and were uniform for positive rates. There were two important differences between this alternate analysis and the maximum likelihood method described above. Firstly, because the systematic uncertainties were varied in each step of the MCMC, the uncertainties included all systematic and statistical uncertainties. Secondly, this method samples the entire posterior probability distribution instead of identifying the maximum likelihood. Reference~\cite{thesisShahnoor} provides details of this method. As shown in Section~\ref{sec:results} the results of the Bayesian and maximum likelihood fits agreed. An alternate Bayesian analysis was also performed with details provided in Reference~\cite{thesisRich}. Though this analysis was not used as a direct crosscheck, its results were consistent with the analysis presented here.


\section{NCD array analysis}
\label{sec:phaseIII}

The NCD array observed neutrons, alphas, and events caused by instrumental backgrounds. Because of their low stopping power in the gas of the proportional counters, electrons rarely triggered the NCD array. A series of cuts described in Reference~\cite{ncdlongpaper} removed the instrumental backgrounds. For neutron events $\Encd{}$ was peaked at approximately 0.76\,MeV, with a maximum energy of 0.85\,MeV when including the resolution. $\Encd{}$ was less than 0.76\,MeV if the proton or triton hit the nickel walls before losing all their energy. We identified the following two major categories of alpha events: so-called bulk alphas, which came from radioactive decays occurring throughout the nickel bodies of the proportional counters due to the presence of U and Th and their progeny, and so-called surface alphas, coming from the decay of $\iso{210}{Po}$, which was supported by $\iso{210}{Pb}$ that had plated onto the inner surface of the nickel bodies. Below 1.4\,MeV both types of alpha events produced relatively flat $\Encd{}$ spectra. Due to differences in construction of the strings, the number of alpha events varied from string-to-string.

The previous analysis of data from Phase III~\cite{cite:snoncd} distinguished between neutron and alpha events by fitting the $\Encd{}$ spectrum. The PDF of $\Encd{}$ for neutron events was obtained from calibration data, and for alpha events it was obtained from simulations. Between 0.4\,MeV and 1.4\,MeV the fitted number of alpha and neutron events, respectively, were approximately 5560 and 1170. The large number of alpha events resulted in both a large statistical uncertainty, and a large systematic uncertainty due to difficulties in accurately determining the PDF of $\Encd{}$ for alpha events. 

The waveforms of neutron events could be significantly broader than those from alpha events, depending on the orientation of the proton-triton trajectory. This distinction was lessened by the significant tail in the waveforms caused by the long ion drift times. In an attempt to reduce the number of alpha events, and therefore the uncertainties associated with them, we developed four different particle identification (PID) parameters and a cut on these parameters. As described Section~\ref{sec:PSA}, this cut reduced the number of events in the strings filled with $\iso{4}{He}$ (alpha events) by more than 98\%, while maintaining 74.78\% of the neutron events. Section~\ref{sec:encdfit} describes the fit to the $\Encd{}$ spectrum after this cut.

These analyses rely heavily on two calibration periods with a $\iso{24}{Na}$ source distributed uniformly throughout the detector~\cite{Boudjemline2010171}, which produced neutrons similar to those from \B{} NC reactions. These calibrations were performed in 2005 and 2006, and were respectively referred to as $\iso{24}{Na}$-2005 and $\iso{24}{Na}$-2006. A composite source of $\iso{241}{Am}$ and $\iso{9}{Be}$, referred to as AmBe, produced a point-like source of neutrons. This source was positioned throughout the detector during six significant calibration campaigns spanning Phase III. These data were useful for assessing systematic uncertainties associated with temporal and spatial variations in the neutron detection efficiency and PDF of $\Encd{}$.

\subsection{Particle identification in the NCD array}
\label{sec:PSA}

Before analyzing the waveforms, the effect of the logarithmic amplifier was removed using parameters determined from various calibration pulses in a process referred to as de-logging~\cite{ncdlongpaper}. 

\begin{figure}[tbp]	
\centering
\includegraphics[width=\columnwidth]{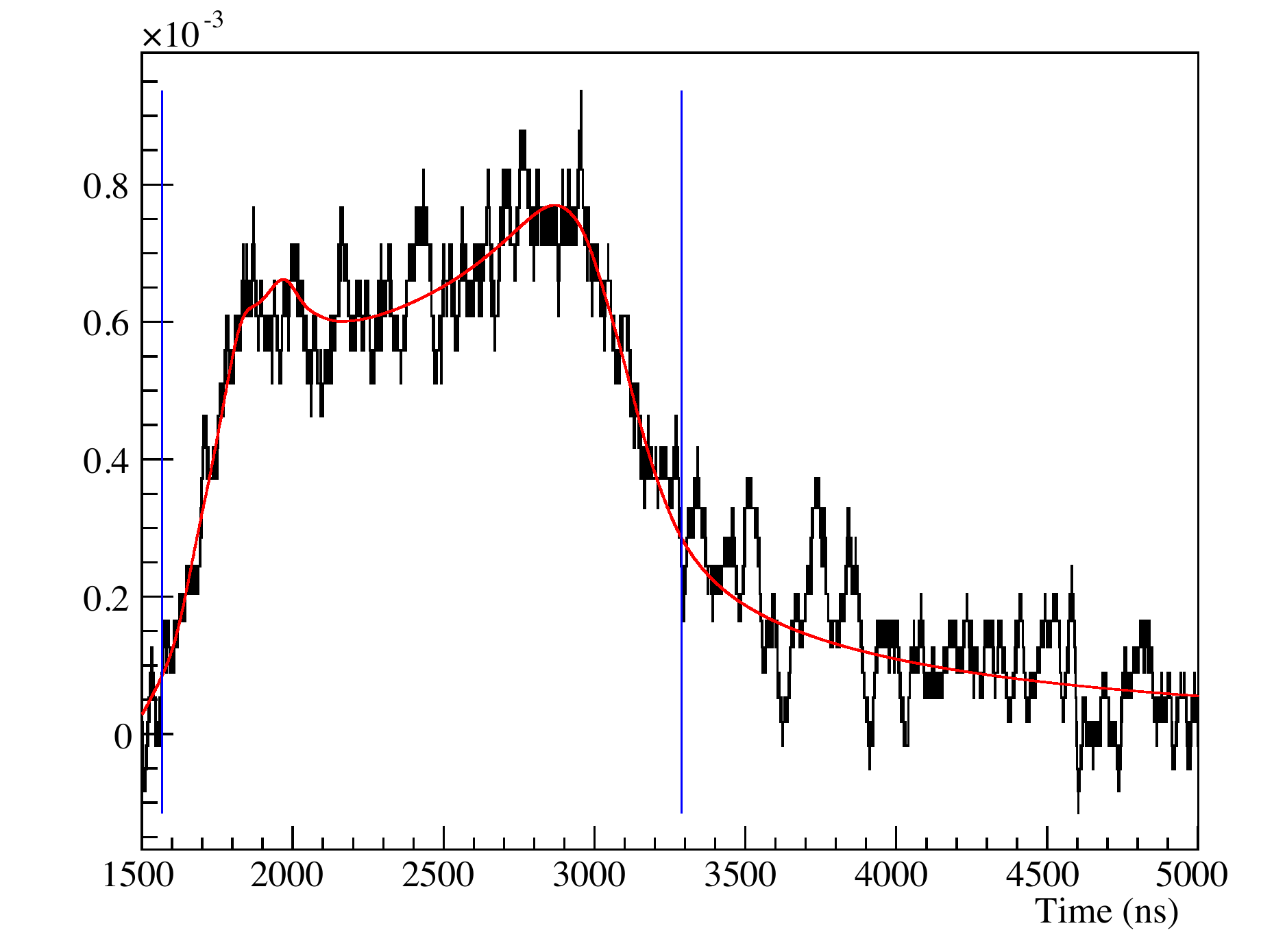}
\includegraphics[width=\columnwidth]{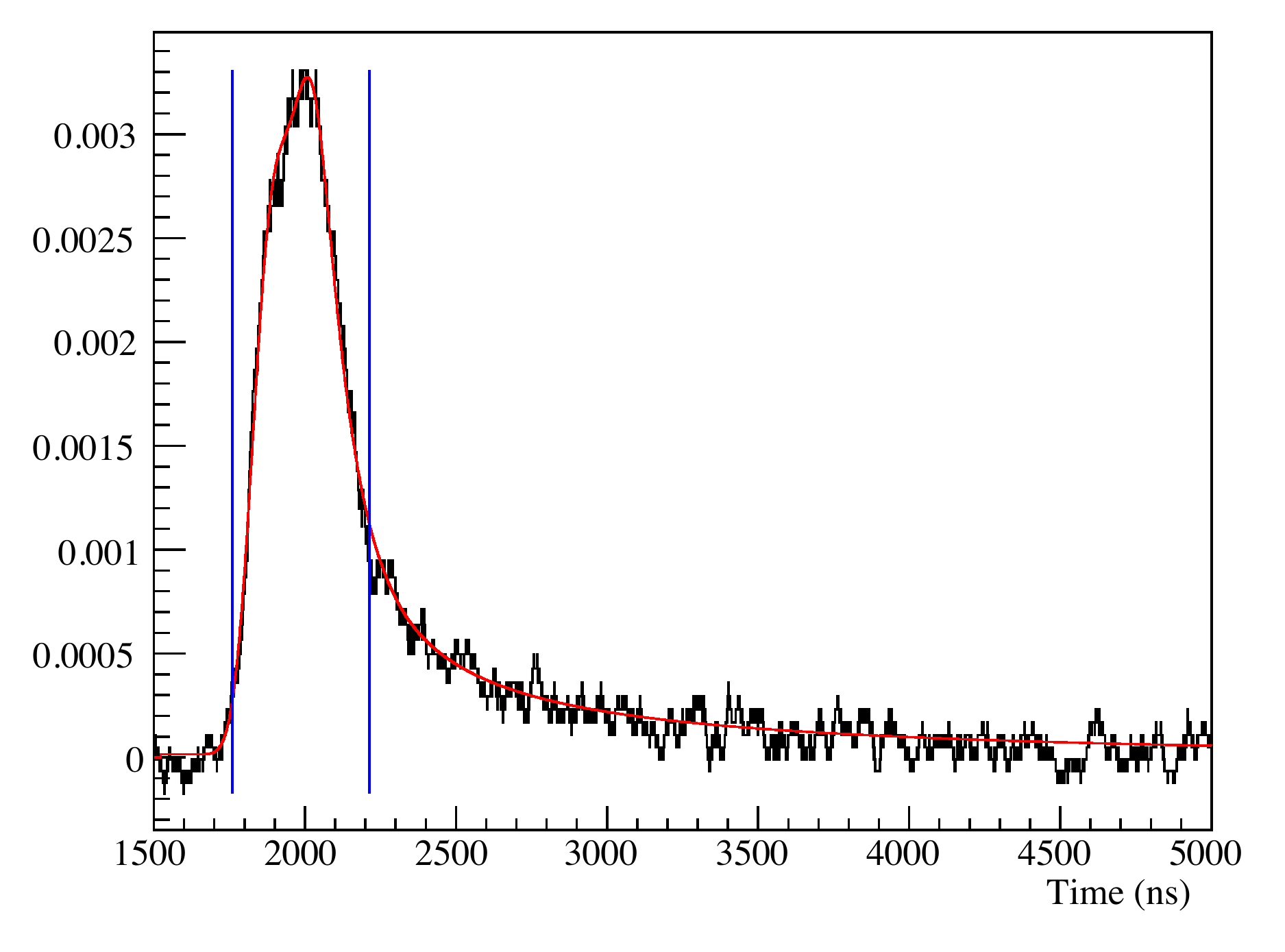}
\caption{Sample waveforms. The top plot shows a neutron waveform (black) obtained from $\iso{24}{Na}$ calibration data with the best fit to the neutron hypothesis (red). The bottom plot shows an alpha waveform (black) obtained from a string filled with $\iso{4}{He}$ with the best fit to the alpha hypothesis (red). The vertical lines represent the fit boundaries.}
\label{fig:NAPfit}
\end{figure}

Two particle identification parameters, \pa{} and \pb{}, were based on fitting the waveforms to libraries of known neutron and alpha waveforms. Each waveform was fitted to each library waveform based on a \chis{} method. The relative start time of the event and library waveforms was varied to find the minimum \chis{}. In both cases the fits did not extend to later times to avoid the effects of ion mobility. Both of these particle identification parameters were defined by
\begin{eqnarray}
\log\left(\frac{\chi^2_\alpha}{\chi^2_n}\right),
\end{eqnarray}
where $\chi^2_\alpha$ and $\chi^2_n$, respectively, were the best \chis{}s from the alpha and neutron hypotheses. The libraries used to calculate \pa{} were primarily based on simulation~\cite{2011NJPh...13g3006B}, and the \chis{} was calculated between where the waveform first crossed a value equal to 10\% of the peak value and where it first returned to 30\% of the peak value~\cite{thesis:Noah}. Figure~\ref{fig:NAPfit} shows some sample fits. This clearly shows the broad waveform for neutrons with a proton-triton trajectory that was roughly perpendicular to the anode, which allows them to be separated from alphas.

To calculate \pb{}, the neutron library was obtained from $\iso{24}{Na}$-2005 data, and the alpha library was obtained from events on the strings filled with $\iso{4}{He}$~\cite{thesis:Ryan}. The \chis{} was calculated between where the waveform first crossed a value equal to 10\% of the peak value and where it first returned to 40\% of the peak value. The libraries for this parameter included events that were used in later studies to evaluate performance. We excluded fitting a waveform to itself because this would result in a \chis{} equal to zero, i.e. a perfect match.

The remaining two particle identification parameters, \pc{} and \pd{}, were respectively based on the kurtosis and skewness of the waveform after smoothing the waveform and deconvolving the effects of ion mobility assuming an ion drift time of 16\,ns. The skewness and kurtosis were calculated using the region between where the waveform first crossed a value equal to 20\% of the maximum and where it first returned to 20\% of the peak value.

\begin{figure}[tbp]
\begin{center}
\includegraphics[height=\columnwidth, angle=90]{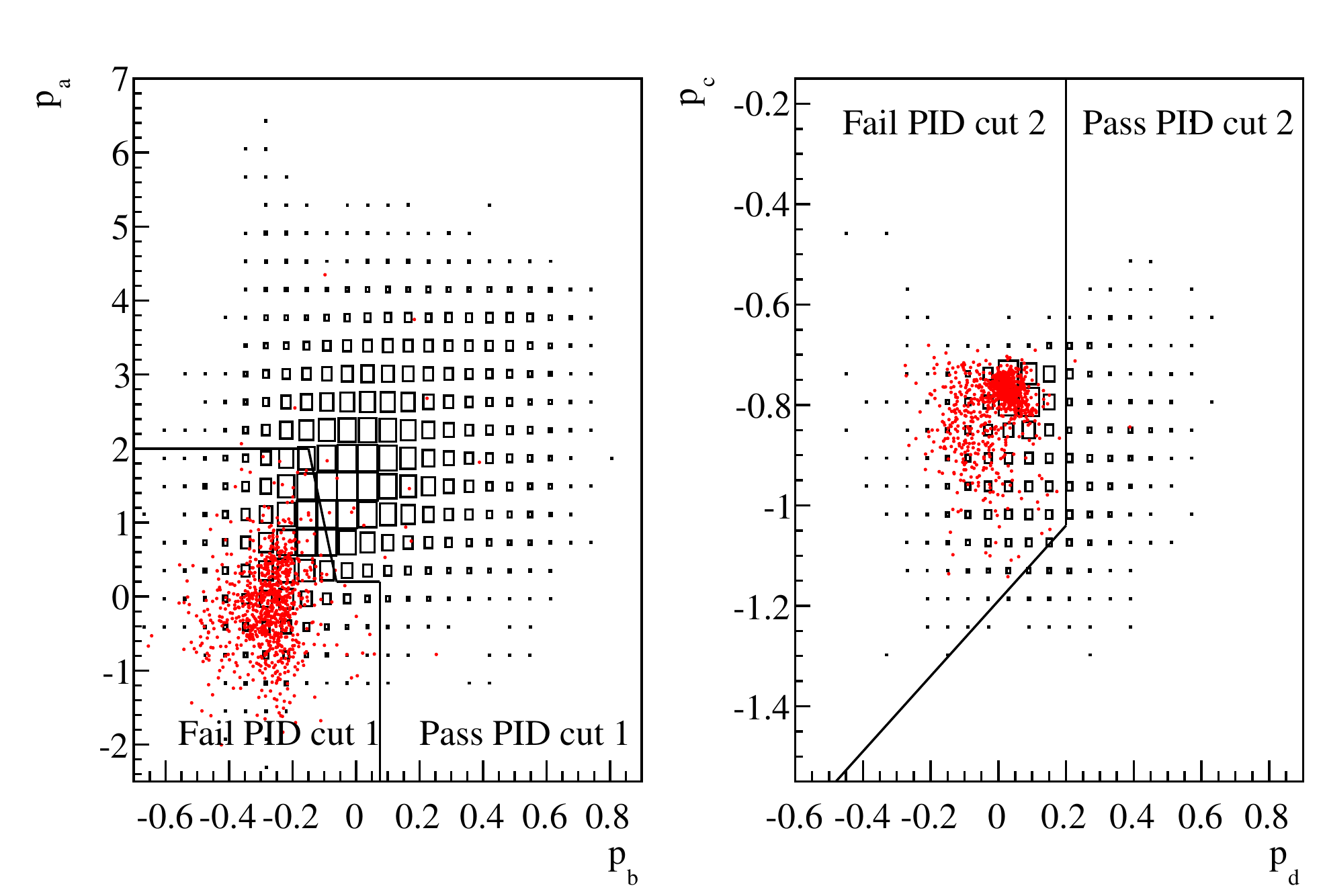}
\caption{Distribution of particle identification parameters for neutron events (boxes, where the area represents the number of events) and alpha events (red marks). The line represents the boundary for cuts. PID cut 1 applies to parameters \pa{} and \pb{}, and PID cut 2 applies to parameters \pc{} and \pd{} for events that failed PID cut 1.}
\label{fig:psa}
\end{center}
\end{figure}

Figure~\ref{fig:psa} shows the distribution of the particle identification parameters for known neutron and alpha events. The left plot shows that \pa{} and \pb{} were highly correlated, which was unsurprising given their similar definitions. This plot also shows that a cut on these two parameters (PID cut 1) removes almost all alpha events while preserving the majority of neutron events. This cut selected events where the alpha hypothesis was significantly worse than the neutron hypothesis. After this cut, we recovered approximately 5\% of the neutron events with a second cut on \pc{} and \pd{} (PID cut 2). PID cut 2 was only applied to events that failed PID cut 1, and selected events with high skewness (\pd{}) or low kurtosis (\pc{}), i.e. the waveforms were not symmetric in time or had a relatively flat peak. This combined cut, selecting events that passed PID cuts 1 or 2, was used for the rest of this analysis.

\begin{figure}[tbp]
\begin{center}
\includegraphics[height=\columnwidth, angle=90]{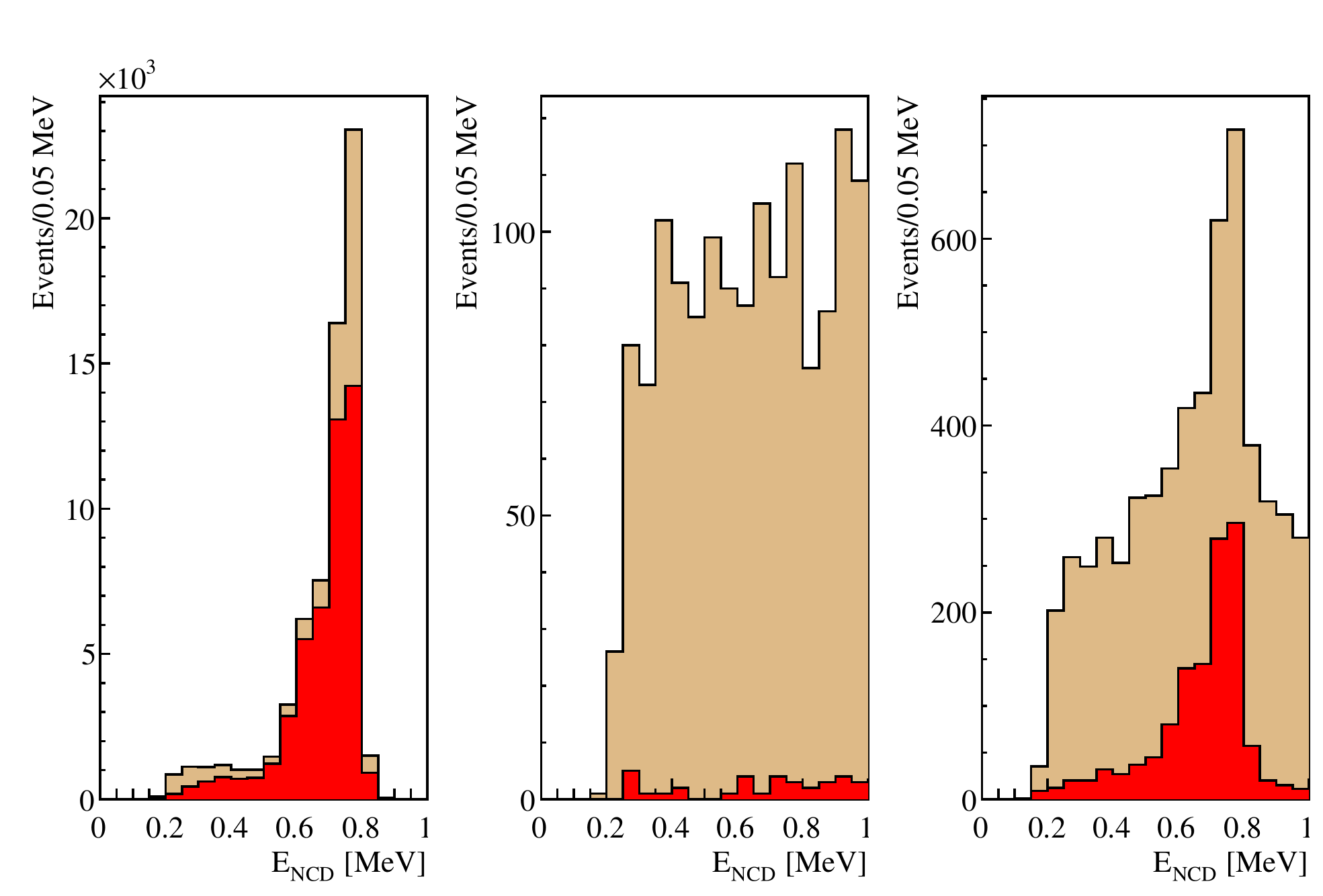}
\caption{$\Encd{}$ spectrum before (brown) and after (red) the particle identification cut. From left to right the plots are for $\iso{24}{Na}$ calibration data (neutrons), data from strings filled with $\iso{4}{He}$ (alphas), and data from strings filled with $\iso{3}{He}$.}
\label{fig:psaResults}
\end{center}
\end{figure}

Figure~\ref{fig:psaResults} shows that the particle identification cut removes almost all the events on the strings filled with $\iso{4}{He}$, i.e. alpha events, while maintaining the majority of the $\iso{24}{Na}$ calibration events, i.e. neutron events. This also shows that the fraction of alpha events removed by the particle identification cut was relatively constant as a function of $\Encd{}$. The right most plot of Figure~\ref{fig:psaResults} shows that the alpha background was significantly reduced, leaving what was clearly mostly neutron events.

\begin{figure}[tbp]
\begin{center}
\includegraphics[width=\columnwidth]{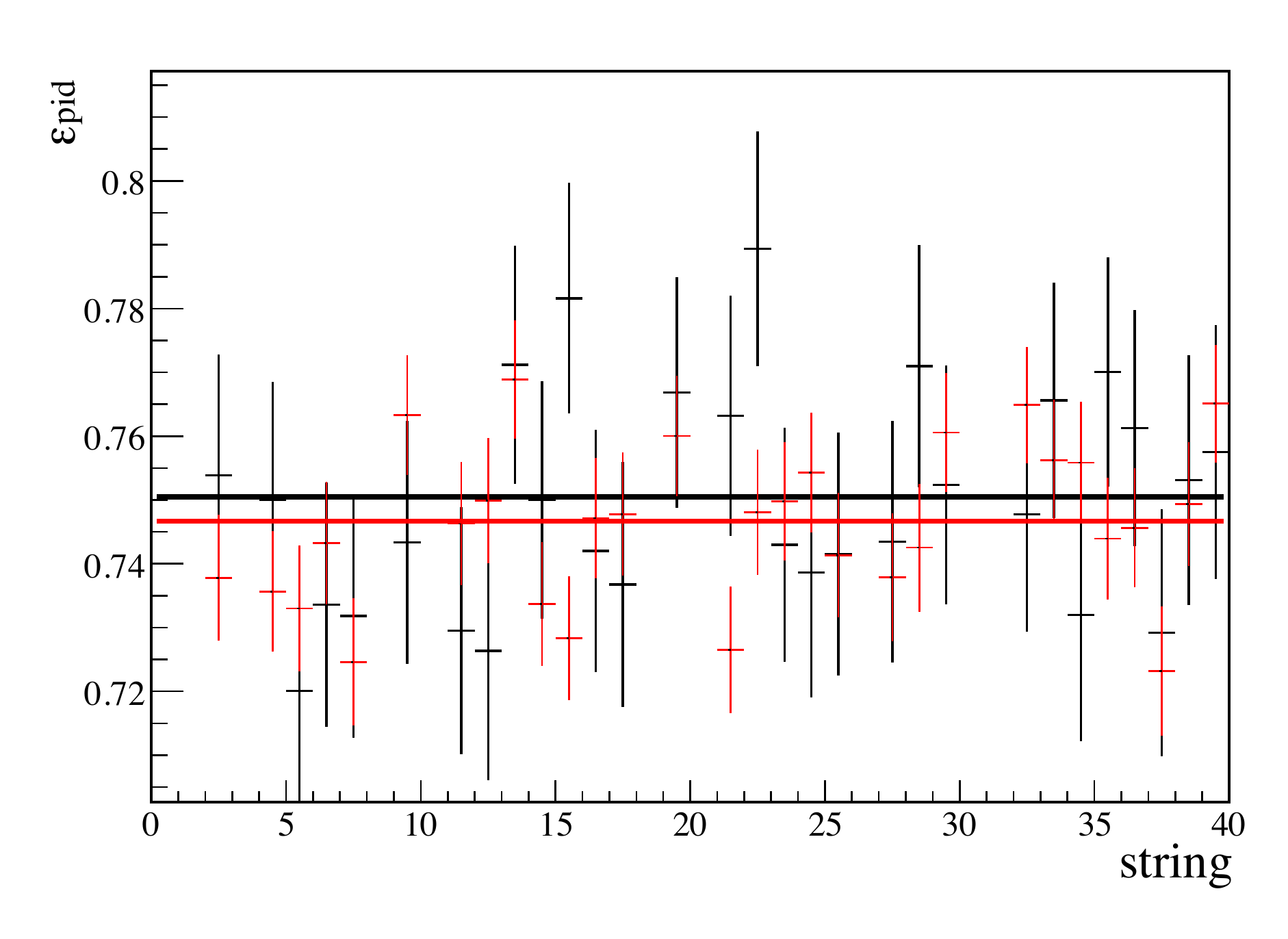}
\caption{$\npsa{}$ as a function of string for the $\iso{24}{Na}$-2005(black) and $\iso{24}{Na}$-2006(red) data. Fits to straight lines give $0.7505\pm0.0035$ with \chis{}/NDF of 24.1/29 and $0.7467\pm0.0018$ with \chis{}/NDF of 49.3/29, respectively.}
\label{fig:Na24_stringbystring_3Cut}
\end{center}
\end{figure}

Figure~\ref{fig:Na24_stringbystring_3Cut} shows the fraction of neutron events surviving the combined particle identification cut, $\npsa{}$, as a function of neutron capture string for $\iso{24}{Na}$-2005 and $\iso{24}{Na}$-2006 data. Table~\ref{tab:neff} shows the average obtained from these measurements. The high \chis{}/NDF obtained with the $\iso{24}{Na}$-2006 data suggests a slight variation in $\npsa{}$ as a function of string; however, the correlation between the $\npsa{}$ calculated for each string between the 2005 and 2006 calibrations was only 0.159, which was so small that it suggested random string-to-string variation instead of a feature of the NCD array. 

\begingroup
\squeezetable
\begin{table}[htdp]
\caption{$\npsa{}$ obtained with the $\iso{24}{Na}$-2005 and $\iso{24}{Na}$-2006 data. The weighted average included a scaling of the uncertainty by $\sqrt{\chi^2{\rm /NDF}}$ for the $\iso{24}{Na}$-2006 data.}
\begin{center}
\begin{tabular}{lrr}
\hline\hline
Data					&$\npsa{}$			&$\chi^2{\rm /NDF}$\\
\hline
$\iso{24}{Na}$-2005		&$0.7505\pm0.0035$	&24.1/29\\
$\iso{24}{Na}$-2006		&$0.7467\pm0.0018$	&49.3/29\\
\hline
Weighted average		&$0.7478\pm0.0019$	&\\
\hline\hline
\end{tabular}
\end{center}
\label{tab:neff}
\end{table}
\endgroup

Table~\ref{tab:summaryAcceptanceSystematics3Cut} summarizes the systematic uncertainties associated with $\npsa{}$. Based on the methods for deriving these systematic uncertainties, we assumed most correlations were zero. A correlation of 0.50 was assumed between the following pairs of systematic uncertainties: de-logging and $\iso{24}{Na}$ uniformity, de-logging and temporal variation, \pa{} correction and $\iso{24}{Na}$ uniformity, \pa{} correction and temporal variation, and \pa{} correction and de-logging. Including these correlations the total absolute systematic uncertainty was 0.0065. Combining the systematic and statistical uncertainties in quadrature led to a total absolute uncertainty of 0.0068.

\begingroup
\squeezetable
\begin{table}[htdp]
\caption{Absolute systematic uncertainties for $\npsa{}$.}
\begin{center}
\begin{tabular}{lr}
\hline\hline
Systematic uncertainty&\\
\hline
$\iso{24}{Na}$ uniformity	&0.0010\\
Temporal variation		&0.0037\\
Contamination			&0.0019\\
De-logging			&0.0018\\
\pa{} correction			&0.0010\\
\pb{} neutron library		&0.0019\\
\hline
Total					&0.0065\\
\hline\hline
\end{tabular}
\end{center}
\label{tab:summaryAcceptanceSystematics3Cut}
\end{table}
\endgroup

\begin{figure}[tbp]
\begin{center}
\includegraphics[width=1.0\columnwidth]{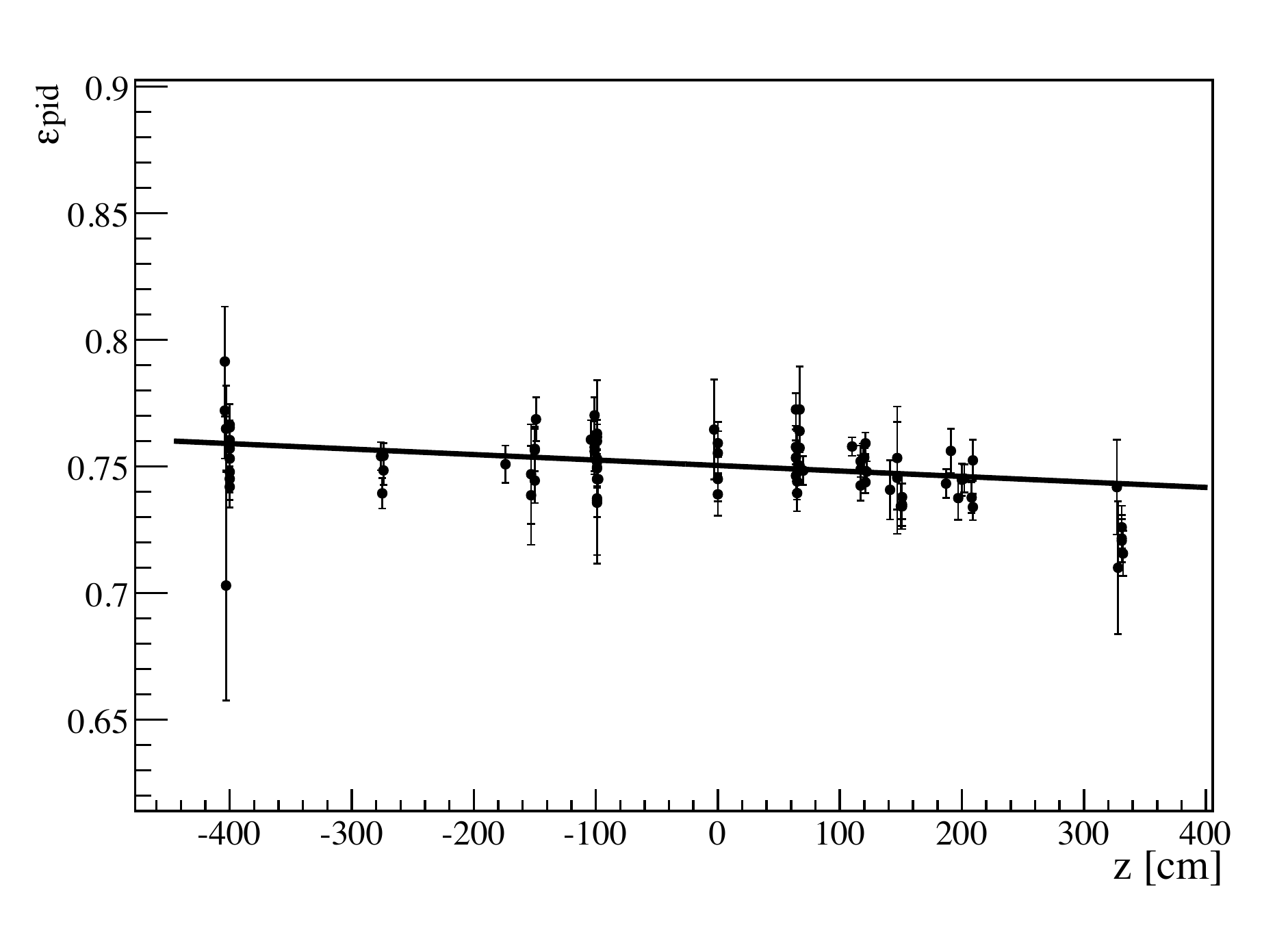}
\caption{$\npsa{}$ as function of $z$ for a single string. This was typical of all strings.}
\label{fig:AmBe_SourceZ_ByScan_0}
\end{center}
\end{figure}

The $\iso{24}{Na}$ calibration data used to calculate $\npsa{}$ had a measured variation in the neutron production rate as a function of $z$ position of less than 10\% between the maximum and the value at $z=0$. Figure~\ref{fig:AmBe_SourceZ_ByScan_0} shows that the dependence of $\npsa{}$ with source position, as measured with the AmBe data, was well approximated by a linear function with a maximum deviation compared to that at $z=0$ of less than 0.01. Combining the possible non-uniformity in the $\iso{24}{Na}$ source distribution with the variation in $\npsa{}$ as a function of $z$ position resulted in an absolute systematic uncertainty in $\npsa{}$ due to $\iso{24}{Na}$ uniformity of 0.0010. The variation due to the $x$ and $y$ position non-uniformity was accounted for in the string averaging used to calculate $\npsa{}$.

The systematic uncertainty in $\npsa{}$ due to temporal variations was estimated based on the standard deviation of $\npsa{}$ calculated from the AmBe data averaged over all strings, and calculated at $z=0$ assuming a linear dependence on $z$. The systematic uncertainty in $\npsa{}$ due to alpha events contaminating the $\iso{24}{Na}$ calibrations was estimated using the number of events with $\Encd{}$ between 0.9\,MeV and 1.4\,MeV as an estimate of the alpha contamination. The systematic uncertainty in $\npsa{}$ due to the de-logging process was estimated by recalculating $\npsa{}$ with the individual de-logging parameters shifted by their estimated uncertainties; because of possible correlations, the magnitude of the maximum shifts with each parameter were added together.

A correction to the \pa{} parameter based on $\iso{24}{Na}$ and AmBe data reduced the spatial and temporal variations in this parameter. A systematic uncertainty to account for the effect of this correction was estimated by calculating $\npsa{}$ assuming a one standard deviation shift in the correction to the \pa{} parameter and then combining the shifts caused by each string in quadrature, which assumed that the corrections from string-to-string were not correlated.

The $\iso{24}{Na}$-2005 data were used in both the neutron library used to calculate \pb{}, and in the determination of $\npsa{}$. Although we did not expect this to bias the calculation of $\npsa{}$, we conservatively included an additional absolute uncertainty of 0.0019, half the difference between $\npsa{}$ calculated with the $\iso{24}{Na}$-2005 and $\iso{24}{Na}$-2006 data.

\subsection{Method for fitting the NCD array data}
\label{sec:encdfit}

After the particle identification cut, the number of neutron events was determined from a likelihood fit to a histogram of $\Encd{}$ with 50 bins uniformly spaced between 0.4 and 0.9\,MeV. 

The PDF of $\Encd{}$ for neutron events was obtained from $\iso{24}{Na}$-2006 data, and for alpha events it was approximated by
\begin{eqnarray}
\nonumber P_\alpha(\Encd{})&=&p_0 \bigg [P_0(\Encd{})+\\
&&\sum_{n=1}^{N_{\rm max}}{p_nP_n(\Encd{})}\bigg ],
\label{eqn:polyFit}
\end{eqnarray}
where $p_0$ and the $p_n$s were fit parameters, $P_n(\Encd{})$ is the Legendre polynomial of order $n$: $P_0=1$, $P_1=x$, $P_2=\nicefrac{1}{2}(3x^2-1)$, $P_3=\nicefrac{1}{2}(5x^3-3x)$, $P_4=\nicefrac{1}{8}(35x^4-30x^2+3)$, with $x=4(\Encd{}{\rm [MeV]}-0.65)$. In order to ensure a well defined PDF, negative values of this function were set to zero. The fit was repeated with different values for the systematic uncertainties associated with the $\Encd{}$ scale, \ncdEscale{}, and resolution, \ncdEres{}, (see Equation~\ref{eqn:encdsys} in Appendix~\ref{apx:systematic}) selected from a 2-dimension scan of these parameters. The point in this 2-dimensional scan with the minimum \chis{} was chosen as the best fit point, and the systematic uncertainty associated with \ncdEscale{} and \ncdEres{} was obtained from the maximum difference in the number of neutron events from the best fit point at the $1\sigma$ contour.

This fit was performed for values of $N_{\rm max}$ up to four, at which point, based on simulations, the polynomial started to fit to fluctuations in the data. We started with the assumption that a zeroth order polynomial was a satisfactory fit to the alpha background. If a higher order polynomial had a significant improvement in \chis{} then this became the new default, and this was tested against higher order polynomials. A significant improvement in \chis{} was defined as a decrease in \chis{} that would result in a 32\% probability for accepting the higher order polynomial when the higher order was not a better model. This calculation included the fact that testing against many different higher order polynomials increases the chances of erroneously choosing a higher order polynomial, so a larger improvement in \chis{} was required. Reference~\cite{ref:thesiswright} gives the changes in \chis{} defined as significant. This method was generic to any type of background, including instrumental backgrounds, provided they did not have features sharper than the assumed background shape.

\begin{figure}[tbp]
\begin{center}
\includegraphics[height=\columnwidth, angle=90]{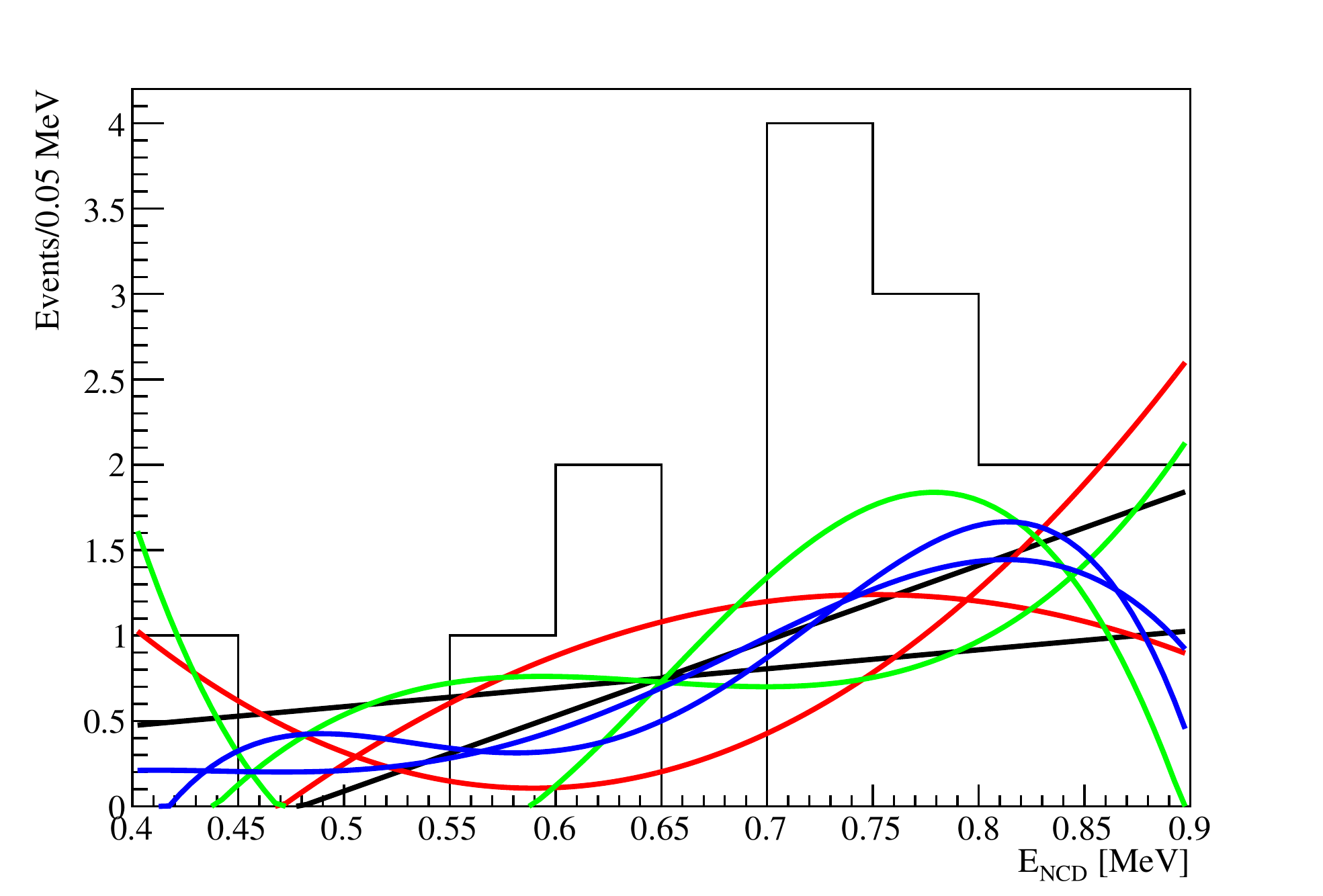}
\caption{$\Encd{}$ spectrum for events on the strings filled with $\iso{4}{He}$ after the particle identification cut. The black, red, green, and blue lines, respectively, show the PDFs used to simulate alpha events for $N_{\rm max}$ equal to one, two, three, and four.}
\label{fig:alphaPolBackgrounds}
\end{center}
\end{figure}

We tested the bias of this method using simulated data. The mean number of neutron events in these sets of simulated data was based on the number of neutrons obtained from the previous analysis of data from Phase III~\cite{cite:snoncd} and $\npsa{}$. The $\Encd{}$ values for these simulated neutron events were obtained from events that passed the particle identification cut in the $\iso{24}{Na}$-2006 data~\footnote{Note that for these tests the PDF of $\Encd{}$ for neutron events was created from the $\iso{24}{Na}$-2005 data, unlike the fit to real data, which used the $\iso{24}{Na}$-2006 data.}. The mean number of alpha events in these sets of simulated data was based on the number of alphas obtained from the previous analysis of data from Phase III~\cite{cite:snoncd} and the approximate fraction of alpha events removed by the particle identification cut. The $\Encd{}$ values for the simulated alpha events were obtained from events that passed the particle identification cut in the strings filled with $\iso{4}{He}$. Because these strings did not have enough events, instead of using these events directly, we fitted the limited data to polynomials of the form in Equation~\ref{eqn:polyFit} with $N_{\rm max}$ varied from 1 to 4, and then used these polynomials to simulate as many $\Encd{}$ values as necessary. In order to test extreme possibilities for the alpha event $\Encd{}$ distributions, the highest order term from the fit was changed by plus and minus $2\sigma$, resulting in the eight different PDFs shown in Figure~\ref{fig:alphaPolBackgrounds}. The bias was less than 2\% for all eight alpha PDFs.

Since only the $\iso{24}{Na}$-2006 data were used to determine the PDF of $\Encd{}$ for neutrons, we included additional systematic uncertainties to account for changes in this PDF due to non-uniformity of the $\iso{24}{Na}$ source and possible temporal variations. The size of these systematic uncertainties were estimated using AmBe calibration data collected at various positions and times to calculate the PDF of $\Encd{}$ for neutron events, and then calculating the size of the shift in the reconstructed number of neutron events. The systematic uncertainties from the temporal and position variation were summed in quadrature to give a total systematic uncertainty of 0.64\% on the number of neutrons obtained from the fit due to the PDF of $\Encd{}$ for neutron events.


\section{Results}
\label{sec:results}

Section~\ref{sec:resultsPSA} presents the results from the analysis of data from the NCD array in Phase III. Because this was a new analysis of this data, we used a statistically-limited and randomly-selected one-third subset of the data to develop the particle identification cut and analysis. Once we had finalized all aspects of this analysis we fitted the entire set of data from the NCD array in Phase III. After completing this full analysis we realized that there was an error in the method to calculate the systematic uncertainty due to \ncdEscale{} and \ncdEres{}, which was corrected in the results presented here.

The total number of neutron events detected in the NCD array obtained from this new analysis of data from Phase III was then used as a constraint in the fits to the combined data presented in Section~\ref{sec:results3phase}. The combined analysis of the three phases also used a statistically-limited and randomly-selected one-third subset of the data to develop the fitting method. Once we had finalized all aspects of this analysis we fitted the entire set of data from all three phases.

\subsection{Results from fit to NCD array data}
\label{sec:resultsPSA}

Table~\ref{tab:allOrdersFullResults} shows the \chis{} and statistical uncertainty from the fit to the $\Encd{}$ spectrum for various values of $N_{\rm max}$ in Equation~\ref{eqn:polyFit}. In general including extra terms in the PDF of $\Encd{}$ for alpha events should not result in best fits with higher \chis{}, but this can occur if the minimization routine finds different local minima. Based on our method for choosing the value of $N_{\rm max}$ representing the point where improvements in fit quality cease, the best fit occurs when $N_{\rm max}=4$. This corresponded to the maximum value of $N_{\rm max}$ considered before performing the fit, so to check that larger values of $N_{\rm max}$ did not produce better fits, we also fitted with $N_{\rm max}$ equal to five and six, as shown in Table~\ref{tab:allOrdersFullResults}. These fits did not produce better results.

\begingroup
\squeezetable
\begin{table}[htdp]
\caption{\chis{} and 1$\sigma$ statistical uncertainty for various values of $N_{\rm max}$ in Equation~\ref{eqn:polyFit}.}
\begin{center}
\begin{tabular}{rrr}
\hline\hline
$N_{\rm max}$&	$\chi^2$/NDF&	Stat. uncertainty\\
\hline
0&		54.92/48&		4.2\%\\
1&		56.72/47&		4.2\%\\
2&		47.63/46&		5.5\%\\
3&		41.78/45&		6.5\%\\
4&		40.20/44&		6.9\%\\
5&		40.34/43&		9.4\%\\
6&		40.41/42&		9.2\%\\
\hline\hline
\end{tabular}
\end{center}
\label{tab:allOrdersFullResults}
\end{table}
\endgroup

\begin{figure}[tbp]
\begin{center}
\includegraphics[width=\columnwidth]{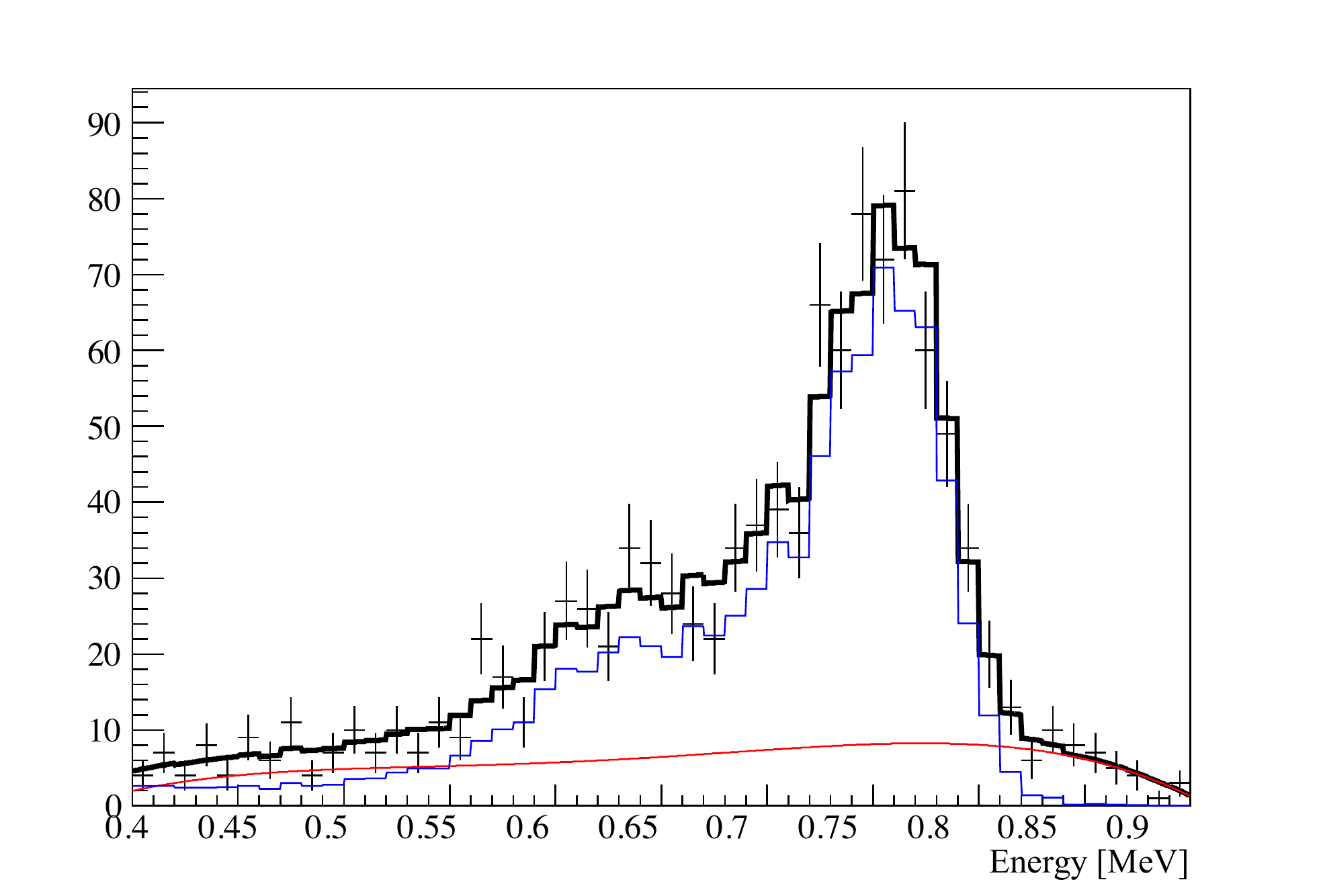}
\caption{The fitted $\Encd{}$ spectrum after the particle identification cut. The thick black line is the best fit. The blue and red lines are the best fitted neutron and alpha spectra, respectively.}
\label{fig:fullenergyFit3Cut}
\end{center}
\end{figure}

Figure~\ref{fig:fullenergyFit3Cut} shows the best fit of the $\Encd{}$ spectrum. Although the best fit turns down at higher values of $\Encd{}$ the parameters were consistent with a flat PDF in that region. This variation in the allowed PDF was reflected in the increased statistical uncertainty with large $N_{\rm max}$.

For the fit with $N_{\rm max}=4$ the systematic uncertainty due to \ncdEscale{} and \ncdEres{} was 5 neutrons. Combining this with the systematic uncertainty in the PDF of $\Encd{}$ for neutrons, the statistical uncertainty in the fit, and dividing by $\npsa$, the total number of neutrons observed in the NCD array equals $1115\pm79$. The previous analysis of data from Phase III gave 1168 neutrons in the NCD array, with similar uncertainty~\cite{cite:snoncd}. That analysis had a large background due to alpha events, which made the assessment of the systematic uncertainty on the fitted number of events more challenging. The result presented here avoids that problem by eliminating most of the background from alpha events and allowing a very general PDF to describe the $\Encd{}$ spectrum for any remaining background events. Since the particle identification cut removed almost all alpha events, the fitted number of neutron events had a small to moderate correlation with the previous analysis of this data.

\subsection{Results from combined fit to all data}
\label{sec:results3phase}

For the combined fit to all data using the maximum likelihood technique, Table~\ref{qsigex_ncd_psa_final_3phase_pee} shows $\fB{}$ and the \nue{} survival probability parameters as defined in Equations~\ref{eqn:peed} and \ref{eqn:aee} of Section~\ref{sec:parameterization}. Table~\ref{qsigex_ncd_psa_final_3phase_pee_cor} shows the correlation between these parameters. The combined fit to all data from SNO yielded a total flux of active neutrino flavors from $\iso{8}{B}$ decays in the Sun of $\fB{}$=\numberSNOBflux{}. During the day the \nue{} survival probability at 10\,MeV was \numberPeea{}, which was inconsistent with the null hypothesis that there were no neutrino oscillations at very high significance. Using the covariance matrix obtained from this combined analysis we can compare the best fit to various null hypotheses. The null hypothesis that there were no spectral distortions of the \nue{} survival probability (i.e. $\Peeb{}=0$, $\Peec{}=0$, $\Aeea{}=0$, $\Aeeb{}=0$), yielded $\Delta\chi^2=1.97$ (26\% C.L) compared to the best fit. The null hypothesis that there were no day/night distortions of the \nue{} survival probability (i.e. $\Aeea{}=0$, $\Aeeb{}=0$), yielded $\Delta\chi^2=1.87$ (61\% C.L.) compared to the best fit. 

\begingroup
\squeezetable
\begin{table}[htdp]
\centering
\setlength{\extrarowheight}{2pt}
\caption{Results from the maximum likelihood fit. Note that $\fB${} is in units of $\flux$. The D/N systematic uncertainties includes the effect of all nuisance parameters that were applied differently between day and night. The MC systematic uncertainties includes the effect of varying the number of events in the Monte Carlo based on Poisson statistics. The basic systematic uncertainties include the effects of all other nuisance parameters.}
\begin{tabular}{lD{.}{.}{4}@{}rrrrr}	
\hline\hline
&\multicolumn{1}{l}{Best fit}&	Stat.&	\multicolumn{4}{c}{Systematic uncertainty}\\
&			&		&						Basic &					D/N &					MC&						Total\\
\hline
$\fB{}$&		5.25&	$\pm0.16$&				${}^{+0.11}_{-0.12}$&			$\pm0.01$&				${}^{+0.01}_{-0.03}$&		${}^{+0.11}_{-0.13}$\\
$\Peea{}$&	0.317&	$\pm0.016$&				${}^{+0.008}_{-0.010}$&		$\pm0.002$&				${}^{+0.002}_{-0.001}$&		$\pm0.009$\\
$\Peeb{}$&	0.0039&	${}^{+0.0065}_{-0.0067}$&	${}^{+0.0047}_{-0.0038}$&	${}^{+0.0012}_{-0.0018}$&	${}^{+0.0004}_{-0.0008}$&	$\pm0.0045$\\
$\Peec{}$&	-0.0010&	$\pm0.0029$&				${}^{+0.0013}_{-0.0016}$&	${}^{+0.0002}_{-0.0003}$&	${}^{+0.0004}_{-0.0002}$&	${}^{+0.0014}_{-0.0016}$\\
$\Aeea{}$&	0.046&	$\pm0.031$&				${}^{+0.007}_{-0.005}$&		$\pm0.012$&				${}^{+0.002}_{-0.003}$&		${}^{+0.014}_{-0.013}$\\
$\Aeeb{}$&	-0.016&	$\pm0.025$&				${}^{+0.003}_{-0.006}$&		$\pm{0.009}$&				$\pm0.002$&				${}^{+0.010}_{-0.011}$\\
\hline\hline
\end{tabular}
\label{qsigex_ncd_psa_final_3phase_pee}
\end{table}
\endgroup

\begingroup
\squeezetable
\begin{table}[htdp]
\centering
\caption{Correlation matrix from the maximum likelihood fit.}
\begin{tabular}{lD{.}{.}{3}D{.}{.}{3}D{.}{.}{3}D{.}{.}{3}D{.}{.}{3}D{.}{.}{3}}
\hline\hline
&	\multicolumn{1}{c}{$\fB{}$}& \multicolumn{1}{c}{$\Peea{}$}& \multicolumn{1}{c}{$\Peeb{}$}& \multicolumn{1}{c}{$\Peec{}$}& \multicolumn{1}{c}{$\Aeea{}$}& \multicolumn{1}{c}{$\Aeeb{}$}\\
\hline
$\fB{}$& 1.000& -0.723& 0.302& -0.168& 0.028& -0.012\\
$\Peea{}$& -0.723& 1.000& -0.299& -0.366& -0.376& 0.129\\
$\Peeb{}$& 0.302& -0.299& 1.000& -0.206& 0.219& -0.677\\
$\Peec{}$& -0.168& -0.366& -0.206& 1.000& 0.008& -0.035\\
$\Aeea{}$& 0.028& -0.376& 0.219& 0.008& 1.000& -0.297\\
$\Aeeb{}$& -0.012& 0.129& -0.677& -0.035& -0.297& 1.000\\
\hline\hline
\end{tabular}
\label{qsigex_ncd_psa_final_3phase_pee_cor}
\end{table}
\endgroup

\begin{figure}[tbp]
\centering
\includegraphics[width=\columnwidth]{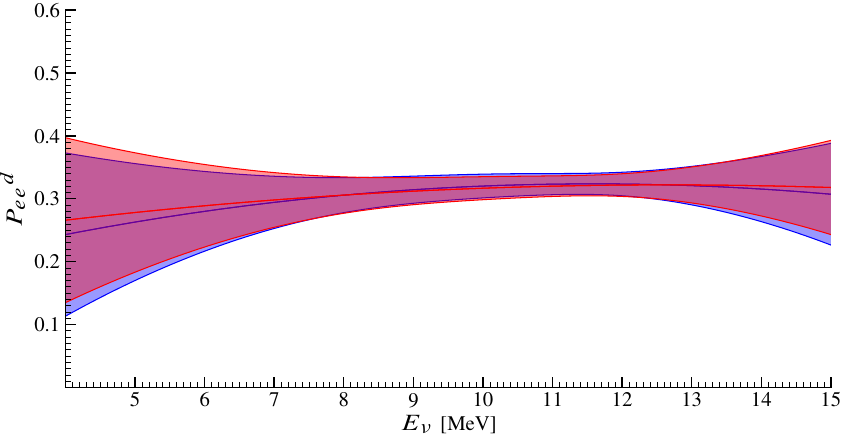}\\
\includegraphics[width=\columnwidth]{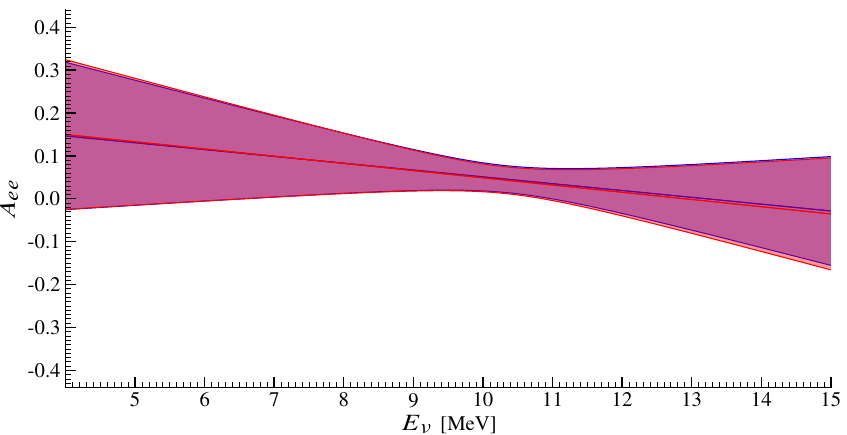}
\caption{RMS spread in $\Peed{}$ and $\Aee{}$, taking into account the parameter uncertainties and correlations. The red band represents the results from the maximum likelihood fit, and the blue band represents the results from the Bayesian fit. The red and blue solid lines, respectively, are the best fits from the maximum likelihood and Bayesian fits.}
\label{ncd_psa_final_3phase_pee_cmp}
\end{figure}

Figure~\ref{ncd_psa_final_3phase_pee_cmp} shows the RMS spread in $\Peed{}$ and $\Aee{}$, taking into account the parameter uncertainties and correlations. This also shows that the maximum likelihood analysis was consistent with the alternative Bayesian analysis. Reference~\cite{supplemental} contains all steps of the MCMC fit after the fit had converged.

\begin{figure}[tbp]
 \centering
 \includegraphics[width=1.0\columnwidth]{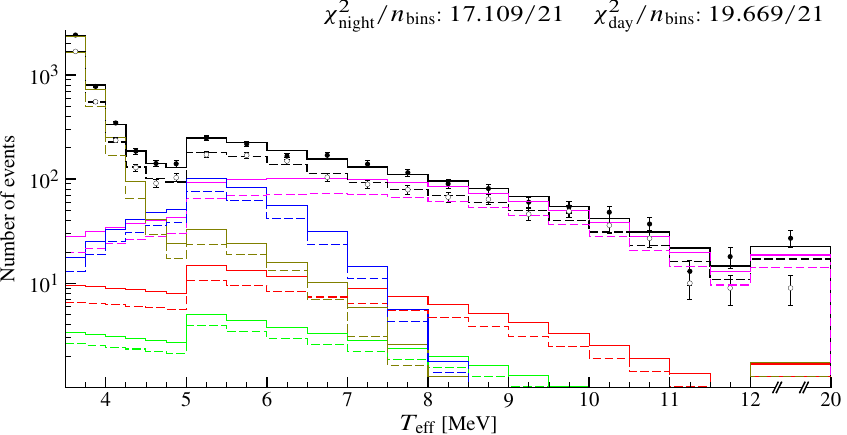}
 \includegraphics[width=1.0\columnwidth]{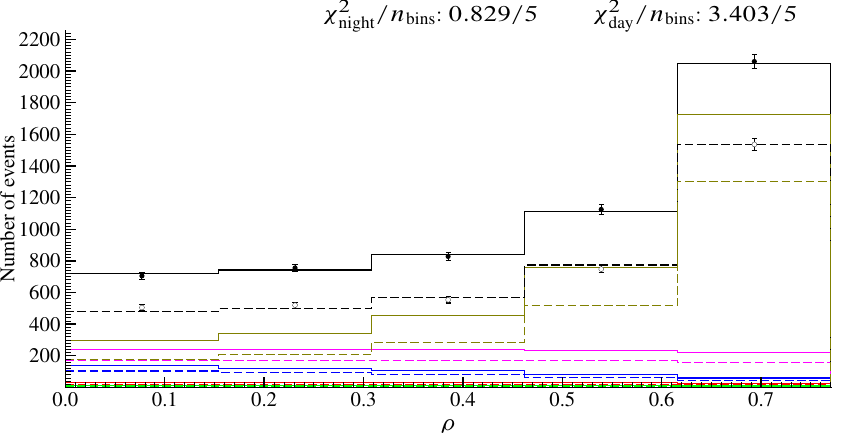}
 \includegraphics[width=1.0\columnwidth]{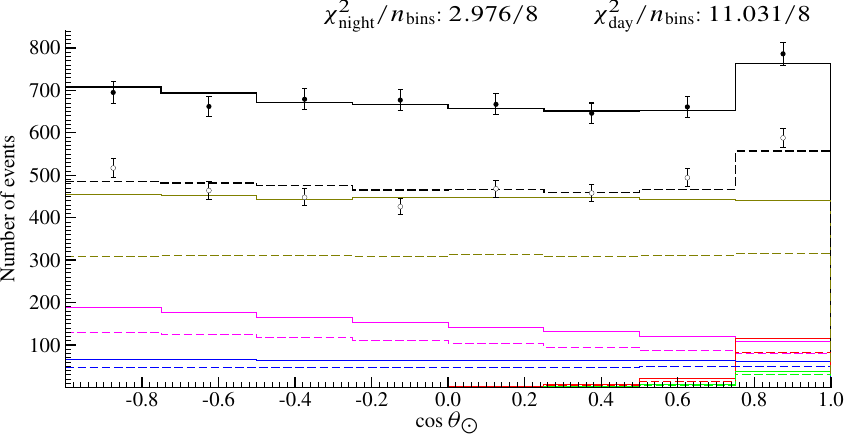}
 \includegraphics[width=1.0\columnwidth]{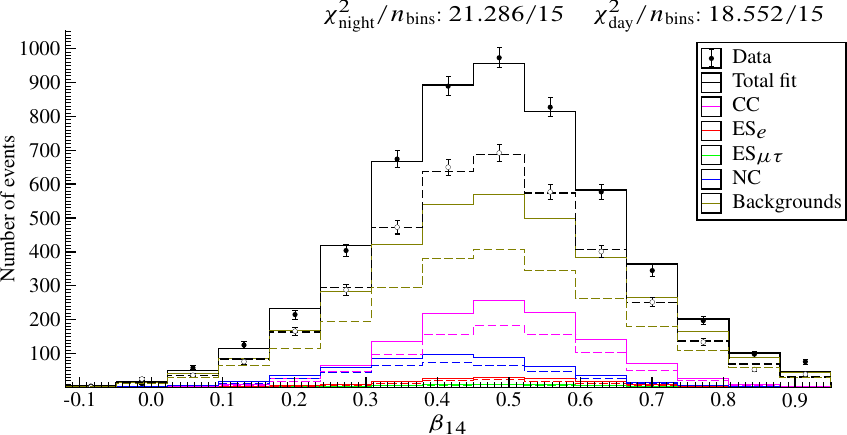}
\caption{Projection of the $\teff{}$, $\rho$, $\cts{}$, and $\be{}$ for the Phase I data. Day events hollow circles and dashed lines. Night events filled circle and solid lines. Note that the sharp break in the data in the top panel at 5\,MeV arises from change of bin width.}
\label{ncd_psa_final_3phase_pee_proj_I}
\end{figure}

\begin{figure}[tbp]
 \centering
 \includegraphics[width=1.0\columnwidth]{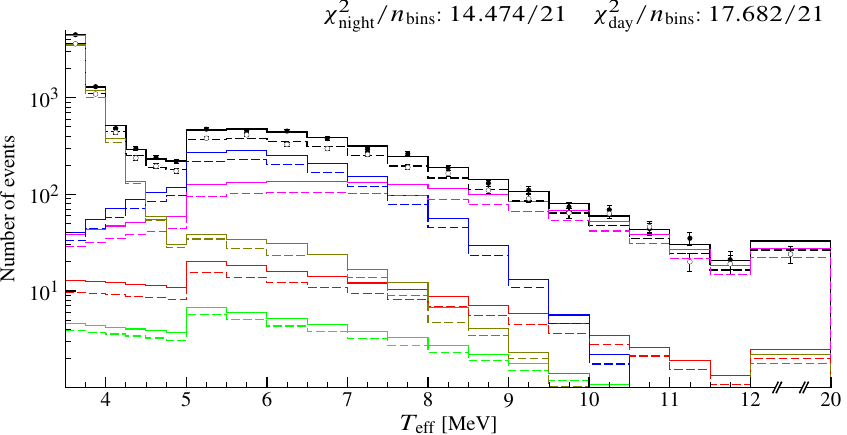}
 \includegraphics[width=1.0\columnwidth]{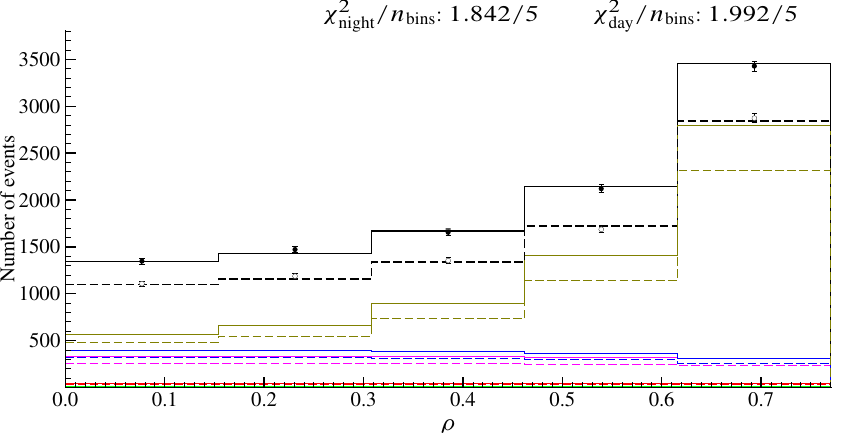}
 \includegraphics[width=1.0\columnwidth]{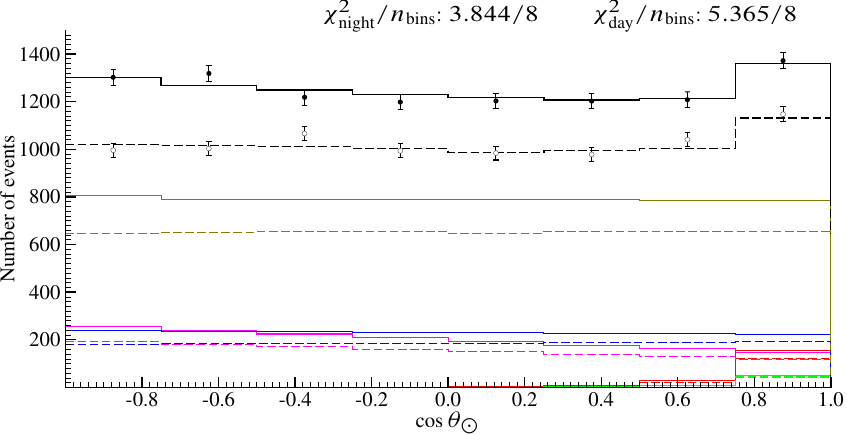}
 \includegraphics[width=1.0\columnwidth]{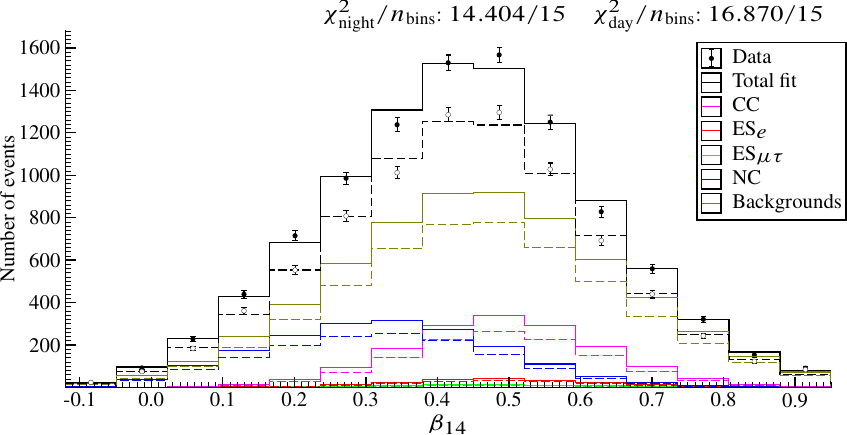}
\caption{Projection of the $\teff{}$, $\rho$, $\cts{}$, and $\be{}$ for the Phase II data. Day events hollow circles and dashed lines. Night events filled circle and solid lines. Note that the sharp break in the data in the top panel at 5\,MeV arises from change of bin width.}
\label{ncd_psa_final_3phase_pee_proj_II}
\end{figure}

\begin{figure}[tbp]
 \centering
 \includegraphics[width=1.0\columnwidth]{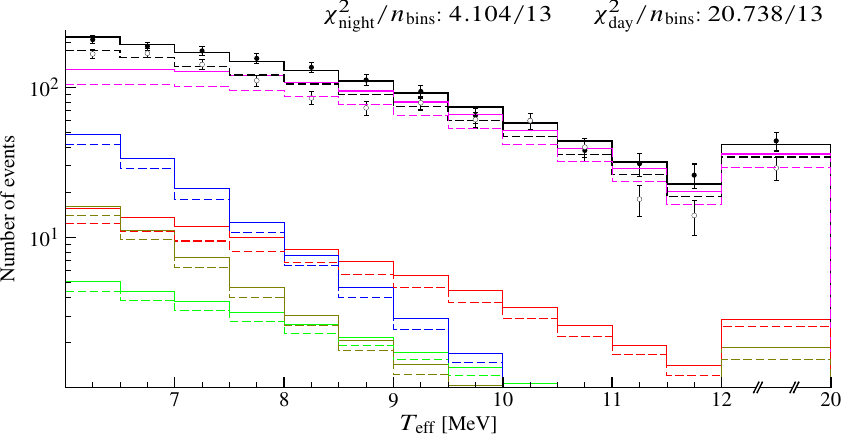}
 \includegraphics[width=1.0\columnwidth]{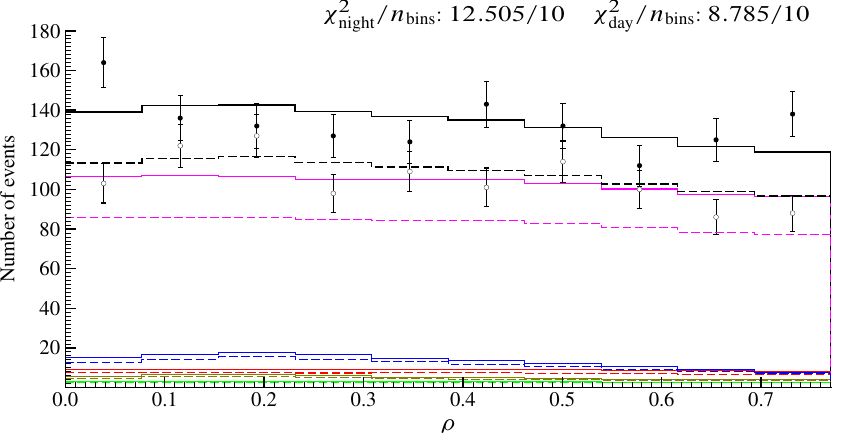}
 \includegraphics[width=1.0\columnwidth]{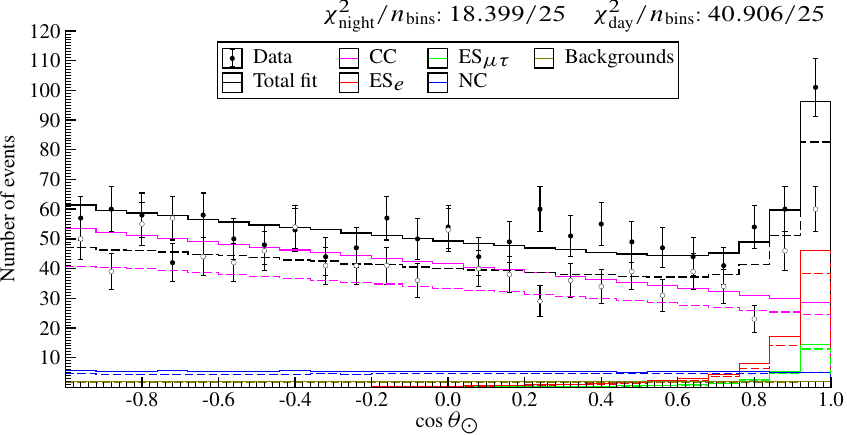}
\caption{Projection of the $\teff{}$, $\rho$, and $\cts{}$ for the data from Phase III. Day events hollow circles and dashed lines. Night events filled circle and solid lines.}
\label{ncd_psa_final_3phase_pee_proj_III}
\end{figure}

Figures~\ref{ncd_psa_final_3phase_pee_proj_I}, \ref{ncd_psa_final_3phase_pee_proj_II}, and \ref{ncd_psa_final_3phase_pee_proj_III}, respectively, show one-dimensional projections of the fit for Phase I, II, and III.

\section{Neutrino oscillations}
\label{sec:oscillations}

Solar models predict the fluxes of neutrinos at the Earth, but the flavors of those neutrinos depends on the neutrino oscillation parameters and the details of where the neutrinos were produced in the Sun. Section~\ref{sec:solarPredictions} describes how the flavor components of the neutrino fluxes were calculated, and Section~\ref{sec:SNOoscillations} describes how these predictions were compared to results for $\fB{}$, $\Peed{}$, and $\Aee{}$ presented here, and with other solar neutrino experiments. Reference~\cite{thesisNuno} provides further details on the neutrino oscillation analysis presented here.

We considered two different neutrino oscillation hypotheses in this analysis. For comparison with previous SNO analyses, Section~\ref{sec:twoneu} presents the so-called two-flavor neutrino oscillations, which assumed $\thetaonethree{}=0$ and had two free neutrino oscillation parameters, $\thetaonetwo{}$, and $\Dmonetwo{}$. In Section~\ref{sec:threeneu} we also considered the so-called three-flavor neutrino oscillations, which had the following three free neutrino oscillation parameters: $\thetaonetwo{}$, $\thetaonethree{}$, and $\Dmonetwo{}$. Note that the mixing angle, $\thetatwothree{}$, and the CP-violating phase, $\delta$, are irrelevant for the neutrino oscillation analysis of solar neutrino data. The solar neutrino data considered here was insensitive to the exact value $\Dmonethree{}$, so we used a fixed value of $\pm2.45\times10^{3}\,{\rm eV^{2}}$ obtained from long-baseline accelerator experiments and atmospheric neutrino experiments~\cite{1367-2630-13-6-063004}.

\subsection{Solar neutrino predictions}
\label{sec:solarPredictions}

Predicting the solar neutrino flux and $E_\nu$ spectrum for all neutrino flavors requires a model of the neutrino production rates as a function of location within the Sun, and a model of the neutrino survival probabilities as they propagate through the Sun, travel to the Earth, and then propagate through the Earth. For consistency with previous calculations, and because of the conservative model uncertainties, we used the BS05(OP) model~\cite{bs05} to predict the solar neutrino production rate within the Sun rather than the more recent BPS09(GS) or BPS09(AGSS09) models~\cite{ssh:bps09}. Reference~\cite{supplemental} provides the results presented below assuming these other solar models. We used the $E_\nu$ spectrum for \B{}s from Reference~\cite{b8winter}, and all other spectra were from Reference~\cite{Bahcall}.

In previous analyses we used numerical calculations to construct a lookup table of neutrino survival probability as a function of the neutrino oscillation parameters. Such a table was still used to study the entire region of neutrino oscillation parameters. Previous analyses of SNO data combined with other solar neutrino experiments left only the region referred to as LMA. This analysis used an adiabatic approximation when calculating neutrino oscillation parameters in that region. We verified that these two calculations gave equivalent results for a fixed set of neutrino oscillation parameters in the LMA region. Due to the improved speed of the adiabatic calculation we could scan discrete values of both $\Dmonetwo{}$ and $E_\nu$, whereas the lookup table used previously was calculated at discrete values of $\Dmonetwo{}/E_\nu$, which resulted in small but observable discontinuities.

We also updated the electron density as a function of Earth radius, which affects the survival probability of neutrinos propagating through the Earth. Previous SNO analyses used the PREM~\cite{PREM} model, which averages over the continental and oceanic crust. When neutrinos enter the SNO detector from above they must have propagated through continental crust; therefore, we modified the Earth density profile to use PEM-C~\cite{Dziewonski:1975ih}, which assumes continental crust for the outer most layer of the Earth. Because this significantly affected neutrinos only during a short period of each day, this had a negligible effect on the calculated neutrino survival probability.

\subsection{Analysis of solar neutrino and KamLAND data}
\label{sec:SNOoscillations}

To compare the \nue{} survival probability parameters calculated in Section~\ref{sec:results} with a neutrino oscillation prediction it was necessary to account for the sensitivity of the SNO detector. Table~\ref{table:neutrinoSensitiivty} in Appendix~\ref{apx:neutrinoSensitivity} gives $S(E_\nu)$, the predicted spectrum of $E_\nu$ detectable by the SNO detector after including all effects such as the energy dependence of the cross-sections, reaction thresholds, and analysis cuts, but not including neutrino oscillations. We multiplied $S(E_\nu)$ by the predicted neutrino oscillation hypothesis distortions, and fitted the resulting spectrum to $S(E_\nu)$ distorted by Equations~\ref{eqn:peed} and \ref{eqn:aee}. We then calculated the \chis{} between the results from this fit and our fit to the SNO data presented in Section~\ref{sec:results3phase}. This calculation used the uncertainties and correlation matrix from the fit to SNO data, but did not include the uncertainties from the fit to the distorted $S(E_\nu)$ as this does not represent a measurement uncertainty.

The \chis{} was calculated as a function of the neutrino oscillation parameters. The best fit was determined from the parameters resulting in the minimum \chis{}, and the uncertainties were calculated from the change in \chis{} from this minimum. Tests with simulated data revealed that this method produced neutrino oscillation parameters that were unbiased and produced uncertainties consistent with frequentist statistics.

The following additional solar neutrino results were used in calculating the results from solar neutrino experiments: the solar neutrino rates measured in Ga~\cite{Gallium2009, *Altmann2005174, *thesis:Kaether}, and Cl~\cite{Cleveland:1998nv} experiments, the rate of $\iso{7}{Be}$ solar neutrinos measured in Borexino~\cite{:1343896}, and the rates and recoil electron energy spectra of \B{} ES reactions measured in Borexino~\cite{borexino_b8}, SuperKamiokande-I~\cite{sk1}, SuperKamiokande-II~\cite{sk2}, and SuperKamiokande-III~\cite{Abe:2011fk}. The last two SuperKamiokande results were split into day and night, and the first SuperKamiokande result was split into multiple periods of the day. The difference in the day and night rate of $\iso{7}{Be}$ solar neutrinos recently measured in Borexino~\cite{2011arXiv1104.2150T}, and the recent measurement of the \B{} spectrum in KamLAND~\cite{2011arXiv1106.0861A}, were not included, but these results would not significantly change the results reported here. For a given set of neutrino oscillation parameters and $\fB{}$, the predictions for the set of experiments were calculated and compared to their results. This comparison was added to the \chis{} described above, and the resulting \chis{} was then minimized with respect to $\fB$. The same procedure as above was used to determine the best fit values and uncertainties.

The KamLAND experiment observed neutrino oscillations in \nuebar{}s from nuclear reactors. By assuming CPT invariance we can directly compare these results with the neutrino oscillations observed with solar neutrinos. Because this was a completely independent result, the lookup table of \chis{} as a function of $\thetaonetwo{}$, $\thetaonethree{}$, and $\Dmonetwo{}$ published by the KamLAND collaboration~\cite{PhysRevD.83.052002} was added directly to the \chis{} values calculated from the solar neutrino analysis, and the same procedure was used to determine the best fit values and uncertainties.

\subsection{Two-flavor neutrino oscillation analysis}
\label{sec:twoneu}

\begin{figure}[tbp]	
\centering
\includegraphics[width=1.0\columnwidth]{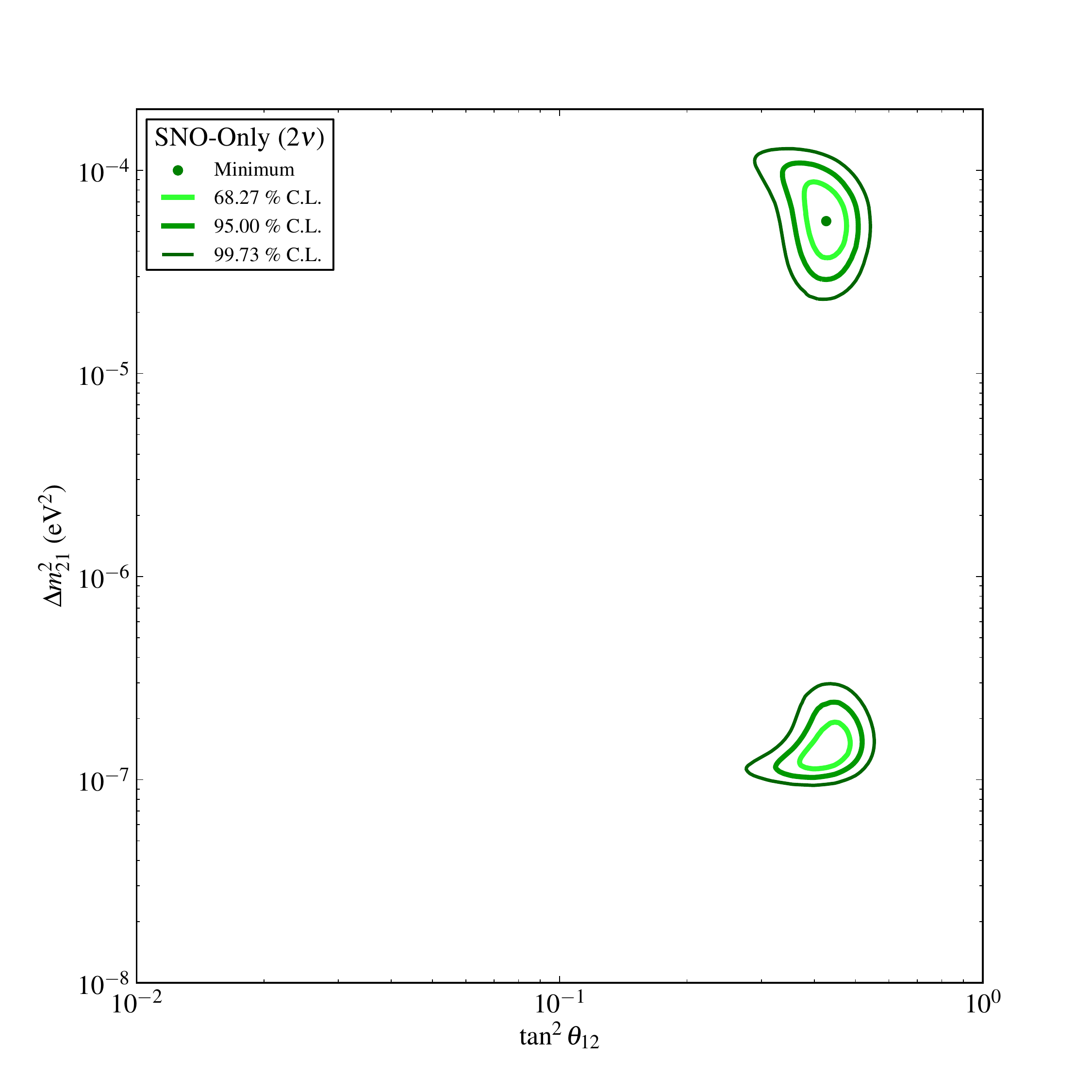}
\caption{Two-flavor neutrino oscillation analysis contour using only SNO data.}
\label{fig:final:sno}
\end{figure}

Figure~\ref{fig:final:sno} shows the allowed regions of the $(\tanthetaonetwo{},\Dmonetwo{})$ parameter space obtained with the results in Tables~\ref{qsigex_ncd_psa_final_3phase_pee} and \ref{qsigex_ncd_psa_final_3phase_pee_cor}. SNO data alone cannot distinguish between the upper (LMA) region, and the lower (LOW) region, although it slightly favors the LMA region.

\begin{figure}[tbp]
\centering
\includegraphics[width=1.0\columnwidth]{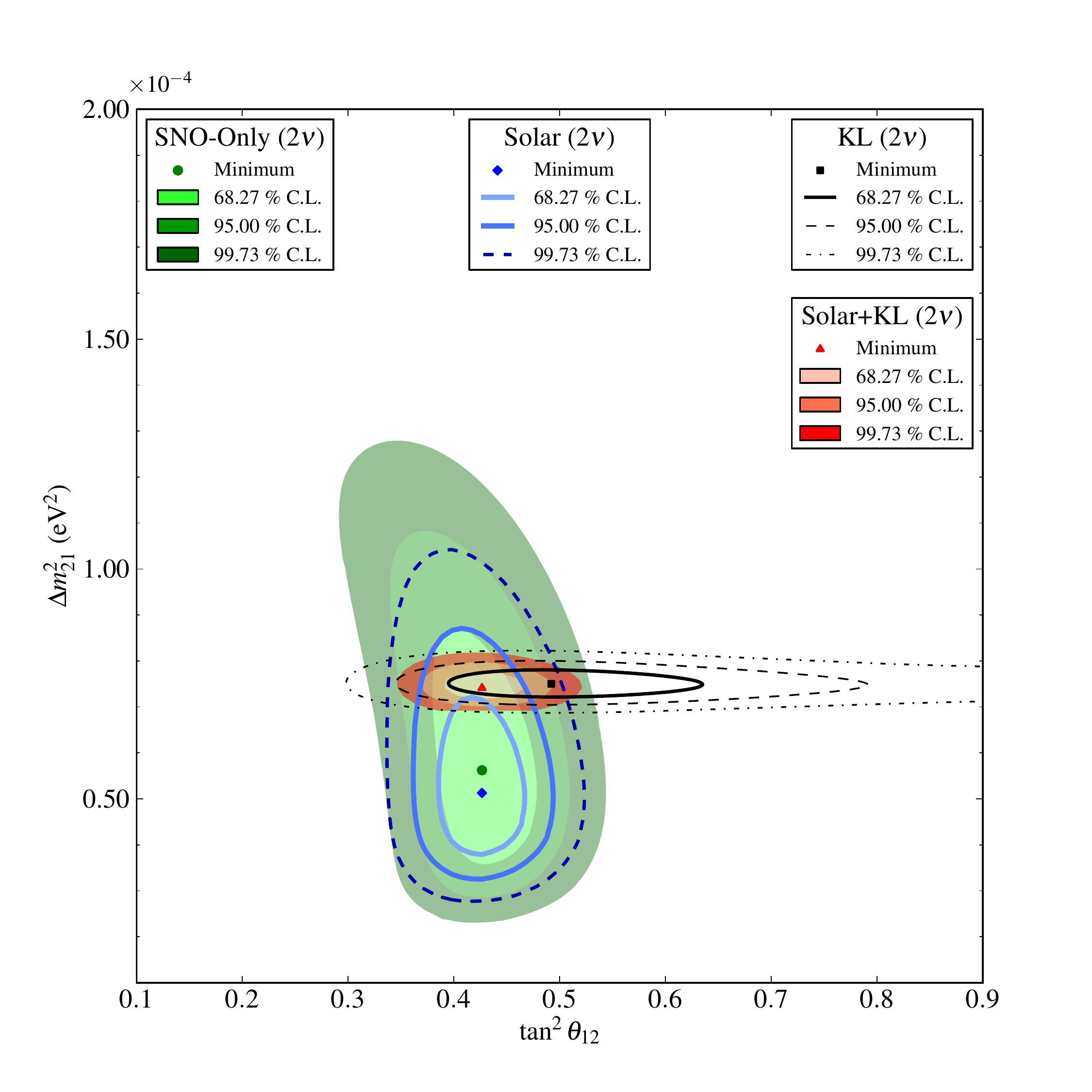}
\caption{Two-flavor neutrino oscillation analysis contour using both solar neutrino and KamLAND (KL) results.}
\label{fig:final:global:2nu}
\end{figure}

Figure~\ref{fig:final:global:2nu} shows the allowed regions of the $(\tanthetaonetwo{},\Dmonetwo{})$ parameter space obtained when the SNO results were combined with the other solar neutrino experimental results, and when this combined solar neutrino result was combined with the results from the KamLAND (KL) experiment. The combination of the SNO results with the other solar neutrino experimental results eliminates the LOW region, and eliminates the higher values of $\Dmonetwo{}$ in the LMA region.

Table~\ref{tbl:final:sno} summarizes the results from these two-flavor neutrino analyses.

\begingroup
\squeezetable
\begin{table}[htdp]
\centering
\caption{Best-fit neutrino oscillation parameters from a two-flavor neutrino oscillation analysis. Uncertainties listed are $\pm 1\sigma$ after the \chis{} was minimized with respect to all other parameters.}
\begin{tabular}{lccc}
\hline\hline
Oscillation analysis & $\tanthetaonetwo{}$ & $\Dmonetwo [{\rm eV^{2}}]$ & $\nicefrac{\chi^{2}}{\rm NDF}$\\
\hline
SNO only (LMA)	&$0.427^{+0.033}_{-0.029}$	& $5.62^{+1.92}_{-1.36}\times 10^{-5}$	& $\nicefrac{1.39}{3}$\\
SNO only (LOW) 	&$0.427^{+0.043}_{-0.035}$	& $1.35^{+0.35}_{-0.14}\times 10^{-7}$	& $\nicefrac{1.41}{3}$ \\
Solar 			&$0.427^{+0.028}_{-0.028}$	& $5.13^{+1.29}_{-0.96}\times 10^{-5}$	& $\nicefrac{108.07}{129}$\\
Solar+KamLAND	&$0.427^{+0.027}_{-0.024}$	& $7.46^{+0.20}_{-0.19}\times 10^{-5}$	& \\
\hline\hline
\end{tabular}
\label{tbl:final:sno}
\end{table}
\endgroup

\subsection{Three-flavor neutrino oscillation analysis}
\label{sec:threeneu}

\begin{figure}[tbp]	
\centering
\subfigure{\includegraphics[width=1.0\columnwidth]{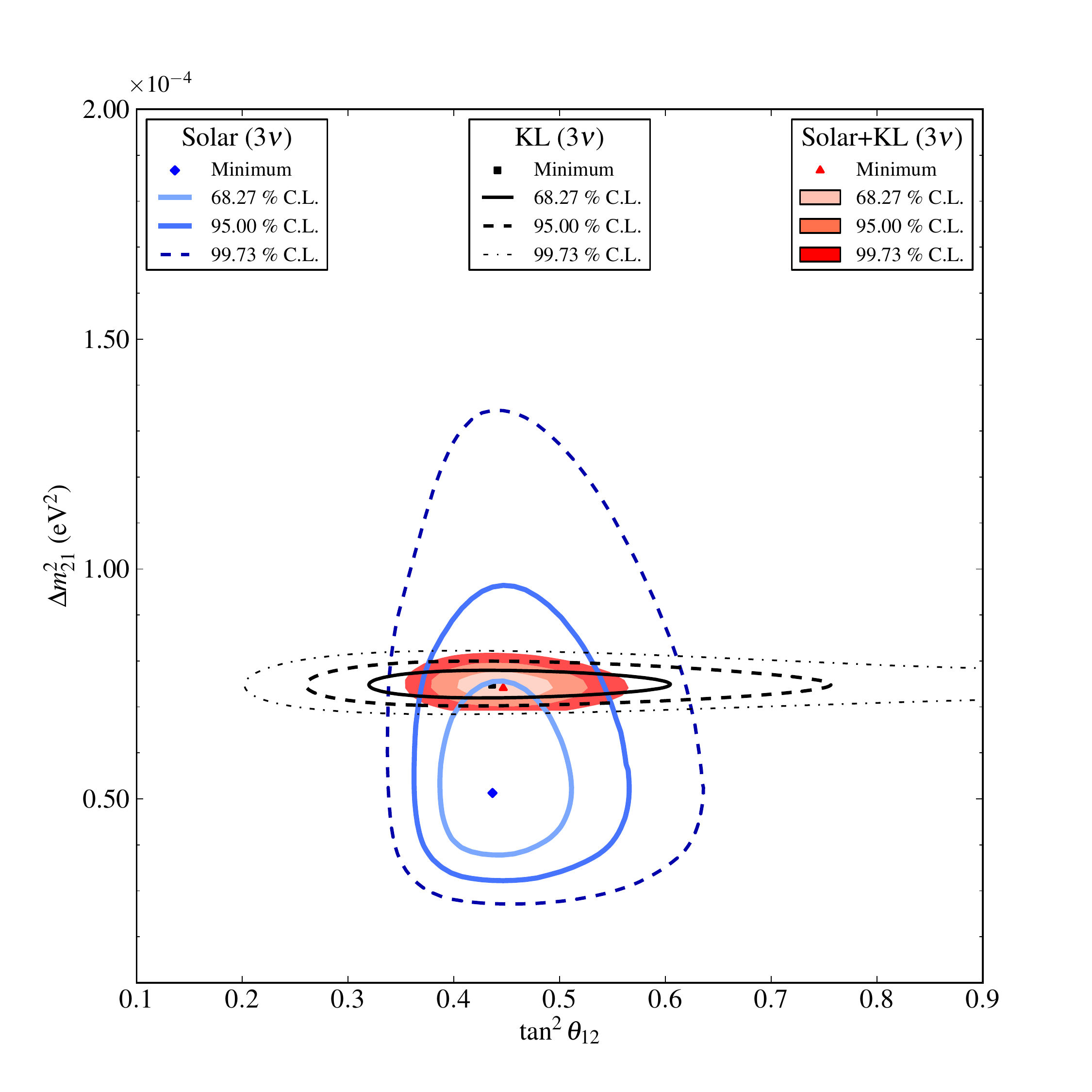}\label{fig:final:global:3nu:12}}
\subfigure{\includegraphics[width=1.0\columnwidth]{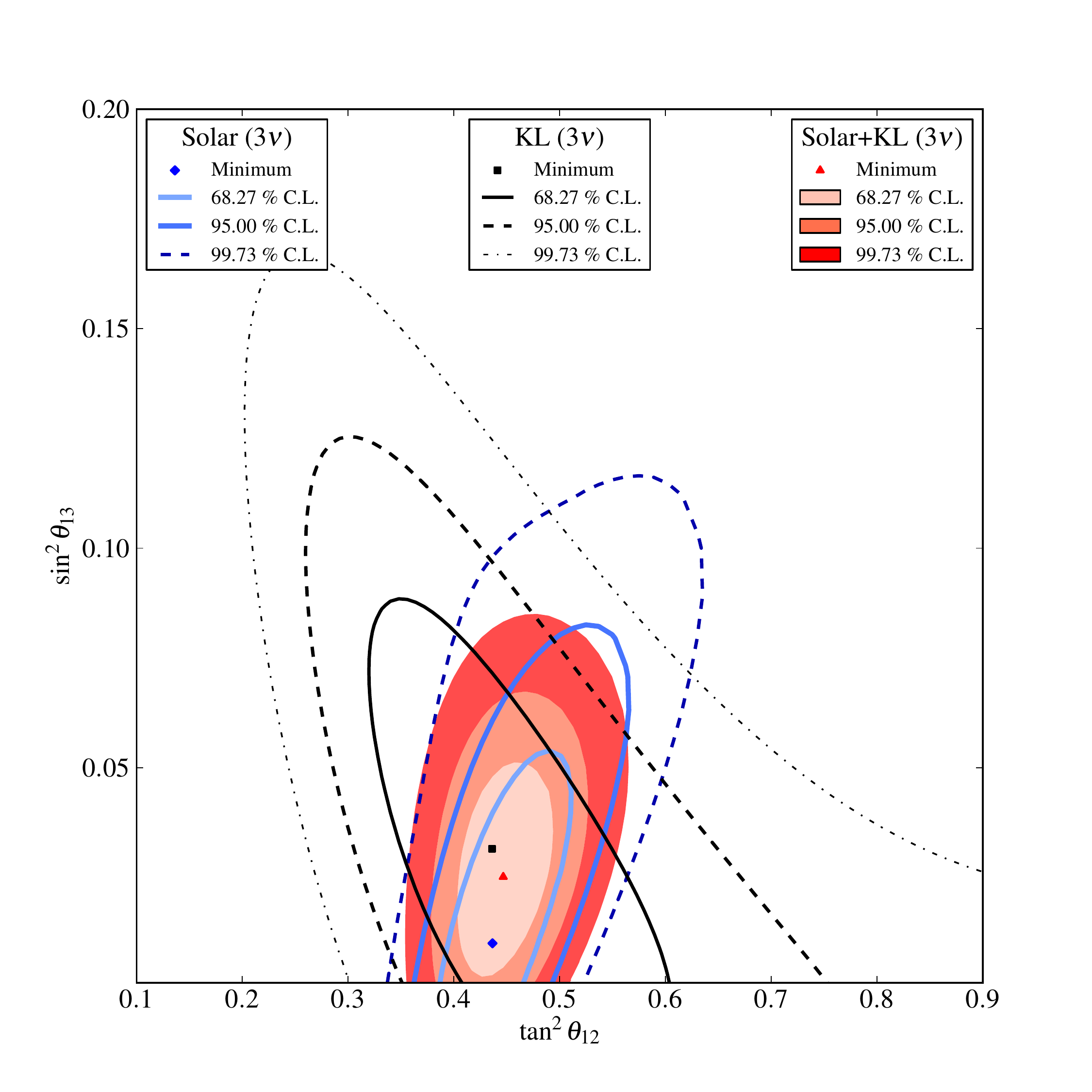}\label{fig:final:global:3nu:13}}
\caption{Three-flavor neutrino oscillation analysis contour using both solar neutrino and KamLAND (KL) results.}
\label{fig:final:global:3nu}
\end{figure}

Figure~\ref{fig:final:global:3nu} shows the allowed regions of the $(\tanthetaonetwo{}, \Dmonetwo{})$ and $(\tanthetaonetwo{}, \sinthetaonethree{})$ parameter spaces obtained from the results of all solar neutrino experiments. It also shows the result of these experiments combined with the results of the KamLAND experiment. Compared to the result in Figure~\ref{fig:final:global:2nu}, this clearly shows that allowing non-zero values of $\thetaonethree$ brings the solar neutrino experimental results into better agreement with the results from the KamLAND experiment.

\begin{figure}[tbp]	
\centering
\subfigure{\includegraphics[width=0.9\columnwidth]{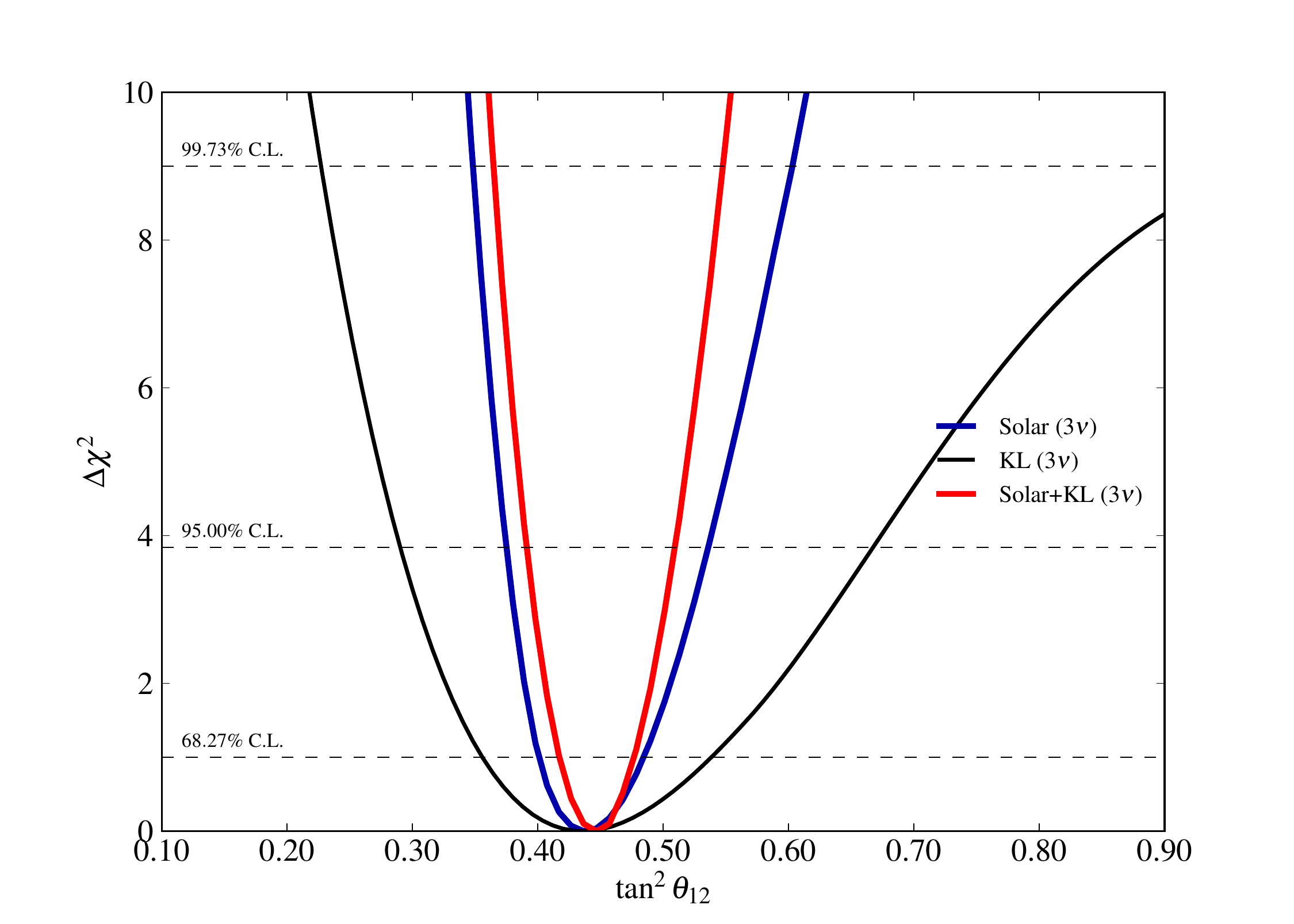}\label{fig:final:global:3nu:proj:th12}}
\subfigure{\includegraphics[width=0.9\columnwidth]{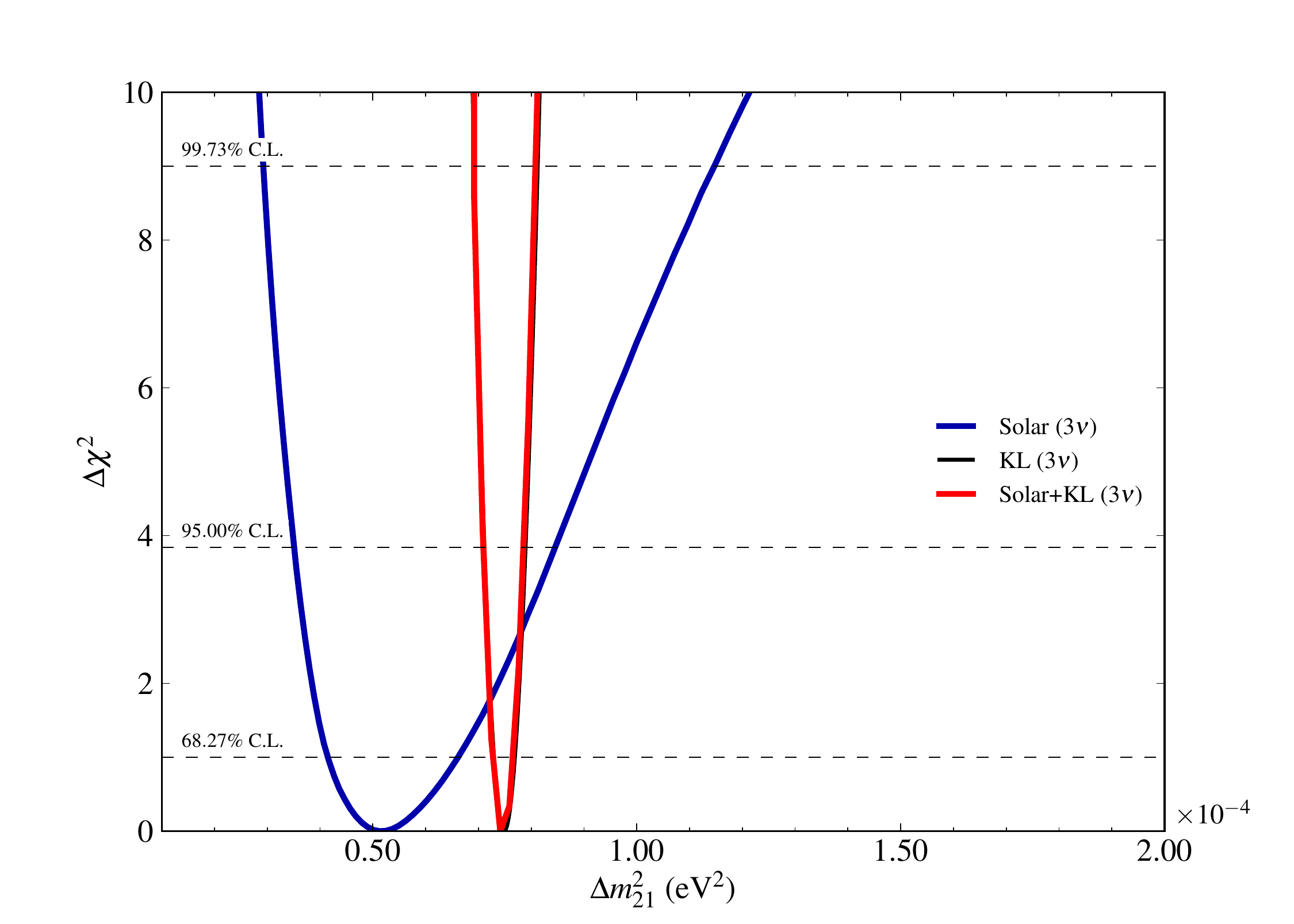}\label{fig:final:global:3nu:proj:dm21}}\\
\subfigure{\includegraphics[width=0.9\columnwidth]{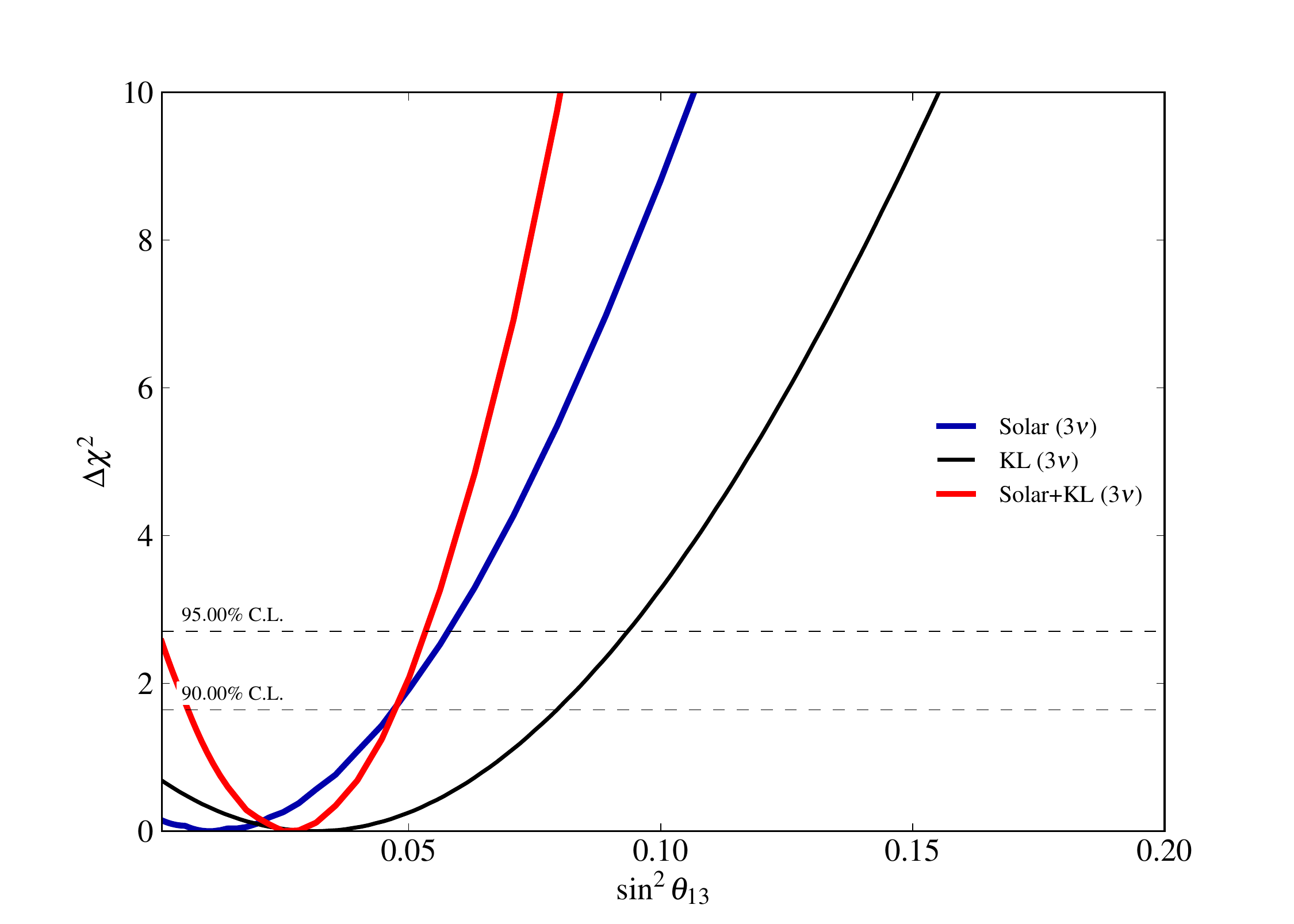}\label{fig:final:global:3nu:proj:th13}}
\caption{Projections of the three-flavor neutrino oscillation parameters determined from Figure \ref{fig:final:global:3nu}. The horizontal lines representing the $\Delta\chi^2$ for a particular confidence level, are for two-sided intervals in plot (a) and (b), and one-sided intervals in plot (c).}
\label{fig:final:global:3nu:proj}
\end{figure}

Figure~\ref{fig:final:global:3nu:proj} shows the projection of these results onto the individual oscillation parameters. This result shows that due to the different dependence between $\tanthetaonetwo{}$ and $\sinthetaonethree{}$ for the solar neutrino experimental results and the KamLAND experimental results, the combined constraint on $\sinthetaonethree{}$ was significantly better than the individual constraints.

Table~\ref{tbl:final:global} summarizes the results from these three-flavor neutrino oscillation analyses. Tests with the inverted hierarchy, i.e. negative values of $\Dmonethree{}$, gave essentially identical results~\cite{thesisNuno}.

\begingroup
\squeezetable
\begin{table}[htpd]
\centering
\caption{Best-fit neutrino oscillation parameters from a three-flavor neutrino oscillation analysis. Uncertainties listed are $\pm 1\sigma$ after the \chis{} was minimized with respect to all other parameters. The global analysis includes Solar+KL+ATM+LBL+CHOOZ.}
\begin{tabular}{lcccc}
\hline\hline
Analysis			&$\tanthetaonetwo{}$ & $\Dmonetwo [{\rm eV^{2}}]$ & $\sinthetaonethree (\times 10^{-2})$\\
\hline
Solar			&$0.436^{+0.048}_{-0.036}$	&$5.13^{+1.49}_{-0.98}\times 10^{-5}$	&$<5.8$ (95\% C.L.)\\
Solar+KL			&$0.446^{+0.030}_{-0.029}$	&$7.41^{+0.21}_{-0.19}\times 10^{-5}$	&$2.5^{+1.8}_{-1.5}$\\
				&						&								&$<5.3$ (95\% C.L.)\\
Global			&						&								&$2.02^{+0.88}_{-0.55}$\\
\hline\hline
\end{tabular}
\label{tbl:final:global}
\end{table}
\endgroup

\begin{figure}[tbp]	
\centering
\includegraphics[width=1.0\columnwidth]{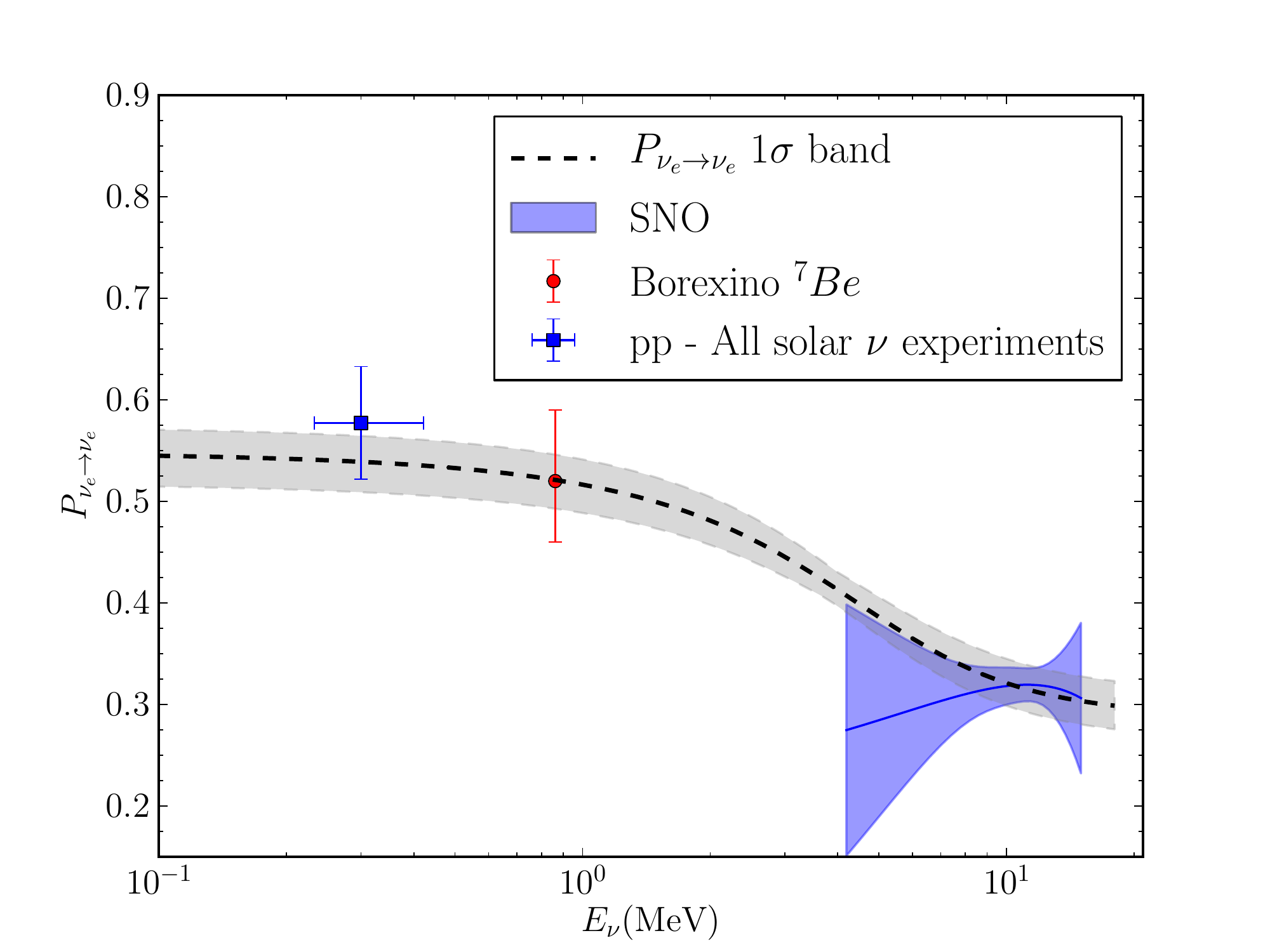}
\caption{Various solar \nue{} survival probability measurements compared to the LMA prediction for \B{}. Using the results from Section~\ref{sec:oscillations} of this paper, the dashed line is the best fit LMA solution for \B{}s and the gray shaded band is the 1$\sigma$ uncertainty. The corresponding bands for \nue{}s from the pp and $\iso{7}{Be}$ reactions (not shown) are almost identical in the region of those measurements. The blue shaded band is the result of the measurement the \B{} \nue{} survival probability reported here. The red point is the result of the Borexino measurement~\cite{:1343896} of the survival probability for \nue{}s produced by $\iso{7}{Be} + e^- \rightarrow \iso{7}{Li} + \nu_e$ reactions in the Sun. The blue point is the result of various measurements~\cite{Gallium2009, *Altmann2005174, *thesis:Kaether} of the survival probability for \nue{}s produced by $p+p \rightarrow \iso{2}{H} + e^+ + \nu_e$ reactions in the Sun; note that these measurements did not exclusively measure this reaction, so the contribution from other reactions were removed assuming the best fit LMA solution, and so actually depends on all solar neutrino results. The uncertainty in absolute flux of the subtracted reactions was included in the calculation of the total uncertainty of this point, but the uncertainty due to the neutrino oscillation probability of these reactions was not. The uncertainty due to the normalization of the two points by the expected flux was included. For clarity, this plot illustrates the LMA solution relative to only a subset of the solar neutrino experimental results.}
\label{fig:survivalProb}
\end{figure}

Figure~\ref{fig:survivalProb} shows the measured solar \nue{} survival probability as a function of $E_\nu$. At higher $E_\nu$ the results of this analysis provide the best constraints on the survival probability. All solar results are consistent with the LMA neutrino oscillation hypothesis.

\begin{figure}[tbp]	
\centering
\includegraphics[width=1.0\columnwidth]{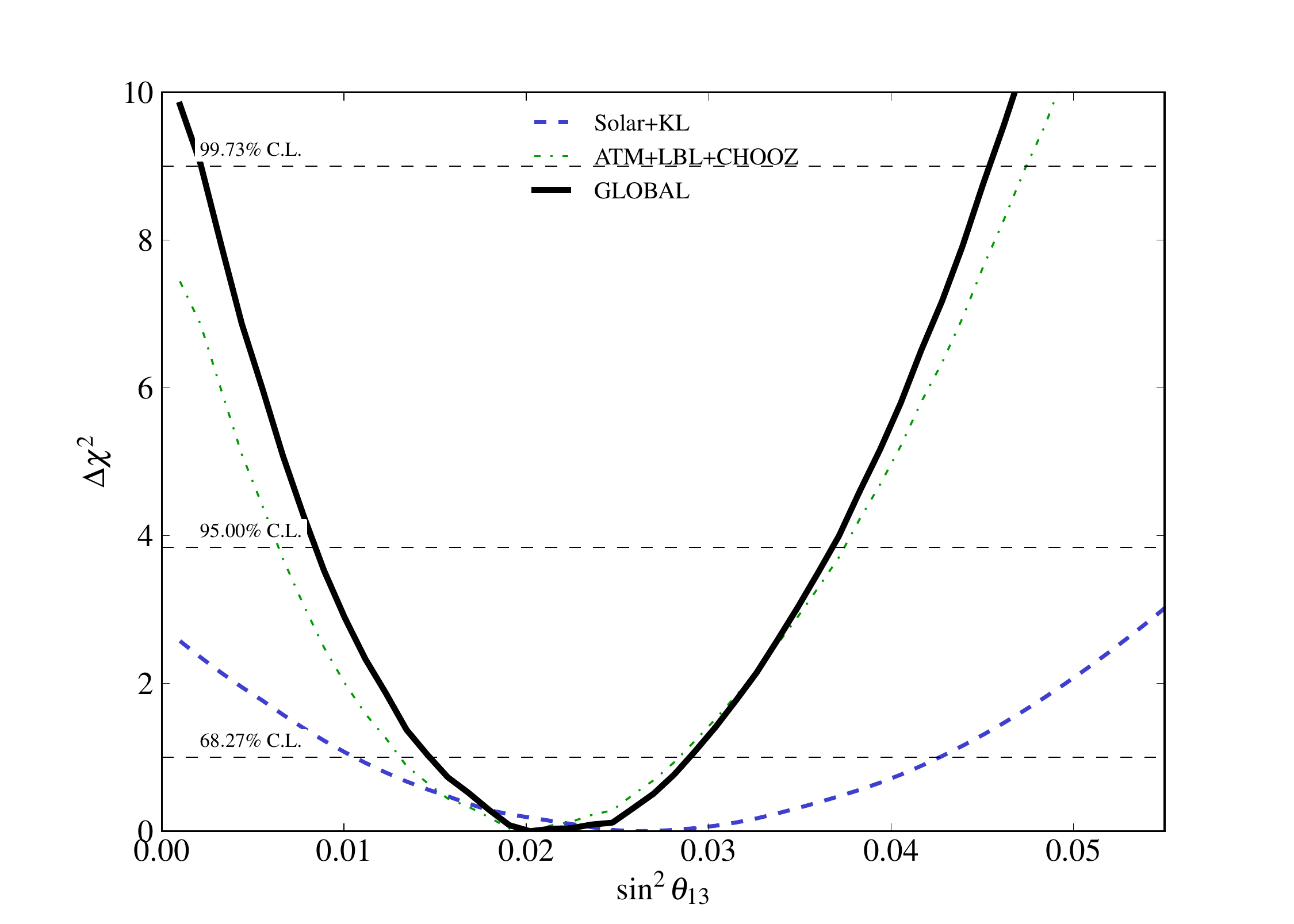}
\caption{Projection over $\sin^{2}\theta_{13}$ combining the projections obtained by analyzing data from all neutrino sources. The data from atmospherics, short-baseline experiments and long-baseline experiments (ATM+LBL+CHOOZ) was determined from Figure 2-left of Reference~\cite{Fogli:2011fk} which already includes the latest T2K~\cite{Collaboration:2011uq} and MINOS~\cite{2011arXiv1108.0015M} results.}
\label{fig:final:th13:global}
\end{figure}

Recent results from the T2K~\cite{Collaboration:2011uq} and MINOS~\cite{2011arXiv1108.0015M} long-baseline (LBL) experiments indicate a non-zero $\thetaonethree{}$ with a significance of approximately $2.7\sigma$. A combined analysis of all LBL and atmospheric (ATM) results, and the results from the CHOOZ~\cite{springerlink:10.1140/epjc/s2002-01127-9} experiment was performed by Fogli~{\it et al.}~\cite{Fogli:2011fk}. Because the LBL+ATM+CHOOZ analysis was insensitive to $\thetaonetwo{}$, and because the solar neutrino+KamLAND analysis was insensitive to $\Dmonethree{}$ we can simply add their projections of $\Delta\chi^2$ onto $\thetaonethree{}$. Table~\ref{tbl:final:global} and Figure~\ref{fig:final:th13:global} show the results of that combination. This shows that the LBL+ATM+CHOOZ experiments currently have better sensitivity to $\thetaonethree{}$ than the combined solar and KamLAND experiments, but the combination of all experiments gives a slightly improved determination of $\thetaonethree{}$, hinting at a non-zero value.

\section{Discussion and Conclusions}

By developing a particle identification cut to analyze data from Phase III of the SNO experiment we measured $1115\pm79$ neutrons. Eliminating most of the background from alpha events and allowing a very general PDF to describe the $\Encd{}$ spectrum for any remaining background events made this analysis less sensitive to background uncertainties than our previous analysis of these data.

Combining data from all phases of the SNO experiment we measured a total flux of active flavor neutrinos from $\iso{8}{B}$ decays in the Sun of \numberSNOBflux{}. We improved the handling of a number of systematic uncertainties in this analysis compared with our previous analyses of these data. This result was consistent with but more precise than both the BPS09(GS), \bpsFlux{}, and BPS09(AGSS09), \bpsAGSflux{}, solar model predictions~\cite{ssh:bps09}.

The precision of the \nue{} survival probability parameters was improved by approximately 20\% compared to our previously reported results due to the additional constraint provided by the data from Phase III. During the day the \nue{} survival probability at 10\,MeV was \numberPeea{}, which was inconsistent with the null hypothesis that there were no neutrino oscillations at very high significance. The null hypotheses that there were no spectral distortions of the \nue{} survival probability (i.e. $\Peeb{}=0$, $\Peec{}=0$, $\Aeea{}=0$, $\Aeeb{}=0$), and that there were no day/night distortions of the \nue{} survival probability (i.e. $\Aeea{}=0$, $\Aeeb{}=0$) could not be rejected at the 95\% C.L.

 A two-flavor neutrino oscillation analysis yielded \numberSNODmonetwo{} and \numberSNOThetaonetwo{}. A three-flavor neutrino oscillation analysis combining this result with results of all other solar neutrino experiments and the KamLAND experiment yielded \numberGlobalDmonetwo{}, \numberGlobalThetaonetwo{}, and \numberGlobalThetaonethree{}. This implied an upper bound of \numberGlobalThetaonethreeLimit{} at the 95\% C.L.

\section{Acknowledgments}

This research was supported by: Canada: Natural Sciences and Engineering Research Council, Industry Canada, National Research Council, Northern Ontario Heritage Fund, Atomic Energy of Canada, Ltd., Ontario Power Generation, High Performance Computing Virtual Laboratory, Canada Foundation for Innovation, Canada Research Chairs; US: Department of Energy, National Energy Research Scientific Computing Center, Alfred P. Sloan Foundation; UK: Science and Technology Facilities Council; Portugal: Funda\c{c}\~{a}o para a Ci\^{e}ncia e a Tecnologia. We thank the SNO technical staff for their strong contributions. We thank Vale (formerly Inco, Ltd.) for hosting this project.

\bibliographystyle{apsrev4-1}
\bibliography{3PhasePaper}

\appendix{}

\section{Sterile neutrinos}
\label{apx:sterile}

If we assume a sterile neutrino, where the probability of an electron neutrino oscillating into a sterile neutrino, $\Pes{}$, was the same during the day and night, then the scaling factors given in Table~\ref{tab:scalings}
of Section~\ref{sec:parameterization} are replaced with those in Table~\ref{tab:scalings2}.

\begingroup
\squeezetable
\begin{table}[htdp]
\caption{\B{} interactions scaling factors including a probability of an electron neutrino oscillating into a sterile neutrino, which was the same during the day and night. $\Peen{}=\Peed{}\frac{2+\Aee{}}{2-\Aee{}}$, and $f(E_\nu)$ was the predicted spectrum of $E_\nu$ detectable by the SNO detector after including the energy dependence of the cross-section.}
\begin{center}
\begin{tabular}{lll}
\hline\hline
Interaction&			Day/Night&	Scaling factor\\
\hline
$\rm {CC,ES_e}$&		Day&		$\fB{}\Peed{}$\\
$\rm {ES_{\mu\tau}}$&	Day&		$\fB{}[1-\Peed{}-\Pes{}]$\\
$\rm {CC,ES_e}$&		Night&		$\fB{}\Peen{}$\\
$\rm {ES_{\mu\tau}}$&	Night&		$\fB{}[1-\Peen{}-\Pes{}]$\\
$\rm {NC}$&			Day+Night&	$\fB{}\frac{\int{(1-\Pes{})f(E_\nu)dE_\nu}}{\int{f(E_\nu)dE_\nu}}$\\
\hline\hline
\end{tabular}
\end{center}
\label{tab:scalings2}
\end{table}
\endgroup

If $\Pes{}$ was a constant as a function of $E_\nu$, and defining $\fB{}' = \fB{}(1-\Pes{})$, and $\Peed{}' = \frac{\Peed{}}{1-\Pes{}}$ we obtain the scaling factors given in Table~\ref{tab:scalings3}.

\begingroup
\squeezetable
\begin{table}[htdp]
\caption{\B{} interactions scaling factors. $\Peen{}'=\Peed{}'\frac{2+\Aee{}}{2-\Aee{}}$.}
\begin{center}
\begin{tabular}{lll}
\hline\hline
Interaction&			Day/Night&	Scaling factor\\
\hline
$\rm {CC,ES_e}$&		Day&		$\fB{}'\Peed{}'$\\
$\rm {ES_{\mu\tau}}$&	Day&		$\fB{}'[1-\Peed{}']$\\
$\rm {CC,ES_e}$&		Night&		$\fB{}'\Peen{}'$\\
$\rm {ES_{\mu\tau}}$&	Night&		$\fB{}'[1-\Peen{}']$\\
$\rm {NC}$&			Day+Night&	$\fB{}'$\\
\hline\hline
\end{tabular}
\end{center}
\label{tab:scalings3}
\end{table}
\endgroup

Notice that scaling factors in Table~\ref{tab:scalings3} are equivalent to those in Table~\ref{tab:scalings}, except our measurement of the \B{} flux would be the true flux scaled by $(1-\Pes{})$ and our measurement of $\Peed{}$ would be scaled by $1/(1-\Pes{})$.

We note that the approximations made in this analysis were also valid for effects involving sterile neutrinos with spectral distortion that are significant only below about 4\,MeV and with small day-night effects for the NC detection process. In this case the principal additional effect would be a further renormalization of the NC interaction rate.

\section{Constraints on the likelihood fit}
\label{apx:systematic}

Systematic uncertainties were propagated on the PDFs as described in this section.

\subsection{{\boldmath $\teff{}$} PDFs}

Table~\ref{tab:Esys} lists the systematic uncertainties on the PDFs of $\teff{}$. For Phases I and II the modified $\teff{}$, $\teff{}'$, during the day was obtained from
\begin{eqnarray}
\teff{}' &=& (1+{a_0^E}_c+a_0^E-A_{dn}^E/2-A_{dir}^E/2)\teff{}.
\end{eqnarray}
The signs of the $A_{dn}^E$ and $A_{dir}^E$ terms were reversed for night. The uncertainty due to $\teff{}$ resolution was obtained by convolving $\teff{}'$ with Gaussians centered at zero with widths of $\sigma^E$, and $\sigma_{dn}^E$ and $\sigma_{dir}^E$ scaled by a parameterized detector resolution. This resolution was applied first to just day events and then to just night events. Differences in the neutrino parameters between the shifted fits and the central fit were taken as the resulting uncertainties.

During Phase III the modified $\teff{}$, $\teff{}''$, was obtained from
\begin{eqnarray}
\teff{}'' &=& \teff{}'+b_0^E(1-B_{dir}^E/2)(\teff{}'-T_g),
\end{eqnarray}
where $T_g$ was the true MC energy. For the NC and external neutrons PDFs $T_g$ was constant and equal to the mean fitter energy 5.65 MeV. $\teff{}'$, was obtained from
\begin{eqnarray}
\teff{}' &=& (1+{a_0^E}_c+a_0^E)(1-A_{dn}^E/2-A_{dir}^E/2)\teff{},
\end{eqnarray}
where the signs of the $A_{dn}^E$ and $A_{dir}^E$ terms were reversed for night.

\begingroup
\squeezetable
\begin{table*}[htdp]
\caption{$\teff{}$ PDF systematic uncertainties.}
\begin{center}
\begin{tabular}{llllrrrr}
\hline\hline
Parameter&			Description&						Events&			Phase&	Nominal&		Variation&				Fit Value&		Application\\
\hline
${a_0^E}_c$&			$\teff{}$ scale&						all&				I, II, III &	0&			$\pm$0.0041&				$0.0004^{+0.0033}_{-0.0024}$&	scanned\\
$a_0^E$&			$\teff{}$ scale&						all&				I&		0&			$^{+0.0039}_{-0.0047}$&		$-0.0007^{+0.0038}_{-0.0030}$&	scanned\\
$a_0^E$&			$\teff{}$ scale&						all&				II&		0&			$^{+0.0034}_{-0.0032}$&		$0.0001^{+0.0026}_{-0.0027}$&	scanned\\
$a_0^E$&			$\teff{}$ scale&						all&				III&		0&			$\pm0.0081$&				$0.0065^{+0.0042}_{-0.0084}$&	scanned\\
$c_0^E$&				$\teff{}$ scale non-linearity with $\teff{}$&	all&				I, II, III&	0&			$\pm0.0069$&				N/A&							shift-and-refit\\
$A_{dn}^E$&			$\teff{}$ scale diurnal variation&		all&				I&		0&			$\pm0.0032$&				N/A&							shift-and-refit\\
$A_{dir}^E$&			$\teff{}$ scale directional variation&		CC&				I&		0&			$\pm0.0009$\footnotemark[1]&	N/A&							shift-and-refit\\
$A_{dir}^E$&			$\teff{}$ scale directional variation&		ES&				I&		0&			$\pm0.0092$\footnotemark[1]&	N/A&							N/A\\
$A_{dn}^E$&			$\teff{}$ scale diurnal variation&		all&				II&		0&			$\pm0.004$&				N/A&							shift-and-refit\\
$A_{dir}^E$&			$\teff{}$ scale directional variation&		CC&				II&		0&			$\pm0.0009$\footnotemark[2]&	N/A&							shift-and-refit\\
$A_{dir}^E$&			$\teff{}$ scale directional variation&		ES&				II&		0&			$\pm0.0079$\footnotemark[2]&	N/A&							N/A\\
$A_{dn}^E$&			$\teff{}$ scale diurnal variation&		all&				III&		0&			$\pm0.0038$&				$0.0005^{+0.0037}_{-0.0035}$&	scanned\\
$A_{dir}^E$&			$\teff{}$ scale directional variation&		ES&				III&		0&			$\pm0.0099$&				$-0.0038^{+0.0099}_{-0.0096}$&	scanned\\
$\sigma^E$ [MeV]&		$\teff{}$ resolution&					all&				I&		0.155&		$^{+0.041}_{-0.080}$&		$0.214^{+0.023}_{-0.034}$&		scanned\\ 
$\sigma^E$ [MeV]&		$\teff{}$ resolution&					e$^-$, $\gamma$&	II&		0.168&		$^{+0.041}_{-0.080}$&		$0.203^{+0.033}_{-0.041}$&		scanned\\
$\sigma^E$ [MeV]&		$\teff{}$ resolution&					n&				II&		0.154&		$\pm0.018$&				$0.155^{+0.017}_{-0.019}$&		scanned\\
$b_0^E$&			$\teff{}$ resolution&					n&				III&		0.0119&		$\pm0.0104$\footnotemark[3]&	$0.0109^{+0.0107}_{-0.0100}$&	scanned\\
$b_0^E$&			$\teff{}$ resolution&					e$^-$, $\gamma$&	III&		0.016184&	$\pm0.0141$\footnotemark[3]&	N/A&							N/A\\
$\sigma_{dn}^E$&		$\teff{}$ resolution diurnal variation&	all&				I&		0&			$\pm0.003$&				N/A&							shift-and-refit\\
$\sigma_{dir}^E$&		$\teff{}$ resolution directional variation&	CC&				I&		0&			$\pm0.0014$\footnotemark[4]&	N/A&							shift-and-refit\\
$\sigma_{dir}^E$&		$\teff{}$ resolution directional variation&	ES&				I&		0&			$\pm0.0064$\footnotemark[4]&	N/A&							N/A\\
$\sigma_{dn}^E$&		$\teff{}$ resolution diurnal variation&	all&				II&		0&			$\pm0.005$&				N/A&							shift-and-refit\\
$\sigma_{dir}^E$&		$\teff{}$ resolution directional variation&	CC&				II&		0&			$\pm0.0013$\footnotemark[5]&	N/A&							shift-and-refit\\
$\sigma_{dir}^E$&		$\teff{}$ resolution directional variation&	ES&				II&		0&			$\pm0.013$\footnotemark[5]&	N/A&							N/A\\
$B_{dir}^E$&			$\teff{}$ resolution directional variation&	ES&				III&		0&			$\pm0.012$&				$0.000\pm0.012$&				scanned\\
\hline\hline
\end{tabular}
\footnotetext[1]{Correlation of -1.}
\footnotetext[2]{Correlation of -1.}
\footnotetext[3]{Correlation of 1.}
\footnotetext[4]{Correlation of -1.}
\footnotetext[5]{Correlation of -1.}
\end{center}
\label{tab:Esys}
\end{table*}
\endgroup

\subsection{{\boldmath $\be{}$} PDFs}

Table~\ref{tab:besys} lists the systematic uncertainties on the PDFs of $\be{}$. During the day these uncertainties were applied using
\begin{eqnarray}
\nonumber \be{}'&=&\be{}(1+a_0^{\be}+c_0^\beta(\teff{}-5.589[{\rm MeV}]))\\
&&-A_{dn}^\beta/2-A_{dir}^\beta/2.
\end{eqnarray}
The signs of the $A_{dn}^\beta$ and $A_{dir}^\beta$ terms were reversed for night. For electrons and $\gamma$-rays the systematic uncertainties associated with $\be{}$ resolution were applied as
\begin{eqnarray}
\be{}''=\be{}'+(\be{}'-\bar{\beta}_{14})b_0^\beta,
\end{eqnarray}
where $\be{}''$ was the value of $\be{}$ including all of the systematic uncertainties, and $\bar{\beta}_{14}$ a parameterized average value of $\be{}$ for the PDF. For neutrons in Phase II the systematic uncertainties associated with resolution were applied as a convolution with a Gaussian centered at zero with a width of $\sigma^{\be}$. This correction can only be applied in the positive direction, and the negative fit uncertainties were inferred to be the same as the positive ones.

\begingroup
\squeezetable
\begin{table*}[htdp]
\caption{$\be{}$ PDF systematic uncertainties.}
\begin{center}
\begin{tabular}{llllrrrr}
\hline\hline
Parameter&		Description&						Events&			Phase&	Nominal&				Variation&			Fit Value&		Application\\
\hline
$a_0^\beta$&		$\be{}$ scale&						all&				I&		-0.0081&				$\pm0.0042$\footnotemark[1]&		N/A&								scanned\\
$a_0^\beta$&		$\be{}$ scale&						e$^-$&			II&		0&					$\pm0.0024$\footnotemark[1]&		$0.00102^{+0.00112}_{-0.00205}$&	N/A\\
$a_0^\beta$&		$\be{}$ scale&						n&				II&		-0.0144&				$^{+0.0038}_{-0.0022}$&			$-0.0138^{+0.0036}_{-0.0025}$&		scanned\\
$c_0^\beta$&		$\be{}$ scale non-linearity with $\teff{}$&	all&				I, II&		0.00275597&			$\pm0.00069$&				$0.00201^{+0.00058}_{-0.00044}$&	scanned\\
$A_{dn}^\beta$&	$\be{}$ offset diurnal variation&		all&				I&		0&					$\pm0.0043$&					N/A&								shift-and-refit\\
$A_{dir}^\beta$&	$\be{}$ offset directional variation&		CC&				I&		0&					$\pm0.00038$\footnotemark[2]&	N/A&								shift-and-refit\\
$A_{dir}^\beta$&	$\be{}$ offset directional variation&		ES&				I&		0&					$\pm0.0034$\footnotemark[2]&		N/A&								N/A\\
$A_{dn}^\beta$&	$\be{}$ offset diurnal variation&		all&				II&		0&					$\pm0.0043$&					N/A&								shift-and-refit\\
$A_{dir}^\beta$&	$\be{}$ offset directional variation&		CC&				II&		0&					$\pm0.00038$\footnotemark[3]&	N/A&								shift-and-refit\\
$A_{dir}^\beta$&	$\be{}$ offset directional variation&		ES&				II&		0&					$\pm0.0034$\footnotemark[3]&		N/A&								N/A\\
$b_0^\beta$&		$\be{}$ resolution&					all&				I&		0&					$\pm0.0042$\footnotemark[4]&		N/A&								shift-and-refit\\
$b_0^\beta$&		$\be{}$ resolution&					e$^-$&			II&		0&					$\pm0.0054$\footnotemark[4]&		N/A&								N/A\\
$\sigma^\beta$&	$\be{}$ resolution&					n&				II&		0.0150&				$\pm0.0045$&					N/A&								shift-and-refit\\ 
\hline\hline
\end{tabular}
\footnotetext[1]{Correlation of 1.}
\footnotetext[2]{Correlation of -1.}
\footnotetext[3]{Correlation of -1.}
\footnotetext[4]{Correlation of 1.}
\end{center}
\label{tab:besys}
\end{table*}
\endgroup

\subsection{{\boldmath $\rho$} PDFs}

Table~\ref{tab:possys} lists the systematic uncertainties on the PDFs of $\rho$. The radius was modified by
\begin{eqnarray}
\rho '' &=& \frac{\left [(x''[{\rm cm}])^2+(y''[{\rm cm}])^2+(z''[{\rm cm}])^2\right ]^{3/2}}{600^3},
\end{eqnarray}
where $x''$, $y''$, and $z''$ were the modified cartesian coordinates, as described below. Each event was weighted by a factor $1+c^{\rho} \times (\teff{}-5.05[{\rm MeV}])$.

\begingroup
\squeezetable
\begin{table*}[htdp]
\caption{$\rho$ PDF systematic uncertainties.}
\begin{center}
\begin{tabular}{llllrrrr}
\hline\hline
Parameter&				Description&						Events&		Phase&	Nominal&				Variation&						Fit Value&			Application\\
\hline
$a_1^\rho$&				$\rho$ scale&						all&			I&		0&					$^{+0.0010}_{-0.0057}$&				N/A&				shift-and-refit\\
$a_1^{z}$&				z scale&							all&			I&		0&					$^{+0.0040}_{-0.0}$&				N/A&				shift-and-refit\\
$a_1^\rho$&				$\rho$ scale&						all&			II&		0&					$^{+0.0004}_{-0.0034}$&				N/A&				shift-and-refit\\
$a_1^{z}$&				z scale&							all&			II&		0&					$^{+0.0003}_{-0.0025}$&				N/A&				shift-and-refit\\
$a_1^\rho$&				$\rho$ scale&						all&			III&		0&					$^{+0.0029}_{-0.0077}$&				$0.0004^{+0.0027}_{-0.0051}$&	scanned\\
$a_1^{z}$&				z scale&							all&			III&		0&					$^{+0.0015}_{-0.0012}$&				N/A&				shift-and-refit\\
$c^{\rho}$&				$\rho$ scale non-linerity with $\teff{}$&	all&			I&		0&					$^{+0.0085}_{-0.0049}$&				N/A&				shift-and-refit\\
$c^{\rho}$&				$\rho$ scale non-linerity with $\teff{}$&	all&			II&		0&					$^{+0.0041}_{-0.0048}$&				N/A&				shift-and-refit\\
$c^{\rho}$&				$\rho$ scale non-linerity with $\teff{}$&	all&			III&		0&					$^{+0.0088}_{-0.0067}$&				N/A&				shift-and-refit\\
$A_{dn}^\rho$&			$\rho$ scale diurnal variation&			all&			I&		0&					$\pm0.002$&						N/A&				shift-and-refit\\
$A_{dir}^\rho$&			$\rho$ scale directional variation&		CC&			I&		0&					$\pm0.0004$\footnotemark[1]&			N/A&				shift-and-refit\\
$A_{dir}^\rho$&			$\rho$ scale directional variation&		ES&			I&		0&					$\pm0.005$\footnotemark[1]&			N/A&				N/A\\
$A_{dn}^\rho$&			$\rho$ scale diurnal variation&			all&			II&		0&					$\pm0.003$&						N/A&				shift-and-refit\\
$A_{dir}^\rho$&			$\rho$ scale directional variation&		CC&			II&		0&					$\pm0.0002$\footnotemark[2]&			N/A&				shift-and-refit\\
$A_{dir}^\rho$&			$\rho$ scale directional variation&		ES&			II&		0&					$\pm0.0015$\footnotemark[2]&			N/A&				N/A\\
$A_{dn}^\rho$&			$\rho$ scale diurnal variation&			all&			III&		0&					$\pm0.0015$&						N/A&				shift-and-refit\\
$A_{dir}^\rho$&			$\rho$ scale directional variation&		ES&			III&		0&					$\pm0.0018$&						N/A&				shift-and-refit\\
$a_0^{x}$ [cm]&			x shift&							all&			I&		0&					$^{+1.15}_{-0.13}$&					N/A&				shift-and-refit\\
$a_0^{y}$ [cm]&			y shift&							all&			I&		0&					$^{+2.87}_{-0.17}$&					N/A&				shift-and-refit\\
$a_0^{z}$ [cm]&			z shift&							all&			I&		5&					$^{+2.58}_{-0.15}$&					N/A&				shift-and-refit\\
$a_0^{x}$ [cm]&			x shift&							all&			II&		0&					$^{+0.62}_{-0.07}$&					N/A&				shift-and-refit\\
$a_0^{y}$ [cm]&			y shift&							all&			II&		0&					$^{+2.29}_{-0.09}$&					N/A&				shift-and-refit\\
$a_0^{z}$ [cm]&			z shift&							all&			II&		5&					$^{+3.11}_{-0.16}$&					N/A&				shift-and-refit\\
$a_0^{x}$ [cm]&			x shift&							all&			III&		0&					$\pm4.0$&						N/A&				shift-and-refit\\
$a_0^{y}$ [cm]&			y shift&							all&			III&		0&					$\pm4.0$&						N/A&				shift-and-refit\\
$a_0^{z}$ [cm]&			z shift&							all&			III&		5&					$\pm4.0$&						N/A&				shift-and-refit\\ 
$\sigma^{x}$ [cm]&			x resolution&						all&			I&		0&					$\pm3.3$&						N/A&				shift-and-refit\\
$\sigma^{y}$ [cm]&			y resolution&						all&			I&		0&					$\pm2.2$&						N/A&				shift-and-refit\\
$\sigma^{z}$ [cm]&			z resolution&						all&			I&		0&					$\pm1.5$&						N/A&				shift-and-refit\\
$\sigma^{x}$ [cm]&			x resolution&						all&			II&		0&					$\pm3.1$&						N/A&				shift-and-refit\\
$\sigma^{y}$ [cm]&			y resolution&						all&			II&		0&					$\pm3.4$&						N/A&				shift-and-refit\\
$\sigma^{z}$ [cm]&			z resolution&						all&			II&		0&					$\pm5.3$&						N/A&				shift-and-refit\\
$b_0^{x,y}$&				x, y resolution constant term&			all&			III&		0.065&				$\pm0.029$\footnotemark[3]&			N/A&				shift-and-refit\\
$b_1^{x,y}$ [cm$^{-1}$]&		x, y resolution linear term&			all&			III&		$-5.5\times 10^{-5}$&	$\pm6.1\times 10^{-5}$\footnotemark[3]&	N/A&				shift-and-refit\\
$b_2^{x,y}$ [cm$^{-2}$]&		x, y resolution quadratic term&			all&			III&		$3.9\times 10^{-7}$&	$\pm2.0\times 10^{-7}$\footnotemark[3]&	N/A&				shift-and-refit\\
$b_0^z$&					z resolution constant term&			all&			III&		0.0710&				$\pm0.028$\footnotemark[4]&			N/A&				shift-and-refit\\
$b_1^z$ [cm$^{-1}$]&		z resolution linear term&				all&			III&		$1.16\times 10^{-4}$&	$\pm0.83\times 10^{-4}$\footnotemark[4]&N/A&			shift-and-refit\\
$\sigma_{dn}^\rho$ [cm]&		$\rho$ resolution diurnal variation&		all&			I&		0&					$\pm6.82$&						N/A&				shift-and-refit\\
$\sigma_{dn}^\rho$ [cm]&		$\rho$ resolution diurnal variation&		all&			II&		0&					$\pm7.21$&						N/A&				shift-and-refit\\
$\sigma_{dir}^\rho$ [cm]&		$\rho$ resolution directional variation&	CC&			II&		0&					$\pm1.02$\footnotemark[5]&			N/A&				shift-and-refit\\
$\sigma_{dir}^\rho$ [cm]&		$\rho$ resolution directional variation&	ES&			II&		0&					$\pm3.36$\footnotemark[5]&			N/A&				N/A\\
\hline\hline
\end{tabular}
\footnotetext[1]{Correlation of -1.}
\footnotetext[2]{Correlation of -1.}
\footnotetext[3]{Correlation of $\left(
\begin{array}{ccc}
1 &-0.13 &-0.74 \\
-0.13 &1 &0.31 \\
-0.74 &0.31 &1 
\end{array}
\right)$.}
\footnotetext[4]{Correlation of 0.15.}
\footnotetext[5]{Correlation of -1.}
\end{center}
\label{tab:possys}
\end{table*}
\endgroup

During Phases I and II, $x''$, $y''$, and $z''$, respectively, were obtained by convolving $x'$, $y'$, and $z'$ with Gaussians centered at zero with widths of $\sigma^{x}$, $\sigma^{y}$, $\sigma^{z}$, $\sigma_{dn}^\rho$, or $\sigma_{dir}^\rho$. These resolutions were applied first to only day events and then to only night events. $x'$, $y'$, and $z'$ were obtained from
\begin{eqnarray}
x'&=&(1+a_1^\rho-A_{dn}^\rho/2-A_{dir}^\rho/2)(x + a_0^{x})\\
y'&=&(1+a_1^\rho-A_{dn}^\rho/2-A_{dir}^\rho/2)(y + a_0^{y})\\
\nonumber z'&=&(1+a_1^\rho+a_1^{z}-A_{dn}^\rho/2-A_{dir}^\rho/2)\\
&&(z + a_0^{z}).
\end{eqnarray}
The signs of the $A_{dn}^\rho$ and $A_{dir}^\rho$ terms were reversed for night.

During Phase III the uncertainties were applied as
\begin{eqnarray}
x''&=&x' + (b_0^{x,y}+b_1^{x,y}z+b_2^{x,y}z^2)(x'-x_g)\\
y''&=&y' + (b_0^{x,y}+b_1^{x,y}z+b_2^{x,y}z^2)(y'-y_g)\\
z''&=&z' + (b_0^{z}+b_1^{z}z)(z'-z_g),
\end{eqnarray}
where $x_g$, $y_g$, $z_g$ were the true MC positions. $x'$, $y'$, and $z'$ were obtained from
\begin{eqnarray}
x'&=&(1+a_1^\rho)(1-A_{dn}/2-A_{dir}/2)(x + a_0^{x})\\
y'&=&(1+a_1^\rho)(1-A_{dn}/2-A_{dir}/2)(y + a_0^{y})\\
\nonumber z'&=&(1+a_1^\rho+a_1^{z})(1-A_{dn}/2-A_{dir}/2)\\
&&(z+ a_0^{z}).
\end{eqnarray}
The signs of the $A_{dn}^\rho$ and $A_{dir}^\rho$ terms were reversed for night.

\subsection{{\boldmath $\cts{}$} PDFs}

Table~\ref{tab:ctssys} lists the systematic uncertainties on the PDFs of $\cts{}$. For Phases I and II the modified $\cts{}$, $\cts{}'$, was obtained from
\begin{eqnarray}
\cts{}' &=&1+(1+a_0^{\theta}-A_{dir}^{\theta}/2)(\cts{} - 1).
\end{eqnarray}
For Phase III the modified $\cts{}$, $\cts{}'$, was obtained from
\begin{eqnarray}
\cts{}' &=& 1+(1+a_0^{\theta})(1-A_{dir}^{\theta}/2)(\cts{} -1).
\end{eqnarray}
If the transformation moved $\cts{}'$ outside the range $[-1,1]$, $\cts{}'$ was given a random value within this interval.

\begingroup
\squeezetable
\begin{table*}[htdp]
\caption{$\cts{}$ PDF systematic uncertainties.}
\begin{center}
\begin{tabular}{llllrrrr}
\hline\hline
Parameter&		Description&					Events&		Phase&	Nominal&	Variation&					Fit Value&		Application\\
\hline
$a_0^{\theta}$&	$\cts{}$ scale&					ES&			I&		0&		$\pm0.11$&					N/A&						shift-and-refit\\
$a_0^{\theta}$&	$\cts{}$ scale&					ES&			II&		0&		$\pm0.11$&					N/A&						shift-and-refit\\
$a_0^{\theta}$&	$\cts{}$ scale&					ES&			III&		0&		$\pm0.12$&					$0.063^{+0.104}_{-0.099}$&	scanned\\
$A_{dir}^{\theta}$&	$\cts{}$ scale directional variation&	ES&			I&		0&		$\pm0.022$&					N/A&						shift-and-refit\\
$A_{dir}^{\theta}$&	$\cts{}$ scale directional variation&	ES&			II&		0&		$\pm0.052$&					N/A&						shift-and-refit\\
$A_{dir}^{\theta}$&	$\cts{}$ scale directional variation&	ES&			III&		0&		$\pm$0.069&					$-0.015^{+0.073}_{-0.066}$&	scanned\\
\hline\hline
\end{tabular}
\end{center}
\label{tab:ctssys}
\end{table*}
\endgroup

\subsection{{\boldmath $\Encd{}$} PDFs}

Table~\ref{tab:Encdsys} lists the systematic uncertainties on the PDFs of $\Encd{}$. The modified $\Encd{}$, $\Encd{}'$, was obtained from
\begin{eqnarray}
\Encd{}'&=&(1+a_1^{\Encd{}})\Encd{}.
\label{eqn:encdsys}
\end{eqnarray}
For each event in the PDF one hundred random numbers drawn from a Gaussian centered at zero with a width of $b_0^{\Encd{}}\Encd{}$ were used to construct a new PDF.

\begingroup
\squeezetable
\begin{table*}[htdp]
\caption{$\Encd{}$ PDF systematic uncertainties.}
\begin{center}
\begin{tabular}{llllrrrr}
\hline\hline
Parameter&		Description&			Events&		Phase&	Nominal&		Variation\\
\hline
$a_1^{\Encd{}}$&	$\Encd{}$ scale&		n&			III&		0&			$\pm 0.01$\\
$b_0^{\Encd{}}$&	$\Encd{}$ resolution&	n&			III&		0&			$^{+0.01}_{-0.00}$\\
\hline\hline
\end{tabular}
\end{center}
\label{tab:Encdsys}
\end{table*}
\endgroup

\subsection{Background constraints}

Table~\ref{tab:bckphasesIandII} gives the constraints on the backgrounds in Phases I and II. Table~\ref{tab:bckphaseIII} gives the constraints on the backgrounds in Phase III.

\begingroup
\squeezetable
\begin{table*}[htdp]
\caption{Background constraints in Phases I and II. The constraints were all applied to the combined day + night value.}
\begin{center}
\begin{tabular}{llrrrr}
\hline\hline
Background&								Phase&	Constraint&						\multicolumn{2}{c}{Fit Value}&							Application\\
&										&		&								Day&	Night\\
\hline
Internal $\iso{214}{Bi}$ [mBq]&					I&		$126^{+42}_{-25}$&					$64.9^{+7.2}_{-7.1}$&		$96.1^{+6.9}_{-6.9}$&		floated\\
Internal $\iso{208}{Tl}$ [mBq]&					I&		$3.1^{+1.4}_{-1.3}$&				$1.11^{+0.37}_{-0.36}$&		$1.09^{+0.35}_{-0.34}$&		floated\\
External $\iso{214}{Bi}$ [Bq]&					I&		$6.50\pm1.11$&					$11.9^{+4.2}_{-4.2}$&		$2.9^{+3.3}_{-3.4}$&		floated\\
External $\iso{208}{Tl}$ [Bq]&					I&		$0.190^{+0.063}_{-0.054}$&			$0.153^{+0.202}_{-0.199}$&	$0.265^{+0.157}_{-0.157}$&	floated\\
PMT [Arb.]&								I&		N/A&								$0.938^{+0.072}_{-0.071}$&	$1.018^{+0.059}_{-0.058}$&	floated\\
AV surface neutrons [Arb.]&					I&		N/A&								$3.026^{+1.499}_{-1.477}$&	\footnotemark[1]&			floated\\
AV $\iso{214}{Bi}$ [Arb.]&						I&		N/A&								$2.522^{+2.252}_{-2.164}$&	\footnotemark[1]&			floated\\
AV $\iso{208}{Tl}$ [Arb.]&						I&		N/A&								$6.196^{+1.318}_{-1.315}$&	\footnotemark[1]&			floated\\
\hep{} [events]&							I&		15\footnotemark[2]&					N/A&						N/A&						fixed\\
Other $n$ [events]&							I&		$3.2\pm0.8$\footnotemark[3]&			N/A&						N/A&						shift-and-refit\\
Atmospheric $\nu$ [events]&					I&		$21.3\pm4.0$&						N/A&						N/A&						shift-and-refit\\
AV instrumental background [events]&			I&		$0.00^{+24.49}_{-0}$\footnotemark[4]&	N/A&						N/A&						shift-and-refit\\
Internal $\iso{214}{Bi}$ [Arb.]&					II&		N/A&								$0.742^{+0.074}_{-0.074}$&	$0.495^{+0.067}_{-0.067}$&	floated\\
Internal $\iso{208}{Tl}$ [mBq]&					II&		$2.6^{+1.2}_{-1.5}$&				$0.69^{+1.68}_{-1.68}$&		$1.49^{+1.41}_{-1.41}$&		floated\\
Internal $\iso{24}{Na}$ [mBq]&					II&		$0.245\pm 0.060$&					$0.274^{+0.342}_{-0.342}$&	$0.193^{+0.284}_{-0.285}$&	floated\\
External $\iso{214}{Bi}$ [Bq]&					II&		$4.36\pm1.05$&					$4.56^{+3.38}_{-3.35}$&		$5.15^{+2.83}_{-2.86}$&		floated\\
External $\iso{208}{Tl}$ [Bq]&					II&		$0.129\pm0.040$&					$0.216^{+0.159}_{-0.160}$&	$0.071^{+0.135}_{-0.133}$&	floated\\
PMT [Arb.]&								II&		N/A&								$1.093^{+0.053}_{-0.053}$&	$1.244^{+0.049}_{-0.049}$&	floated\\
AV surface neutrons [Arb.]&					II&		N/A&								$-0.359^{+0.473}_{-0.468}$&	\footnotemark[1]&			floated\\
AV $\iso{214}{Bi}$ [Arb.]&						II&		N/A&								$0.821^{+1.486}_{-1.439}$&	\footnotemark[1]&			floated\\
AV $\iso{208}{Tl}$ [Arb.]&						II&		N/A&								$6.218^{+0.981}_{-0.979}$&	\footnotemark[1]&			floated\\
\hep{} [events]&							II&		33\footnotemark[2]&					N/A&						N/A&						fixed\\
Other $n$ [events]&							II&		$12.0\pm3.1$\footnotemark[3]&		N/A&						N/A&						shift-and-refit\\
Atmospheric $\nu$ [events]&					II&		$29.8\pm5.7$&						N/A&						N/A&						shift-and-refit\\
AV instrumental background [events]&			II&		$0.00^{+36.19}_{-0}$\footnotemark[4]&	N/A&						N/A&						shift-and-refit\\
\hline\hline
\end{tabular}
\footnotetext[1]{The fit was performed with day+night combined, so there is only one fit value for both.}
\footnotetext[2]{Fixed at CC=0.35 SSM, ES=0.47 SSM, NC=1.0 SSM~\cite{bs05}.}
\footnotetext[3]{Correlation 1.}
\footnotetext[4]{One-sided, effect is symmetrized.}
\end{center}
\label{tab:bckphasesIandII}
\end{table*}
\endgroup

\begingroup
\squeezetable
\begin{table*}[htdp]
\caption{Background constraints in Phase III. The constraints were applied on the number of events observed in the NCD array. The number of events observed in the PMT array were obtained from the number of events in the NCD array multiplied by the PMT array ratio.}
\begin{center}
\begin{tabular}{lrrrrrr}
\hline\hline
Background&							\multicolumn{2}{c}{NCD Array}&				PMT Array&	\multicolumn{2}{c}{D/N asymmetry}\\
&									Constraint&			Fit Value&				ratio&		Constraint&			Fit Value\\
\hline
External + AV&							$40.9\pm20.6$&		$38.1\pm19.2$&		0.5037&		-0.020$\pm$0.011&		-0.019$\pm$0.011\\
Internal&								$31.0\pm4.7$&			$30.9\pm4.8$&			0.2677&		-0.034$\pm$0.112&		-0.034$\pm$0.112\\
NCD bulk\footnotemark[1]&				$27.6\pm11.0$&		$27.2\pm9.4$&			0.1667&		0&					N/A\\
K2 hotspot&							$32.8\pm5.3$&			$32.6\pm5.2$&			0.2854&		0&					N/A\\
K5 hotspot&							$45.5^{+7.5}_{-8.4}$&	$45.4^{+7.5}_{-8.3}$&	0.2650&		0&					N/A\\
NCD array cables\footnotemark[1]&			$8.0\pm5.2$&			&					0.1407&		0&					N/A\\
Atmospheric $\nu$ and cosmogenic muons&	$13.6\pm2.7$&			$13.4\pm2.7$&			1.8134&		0&					N/A\\
\hline\hline
\end{tabular}
\footnotetext[1]{In the previous analysis of data from Phase III~\cite{cite:snoncd} these two backgrounds were combined.}
\end{center}
\label{tab:bckphaseIII}
\end{table*}
\endgroup

\subsection{PMT background PDF}
\label{apx:PMTPDF}

Table~\ref{tab:PMTbetagammaApx} shows the constraints on the analytical PDF given by Equation~\ref{eqn:PMTPDF} in Section~\ref{sec:PDFs} for PMT background events.

\begingroup
\squeezetable
\begin{table*}[htdp]
\caption{PMT background PDF parameters for the analytical PDF given by Equation~\ref{eqn:PMTPDF} in Section~\ref{sec:PDFs}.}
\begin{center}
\begin{tabular}{lllrrrr}
\hline\hline
Parameter&				Phase&	Day/Night&	Constraint&		Fit Value&						Application\\
\hline
$\epsilon$&				I&		day&			$-6.73\pm1.29$&	$-6.30^{+0.35}_{-0.56}$&			scanned\\
$\epsilon$&				I&		night&		$-5.64\pm1.02$&	$-6.40^{+0.31}_{-0.46}$&			scanned\\
$\epsilon$&				II&		day&			$-6.26\pm0.91$&	$-6.78^{+0.29}_{-0.37}$&			scanned\\
$\epsilon$&				II&		night&		$-6.98\pm0.91$&	$-6.72^{+0.24}_{-0.33}$&			scanned\\
$\eta_1$\footnotemark[1]&	I&		day&			$0\pm1$&			$-0.74^{+1.10}_{-0.54}$&			scanned\\
$\eta_1$\footnotemark[2]&	I&		night&		$0\pm1$&			$-0.39^{+0.39}_{-0.12}$&			scanned\\
$\eta_1$\footnotemark[3]&	II&		day&			$0\pm1$&			$0.74^{+0.42}_{-0.26}$&			scanned\\
$\eta_1$\footnotemark[4]&	II&		night&		$0\pm1$&			$0.31^{+0.26}_{-0.13}$&			scanned\\
$\eta_2$\footnotemark[1]&	I&		day&			$0\pm1$&			$0.09^{+0.62}_{-0.61}$&			scanned\\
$\eta_2$\footnotemark[2]&	I&		night&		$0\pm1$&			$0.08^{+0.77}_{-0.78}$&			scanned\\
$\eta_2$\footnotemark[3]&	II&		day&			$0\pm1$&			$-2.42^{+0.91}_{-0.39}$&			scanned\\
$\eta_2$\footnotemark[4]&	II&		night&		$0\pm1$&			$-3.73^{+0.47}_{-0.49}$&			scanned\\
$\omega_0$&				I&		day&			$0.533\pm0.014$&	$0.5351^{+0.0090}_{-0.0083}$&	scanned\\
$\omega_0$&				I&		night&		$0.533\pm0.014$&	$0.5469^{+0.0071}_{-0.0072}$&	scanned\\
$\omega_0$&				II&		day&			$0.511\pm0.007$&	$0.5096^{+0.0055}_{-0.0047}$&	scanned\\
$\omega_0$&				II&		night&		$0.511\pm0.007$&	$0.5119^{+0.0049}_{-0.0055}$&	scanned\\
$\omega_1$&				I&		day&			$0.237\pm0.051$&	N/A&							shift-and-refit\\
$\omega_1$&				I&		night&		$0.237\pm0.051$&	N/A&							shift-and-refit\\
$\omega_1$&				II&		day&			$0.182\pm0.095$&	N/A&							shift-and-refit\\
$\omega_1$&				II&		night&		$0.182\pm0.095$&	N/A&							shift-and-refit\\
$\beta_s$&				I&		day&			$0.182\pm0.011$&	N/A&							shift-and-refit\\
$\beta_s$&				I&		night&		$0.182\pm0.011$&	N/A&							shift-and-refit\\
$\beta_s$&				II&		day&			$0.195\pm0.007$&	N/A&							shift-and-refit\\
$\beta_s$&				II&		night&		$0.195\pm0.007$&	N/A&							shift-and-refit\\
\hline\hline
\end{tabular}
\footnotetext[1]{$b=-1.00 + 1.29\eta_1$, $\nu = 6.63+0.93*(0.60*\eta_1+\sqrt{1-0.60^2}\eta_2)$}
\footnotetext[2]{$b=3.27 + 12.04\eta_1$, $\nu = 6.78+1.52*(0.96*\eta_1+\sqrt{1-0.96^2}\eta_2)$}
\footnotetext[3]{$b=-0.33 + 2.08\eta_1$, $\nu = 5.32+1.01*(0.91*\eta_1+\sqrt{1-0.91^2}\eta_2)$}
\footnotetext[4]{$b=0.49 + 3.02\eta_1$, $\nu = 5.66+1.07*(0.94*\eta_1+\sqrt{1-0.94^2}\eta_2)$}
\end{center}
\label{tab:PMTbetagammaApx}
\end{table*}
\endgroup

\subsection{Neutron detection efficiencies}
\label{apx:sysOther}

Table~\ref{tab:effSys} shows the constraints on the neutron detection efficiencies. NC interactions observed with the PMT array in Phases I and II were weighted by
\begin{eqnarray}
1+\epsilon_n^{\rm PMT}+\epsilon_{n,{\rm corr.}}^{\rm PMT}.
\end{eqnarray}
NC interactions observed with the PMT array in Phase III were weighted by
\begin{eqnarray}
1+\epsilon_n^{\rm PMT}.
\end{eqnarray}
NC interactions observed with the NCD array in Phase III were weighted by
\begin{eqnarray}
1+\epsilon_n^{\rm NCD}.
\end{eqnarray}
Background neutrons from photo-disintegration observed with the PMT array in Phases I and II were weighted by
\begin{eqnarray}
(1+\epsilon_n^{\rm PMT}+\epsilon_{n,{\rm corr.}}^{\rm PMT})(1+\epsilon_{\rm PD}).
\end{eqnarray}
This uncertainty was already included in the rates of these backgrounds in Table~\ref{tab:bckphaseIII} for Phase III.

\begingroup
\squeezetable
\begin{table*}[htdp]
\caption{Uncertainties in the neutron detection and photo-disintegration backgrounds.}
\setlength{\extrarowheight}{2pt}
\begin{center}
\begin{tabular}{lllrrr}
\hline\hline
Parameter&					Description&						Phase&	Constraint&		Fit Value&			Application\\
\hline
$\epsilon_n^{\rm PMT}$&		Neutron detection in the PMT array&	I&		$0\pm0.0187$&	N/A&				shift-and-refit\\
$\epsilon_n^{\rm PMT}$&		Neutron detection in the PMT array&	II&		$0\pm0.0124$&	N/A&				shift-and-refit\\
$\epsilon_{n,{\rm corr.}}^{\rm PMT}$&	Correlated neutron detection in the PMT array&	I, II&		$0\pm0.007$&		N/A&				shift-and-refit\\
$\epsilon_n^{\rm PMT}$&		Neutron detection in the PMT array&	III&		$0\pm0.028$&		$-0.003\pm0.028$&	float\\
$\epsilon_n^{\rm NCD}$&		Neutron detection in the NCD array&	III&		$0\pm0.024$&		$-0.001\pm0.023$&	float\\
$\epsilon_{\rm PD}$&		Photo-disintegration&				I, II&		$0\pm0.02$&		N/A&				shift-and-refit\\
\hline\hline
\end{tabular}
\end{center}
\label{tab:effSys}
\end{table*}
\endgroup

\section{Neutrino sensitivity}
\label{apx:neutrinoSensitivity}

Table~\ref{table:neutrinoSensitiivty} gives $S(E_\nu)$, the predicted spectrum of $E_\nu$ detectable by the SNO detector after including all effects such as the energy dependence of the cross-sections, reaction thresholds, and analysis cuts, but not including neutrino oscillations. 

\begingroup
\squeezetable
\begin{table*}[htdp]
\caption{Predicted spectrum of $E_\nu$ detectable by the SNO detector after including all effects such as the energy dependence of the cross-sections, reaction thresholds, and analysis cuts, but not including neutrino oscillations, $S(E_\nu)$. The number of events are for all three phases combined and assumes the BS05(OP) solar neutrino model ($\fB{}$=\BPSfive{})~\cite{bs05}.}
\begin{center}
\begin{tabular}{lrrrrrrlrrrrrr}
\hline\hline
$E_\nu$ range&	\multicolumn{2}{c}{CC}&	\multicolumn{2}{c}{ES$_e$}&	\multicolumn{2}{c}{ES$_{\mu\tau}$}&	$E_\nu$ range&	\multicolumn{2}{c}{CC}&	\multicolumn{2}{c}{ES$_e$}&	\multicolumn{2}{c}{ES$_{\mu\tau}$}\\
${\rm [MeV]}$&	day&	night&	day&	night&	day&	night&	[MeV]&	day&	night&	day&	night&	day&	night\\
\hline
2.9-3.0&	0.0&	0.0&	0.0&	0.1&	0.0&	0.0&	9.1-9.2&	185.3&	231.3&	19.9&	24.8&	3.2&	4.0\\
3.0-3.1&	0.0&	0.0&	0.1&	0.1&	0.0&	0.0&	9.2-9.3&	187.7&	235.1&	20.1&	24.6&	3.2&	4.0\\
3.1-3.2&	0.0&	0.0&	0.1&	0.1&	0.0&	0.0&	9.3-9.4&	191.3&	238.8&	20.0&	24.7&	3.2&	4.0\\
3.2-3.3&	0.0&	0.0&	0.1&	0.2&	0.0&	0.0&	9.4-9.5&	194.0&	241.2&	20.1&	24.8&	3.2&	3.9\\
3.3-3.4&	0.0&	0.0&	0.2&	0.2&	0.0&	0.0&	9.5-9.6&	196.4&	244.4&	19.7&	24.4&	3.2&	3.9\\
3.4-3.5&	0.0&	0.0&	0.3&	0.3&	0.0&	0.1&	9.6-9.7&	198.6&	246.1&	19.9&	24.5&	3.2&	3.9\\
3.5-3.6&	0.0&	0.0&	0.4&	0.5&	0.1&	0.1&	9.7-9.8&	199.7&	249.3&	19.5&	24.2&	3.1&	3.9\\
3.6-3.7&	0.0&	0.0&	0.5&	0.6&	0.1&	0.1&	9.8-9.9&	199.8&	250.2&	19.4&	24.0&	3.1&	3.9\\
3.7-3.8&	0.0&	0.0&	0.6&	0.7&	0.1&	0.1&	9.9-10.0&	200.1&	251.1&	19.3&	23.7&	3.1&	3.8\\
3.8-3.9&	0.0&	0.0&	0.8&	0.9&	0.1&	0.2&	10.0-10.1&	200.9&	251.0&	19.0&	23.5&	3.0&	3.7\\
3.9-4.0&	0.0&	0.1&	0.9&	1.2&	0.2&	0.2&	10.1-10.2&	200.8&	250.8&	18.7&	23.1&	3.0&	3.7\\
4.0-4.1&	0.1&	0.1&	1.1&	1.5&	0.2&	0.2&	10.2-10.3&	199.7&	248.7&	18.5&	22.9&	2.9&	3.6\\
4.1-4.2&	0.1&	0.2&	1.4&	1.6&	0.2&	0.3&	10.3-10.4&	198.3&	246.8&	18.1&	22.3&	2.9&	3.6\\
4.2-4.3&	0.3&	0.3&	1.6&	2.0&	0.3&	0.3&	10.4-10.5&	198.0&	246.5&	17.7&	21.9&	2.9&	3.5\\
4.3-4.4&	0.4&	0.5&	1.9&	2.4&	0.3&	0.4&	10.5-10.6&	195.4&	244.5&	17.3&	21.4&	2.8&	3.4\\
4.4-4.5&	0.6&	0.8&	2.2&	2.7&	0.4&	0.4&	10.6-10.7&	193.9&	241.6&	17.0&	21.1&	2.7&	3.4\\
4.5-4.6&	1.0&	1.2&	2.5&	3.1&	0.4&	0.5&	10.7-10.8&	191.7&	237.5&	16.5&	20.6&	2.6&	3.3\\
4.6-4.7&	1.4&	1.8&	2.8&	3.5&	0.5&	0.6&	10.8-10.9&	187.1&	233.2&	16.2&	19.8&	2.6&	3.2\\
4.7-4.8&	2.0&	2.7&	3.1&	3.9&	0.5&	0.6&	10.9-11.0&	184.4&	229.6&	15.9&	19.4&	2.5&	3.1\\
4.8-4.9&	2.8&	3.6&	3.5&	4.5&	0.6&	0.7&	11.0-11.1&	180.5&	226.4&	15.3&	18.8&	2.4&	3.0\\
4.9-5.0&	3.8&	4.8&	3.9&	4.8&	0.6&	0.8&	11.1-11.2&	176.6&	220.0&	14.7&	18.3&	2.3&	2.9\\
5.0-5.1&	4.9&	6.3&	4.4&	5.3&	0.7&	0.9&	11.2-11.3&	172.2&	214.3&	14.2&	17.5&	2.3&	2.8\\
5.1-5.2&	6.4&	8.0&	4.7&	5.9&	0.8&	1.0&	11.3-11.4&	167.8&	208.2&	13.6&	16.8&	2.2&	2.7\\
5.2-5.3&	8.0&	10.2&	5.2&	6.4&	0.8&	1.0&	11.4-11.5&	162.4&	202.7&	13.1&	16.3&	2.1&	2.6\\
5.3-5.4&	10.0&	12.7&	5.7&	6.9&	0.9&	1.1&	11.5-11.6&	157.5&	195.9&	12.6&	15.6&	2.0&	2.5\\
5.4-5.5&	12.3&	15.5&	6.1&	7.6&	1.0&	1.2&	11.6-11.7&	152.2&	188.9&	12.0&	14.9&	1.9&	2.4\\
5.5-5.6&	14.5&	18.7&	6.6&	8.1&	1.0&	1.3&	11.7-11.8&	146.5&	181.8&	11.4&	14.2&	1.8&	2.3\\
5.6-5.7&	17.4&	22.1&	7.1&	8.8&	1.1&	1.4&	11.8-11.9&	139.9&	173.9&	11.0&	13.4&	1.7&	2.1\\
5.7-5.8&	20.3&	26.1&	7.5&	9.2&	1.2&	1.5&	11.9-12.0&	133.7&	166.4&	10.3&	12.7&	1.7&	2.0\\
5.8-5.9&	23.4&	29.9&	8.0&	9.9&	1.3&	1.6&	12.0-12.1&	127.0&	158.4&	9.7&	12.1&	1.5&	1.9\\
5.9-6.0&	26.9&	34.3&	8.4&	10.4&	1.3&	1.7&	12.1-12.2&	121.2&	149.8&	8.9&	11.5&	1.5&	1.8\\
6.0-6.1&	30.8&	39.0&	9.0&	11.2&	1.4&	1.8&	12.2-12.3&	114.0&	141.5&	8.5&	10.6&	1.4&	1.7\\
6.1-6.2&	34.5&	44.0&	9.6&	11.8&	1.5&	1.9&	12.3-12.4&	107.1&	133.2&	8.0&	9.8&	1.3&	1.6\\
6.2-6.3&	38.5&	49.2&	10.0&	12.4&	1.6&	2.0&	12.4-12.5&	100.3&	124.8&	7.4&	9.2&	1.2&	1.5\\
6.3-6.4&	42.5&	54.0&	10.6&	13.1&	1.7&	2.1&	12.5-12.6&	93.3&	116.6&	6.8&	8.6&	1.1&	1.4\\
6.4-6.5&	47.3&	59.4&	11.0&	13.6&	1.8&	2.2&	12.6-12.7&	87.0&	108.3&	6.3&	7.8&	1.0&	1.3\\
6.5-6.6&	51.5&	65.7&	11.5&	14.3&	1.8&	2.3&	12.7-12.8&	80.7&	99.6&	5.9&	7.1&	0.9&	1.1\\
6.6-6.7&	56.3&	71.8&	12.1&	15.0&	1.9&	2.4&	12.8-12.9&	73.3&	91.6&	5.2&	6.5&	0.8&	1.0\\
6.7-6.8&	61.3&	77.2&	12.5&	15.5&	2.0&	2.5&	12.9-13.0&	66.9&	83.0&	4.7&	5.8&	0.8&	0.9\\
6.8-6.9&	66.2&	83.1&	13.1&	16.3&	2.1&	2.6&	13.0-13.1&	60.3&	75.0&	4.3&	5.3&	0.7&	0.8\\
6.9-7.0&	70.5&	89.9&	13.6&	16.6&	2.2&	2.7&	13.1-13.2&	54.3&	67.3&	3.8&	4.7&	0.6&	0.8\\
7.0-7.1&	75.7&	96.7&	14.1&	17.5&	2.3&	2.8&	13.2-13.3&	48.2&	59.9&	3.3&	4.2&	0.5&	0.7\\
7.1-7.2&	81.1&	103.4&	14.5&	18.2&	2.3&	2.9&	13.3-13.4&	42.4&	52.5&	2.9&	3.7&	0.5&	0.6\\
7.2-7.3&	86.8&	109.3&	15.1&	18.5&	2.4&	3.0&	13.4-13.5&	36.9&	45.8&	2.5&	3.1&	0.4&	0.5\\
7.3-7.4&	92.2&	116.8&	15.3&	19.0&	2.5&	3.1&	13.5-13.6&	31.5&	39.6&	2.1&	2.7&	0.3&	0.4\\
7.4-7.5&	98.6&	123.3&	16.0&	19.8&	2.5&	3.1&	13.6-13.7&	27.0&	33.3&	1.8&	2.2&	0.3&	0.4\\
7.5-7.6&	103.6&	130.4&	16.4&	20.3&	2.6&	3.3&	13.7-13.8&	22.6&	27.9&	1.5&	1.9&	0.3&	0.3\\
7.6-7.7&	108.6&	138.2&	16.9&	20.9&	2.7&	3.3&	13.8-13.9&	18.7&	23.3&	1.2&	1.5&	0.2&	0.2\\
7.7-7.8&	115.1&	145.0&	17.1&	21.6&	2.7&	3.4&	13.9-14.0&	15.2&	19.1&	1.0&	1.2&	0.2&	0.2\\
7.8-7.9&	120.5&	152.6&	17.8&	21.7&	2.8&	3.4&	14.0-14.1&	12.2&	15.2&	0.8&	1.0&	0.1&	0.2\\
7.9-8.0&	126.3&	159.7&	17.8&	22.0&	2.9&	3.5&	14.1-14.2&	9.5&	12.0&	0.6&	0.8&	0.1&	0.1\\
8.0-8.1&	131.9&	166.7&	18.2&	22.5&	2.9&	3.6&	14.2-14.3&	7.4&	9.2&	0.5&	0.6&	0.1&	0.1\\
8.1-8.2&	137.7&	172.1&	18.4&	22.8&	3.0&	3.6&	14.3-14.4&	5.7&	7.0&	0.3&	0.4&	0.1&	0.1\\
8.2-8.3&	143.2&	179.2&	18.9&	23.2&	3.0&	3.7&	14.4-14.5&	4.4&	5.2&	0.3&	0.3&	0.0&	0.1\\
8.3-8.4&	148.8&	186.0&	19.0&	23.5&	3.0&	3.7&	14.5-14.6&	3.2&	4.0&	0.2&	0.2&	0.0&	0.0\\
8.4-8.5&	153.7&	193.3&	19.3&	23.9&	3.1&	3.8&	14.6-14.7&	2.4&	3.1&	0.2&	0.2&	0.0&	0.0\\
8.5-8.6&	159.0&	199.2&	19.6&	24.0&	3.1&	3.9&	14.7-14.8&	1.8&	2.1&	0.1&	0.1&	0.0&	0.0\\
8.6-8.7&	164.0&	205.5&	19.6&	24.3&	3.1&	3.9&	14.8-14.9&	1.3&	1.5&	0.1&	0.1&	0.0&	0.0\\
8.7-8.8&	168.2&	211.5&	19.6&	24.6&	3.1&	3.9&	14.9-15.0&	0.9&	1.1&	0.1&	0.1&	0.0&	0.0\\
8.8-8.9&	172.6&	216.6&	20.0&	24.8&	3.2&	3.9&	15.0-15.1&	0.7&	0.8&	0.0&	0.1&	0.0&	0.0\\
8.9-9.0&	177.2&	222.1&	20.0&	24.7&	3.2&	4.0&	15.1-15.2&	0.4&	0.6&	0.0&	0.0&	0.0&	0.0\\
9.0-9.1&	181.6&	226.8&	20.1&	25.1&	3.2&	3.9&	15.2-15.3&	0.2&	0.3&	0.0&	0.0&	0.0&	0.0\\
\hline\hline
\end{tabular}
\end{center}
\label{table:neutrinoSensitiivty}
\end{table*}
\endgroup

\end{document}